\documentclass{dune}
\pdfoutput=1   

\usepackage[pdftex,bookmarks,hidelinks]{hyperref}
\usepackage{authblk}

\graphicspath{ {graphics/} }

\newif\ifdp
\newif\ifsp

\def\expshort{LBNF\xspace}

\def\explong{The Long-Baseline Neutrino Facility\xspace}

\def\thedoctitle{Cryostats and Cryogenics Infrastructure  for the DUNE Far Detector}
\def\thedocsubtitle{LBNF/DUNE}%

\newcommand{\refsec}[2]{Volume~\csname volnumber#1\endcsname \xspace Section~#2}
\newcommand{\refch}[2]{Volume~\csname volnumber#1\endcsname \xspace Chapter~#2}
\newcommand{\refinch}[2]{#2 in Volume~\csname volnumber#1\endcsname \xspace}

\newcommand{\numu}{\ensuremath{\nu_\mu}\xspace}
\newcommand{\nue}{\ensuremath{\nu_e}\xspace}

\newcommand{\numubartonumubar}{
\ensuremath{\overline{\numu}\rightarrow\overline{\numu}}\xspace
}

\def\argon40{${}^{40}$Ar}       
\def\Ar39{$^{39}$Ar}
\def\Cl40{$^{40}$Cl}
\def\K40{$^{40}$K}
\def\B8{$^{8}$B}

\def\fdfiducialmass{\SI{40}{\kt}\xspace}
\def\lartemp{\SI{88}\,K\xspace}
\def\larmass{\SI{17.5}{\kt}\xspace} 

\def\cryostatht{\SI{17.8}{\meter}\xspace} 
\def\cryostatlen{\SI{65.8}{\meter}\xspace} 
\def\cryostatwdth{\SI{18.9}{\meter}\xspace} 

\def\nominalmodsize{\SI{10}{kt}\xspace} 
 
\def\cooldown{cool-down\xspace} 
\def\Cooldown{Cool-down\xspace} 

\def\spmaxdrift{\SI{3.5}{\m}\xspace}

\def\coldbox{cold box\xspace} 

\newcommand{\efield}{E field\xspace}

\newcommand{\threed}{3D\xspace}
\newcommand{\twod}{2D\xspace}
\newcommand{\phel}{photoelectron\xspace} 
\newcommand{\frfour}{FR-4\xspace} 

\newcommand{\lsim}{{\;\raise0.3ex\hbox{$<$\kern-0.75em\raise-1.1ex\hbox{$\sim$}}\;}}
\newcommand{\gsim}{{\;\raise0.3ex\hbox{$>$\kern-0.75em\raise-1.1ex\hbox{$\sim$}}\;}}
\newcommand{\beq}{\begin{equation}}
\newcommand{\eeq}{\end{equation}}
\newcommand{\bea}{\begin{eqnarray}}
\newcommand{\eea}{\end{eqnarray}}

\mathchardef\minus="002D

\newcommand{\rrt}[1]{\ifthenelse{\equal{#1}{}}{[RT:TBD]}{[RT:#1]}}

\newcommand{\lntwo}{LN$_2$\xspace}  

\DeclareSIUnit \s {\second}
\DeclareSIUnit \MB {\mega\byte}
\DeclareSIUnit \GB {\giga\byte}
\DeclareSIUnit \TB {\tera\byte}
\DeclareSIUnit \PB {\peta\byte}
\DeclareSIUnit \Mbps {\mega\bit/\s}
\DeclareSIUnit \Gbps {\giga\bit/\s}
\DeclareSIUnit \Tbps {\tera\bit/\s}
\DeclareSIUnit \Pbps {\peta\bit/\s}
\DeclareSIUnit \kton {\kilo\tonne} 
\DeclareSIUnit \kt {\kilo\tonne}
\DeclareSIUnit \Mt {\mega\tonne}
\DeclareSIUnit \eV {\electronvolt}
\DeclareSIUnit \keV {\kilo\electronvolt}
\DeclareSIUnit \MeV {\mega\electronvolt}
\DeclareSIUnit \GeV {\giga\electronvolt}
\DeclareSIUnit \m {\meter}
\DeclareSIUnit \cm {\centi\meter}
\DeclareSIUnit \in {\inchcommand}
\DeclareSIUnit \km {\kilo\meter}
\DeclareSIUnit \kV {\kilo\volt}
\DeclareSIUnit \kW {\kilo\watt}
\DeclareSIUnit \MW {\mega\watt}
\DeclareSIUnit \MHz {\mega\hertz}
\DeclareSIUnit \mrad {\milli\radian}
\DeclareSIUnit \year {year}
\DeclareSIUnit \POT {POT}
\DeclareSIUnit \sig {$\sigma$}
\DeclareSIUnit\parsec{pc}
\DeclareSIUnit\lightyear{ly}
\DeclareSIUnit\foot{ft}
\DeclareSIUnit\ft{ft}
\DeclareSIUnit \ppb{ppb}
\DeclareSIUnit \ppt{ppt}
\DeclareSIUnit \samples{S}

\usepackage[toc]{glossaries}
\makeglossaries

\newcommand{\dshort}[1]{\glsentrytext{#1}}  

\newcommand{\dfirst}[1]{\glsfirst{#1}\glsunset{#1}}

\newcommand{\dword}[1]{\gls{#1}}
\newcommand{\dwords}[1]{\glspl{#1}}

\newcommand{\newduneword}[3]{
    \newglossaryentry{#1}{
        text={#2},
        long={#2},
        name={\glsentrylong{#1}},
        first={\glsentryname{#1}},
        firstplural={\glsentrylong{#1}\glspluralsuffix},
        description={#3},
        sort={#2}
    }
}

\newcommand{\newduneabbrev}[4]{
  \newglossaryentry{#1}{
    text={#2},
    long={#3},
    shortplural={{#2}\glspluralsuffix},
    longplural={{#3}\glspluralsuffix{}},
    name={\glsentrylong{#1}{} (\glsentrytext{#1}{})},
    first={#3 (#2)},
    firstplural={#3\glspluralsuffix{} (\glsentrytext{#1}\glspluralsuffix{})},
    description={#4},
    sort={#2}
  }
}

\newcommand{\newduneabbrevs}[5]{
  \newglossaryentry{#1}{
    text={#2},
    long={#3},
    plural={#4},
    shortplural={{#2}\glspluralsuffix},
    longplural={#4},
    name={\glsentrylong{#1}{} (\glsentrytext{#1}{})},
    first={#3 (#2)},
    firstplural={#4 (\glsentrytext{#1}\glspluralsuffix{})},
    description={#5},
    sort={#2}    
  }
}

\newduneword{dword}{DUNE Word}{A term in the DUNE lexicon}

\newduneword{nasa}{NASA}{U.S. National Aereonautics and Space Administration}

\newduneabbrev{nd}{ND}{near detector}{Refers to the collection of \gls{dune} detector components 
 installed close to the neutrino source at \gls{fnal}; also a subproject of \gls{usproj} that  includes  installation, infrastructure, and the cryogenics systems for this detector} 

\newduneabbrev{fd}{FD}{far detector}{The \SI{70}{kt} total (\fdfiducialmass fiducial) mass \gls{lartpc} DUNE detector, composed of four \larmass total (\nominalmodsize fiducial) mass modules,  
  to be installed at the far site at \gls{surf} in
  Lead, SD, USA}

\newduneabbrev{sp}{SP}{single-phase}{Distinguishes a \gls{lartpc} technology by the fact that it operates using argon in its liquid phase only; a legacy \gls{dune} term now replaced by \gls{hd} and \gls{vd}} 

\newduneabbrev{dp}{DP}{dual-phase}{Distinguishes a \gls{lartpc} technology by the fact that it operates using argon 
 in both gas and liquid phases; sometimes called double-phase} 

\newduneabbrev{pds}{PDS}{photon detection system}{The detector 
  subsystem sensitive to light produced in the \gls{lar} }

\newduneabbrev{hvs}{HVS}{high voltage system}{The detector 
  subsystem that provides the \gls{tpc} drift field}

\newduneabbrev{tpc}{TPC}{time projection chamber}{Depending on context: (1) A type of particle detector that uses an \efield together with a sensitive volume of gas or liquid, e.g., \gls{lar}, to perform a \threed reconstruction of a particle trajectory or interaction. The activity is recorded by digitizing the waveforms of current
  induced on the anode as the distribution of ionization charge passes by
  or is collected on the electrode. (2) TPC is also used in \gls{usproj} for ``total project cost''} 

\newduneabbrev{lartpc}{LArTPC}{liquid argon time-projection chamber}{A \gls{tpc} filled with liquid argon; 
the basis for the \gls{dune} \gls{fd} modules} 

\newduneabbrevs{apa}{APA}{anode plane assembly}{anode plane assemblies}{A unit of the \gls{sphd}
  detector module containing the elements sensitive to ionization in the \gls{lar}. 
  Each anode face has three planes of wires (two induction, one collection) to provide a 3D view, and interfaces to the cold electronics and photon detection system} 

\newduneabbrev{awg}{AWG}{American wire gauge} {U.S. standard set of non-ferrous wire conductor sizes}

\newduneabbrev{ufer}{Ufer}{concrete encased electrode} {U.S. National Electrical Code grounding method refered to as Concrete Encased Electrode}

\newduneabbrev{cro}{CRO}{charge readout}{The system for detecting
  ionization charge distributions in a 
  detector module} 

\newduneabbrev{lro}{LRO}{light readout}{The system for detecting
  scintillation photons in a \gls{lartpc} detector module}

\newduneabbrev{shv}{SHV}{safe high voltage}{Type of bayonet mount
connector used on coaxial cables that has additional insulation 
compared to standard BNC and MHV connectors that makes it safer
for handling \gls{hv} by preventing accidental contact with the
live wire connector in an unmated connector or plug}

\newduneabbrev{fe}{FE}{front-end}{The front-end refers to a point that is
  ``upstream'' of the data flow for a particular subsystem. 
 For example the \gls{sphd} front-end electronics is where the cold electronics
  meet the sense wires of the TPC and the front-end \gls{daq} is where the \gls{daq} meets the output of the electronics}

\newduneabbrev{ }{DAQ RU}{DAQ readout unit}{The first element in the data flow of the \gls{daq}}

\newduneabbrev{cots}{COTS}{commercial off-the-shelf}{Items, typically hardware such as 
computers, that may be purchased whole, without any custom design or fabrication and 
thus at normal consumer prices and availability}

\newduneabbrev{i2c}{I2C}{Inter-Integrated Circuit}{I$^2$C or I2C is a synchronous, 
multi-master, multi-slave, packet switched, single-ended, serial computer bus widely used 
for attaching lower-speed peripheral ICs to processors and microcontrollers in short-distance, 
intra-board communication} 

\newduneabbrev{spi}{SPI}{Serial Peripheral Interface}{The Serial Peripheral Interface is a 
synchronous serial communication interface specification used for short distance 
communication, primarily in embedded systems}

\newduneabbrev{miso}{MISO}{master in slave out}{The Master In Slave Out is a logic
signal on the \gls{spi} bus on which the data from the slave are transmitted once
a request from the master is received} 

\newduneabbrev{mosi}{MOSI}{master out slave in}{The Master Out Slave In is a logic
signal on the \gls{spi} bus on which the data from the master is transmitted} 

\newduneabbrev{uart}{UART}{Universal Asynchrous Receiver/Transmitter}{A universal 
asynchronous receiver-transmitter is a computer hardware device for asynchronous 
serial communication in which the data format and transmission speeds are configurable}

\newduneword{cr}{CR}{Capacitance-Resistance} 

\newduneword{dc}{DC}{direct coupling} 

\newduneword{ac}{AC}{Alternating Current; when used in the phrase ``AC coupling'' refers to a circuit element that filters out low-frequency components, such as constant offsets, leaving higher frequency signal components. The frequency filtering is determined both by a resistor and a capacitor}

\newduneabbrev{pll}{PLL}{Phase-Locked Loop}{A control system that generates an
output signal whose phase is related to the phase of an input signal}  

\newduneword{fifo}{FIFO}{First-In-First-Out} 

\newduneword{saci}{SACI}{\gls{slac} \gls{asic} Control Interface}

\newduneword{om3}{OM3}{Type of multi-mode fiber optic cable, typically capable of \SI{10}{Gbps} data transmission at lengths up to \SI{300}{m}}

\newduneword{om4}{OM4}{Type of multi-mode fiber optic cable, typically capable of \SI{10}{Gbps} data transmission at lengths up to \SI{550}{m}}

\newduneword{qfp}{QFP}{Quad Flat Package} 

\newduneabbrev{ams}{AMS}{analog and mixed signal}{Verilog-AMS is a derivative of the Verilog hardware description language that includes analog and mixed-signal extensions (AMS) in order to define the behavior of analog and mixed-signal systems}

\newduneabbrev{hepa}{HEPA}{High Efficiency Particulate Air}{The High Efficiency Particulate Air filters are a type of air filter that remove 99.97\% of particles that have a size greater than or equal to \SI{0.3}{\micro\meter}}  

\newduneabbrev{uvm}{UVM}{universal verification methodology}{The Universal Verification Methodology is a standardized methodology for verifying integrated circuit designs}   

\newduneword{lhc}{LHC}{Large Hadron Collider}

\newduneabbrev{lsb}{LSB}{least significant bit}{The bit with the lowest numerical value in a binary number}

\newduneabbrev{ldo}{LDO}{low-dropout regulator}{A low-dropout or LDO regulator is a \gls{dc} linear voltage regulator that can regulate the output voltage even when the supply voltage is very close to the output voltage}

\newduneabbrev{adc}{ADC}{analog-to-digital converter}{A sampling of a voltage
  resulting in a discrete integer count corresponding in some way to
  the input}

\newduneabbrev{inl}{INL}{integral non-linearity}{A commonly used measure of performance in \glspl{adc}. It is the deviation between the ideal input threshold value and the measured threshold level of a certain output code}

\newduneabbrev{dnl}{DNL}{differential non-linearity}{A commonly used measure of performance in \glspl{adc}. The DNL error is defined as the difference between an actual step width and the ideal value of one \gls{lsb}}

\newduneword{pnp}{PNP}{Type of bipolar junction transistor consistning of a
layer of N-doped semiconductor sandwiched between two layers of P-doped material}

\newduneabbrev{spice}{SPICE}{Simulation Program with Integrated Circuit Emphasis}{a general-purpose, 
open-source analog electronic circuit simulator. It is a program used in integrated 
circuit and board-level design to check the integrity of circuit designs and to 
predict circuit behavior} 

\newduneabbrev{daq}{DAQ}{data acquisition}{The data acquisition system
  accepts data from the detector \gls{fe} electronics, buffers
  the data, performs a \gls{trigdecision}, builds events from the selected
  data and delivers the result to the offline \gls{diskbuffer}}

\newduneabbrev{iov}{IOV}{interval of validity}{Interval over which something is valid}

\newduneword{calci}{CALCI}{Calibration and Cryogenic Instrumentation}

\newduneword{detmodule}{far detector module}{The entire DUNE far detector design calls for segmentation into  four modules,  each with a total/fiducial mass of approximately \SI{17}{\kton}/\SI{10}{\kton}}
  
\newduneword{module}{module}{Many aspects of the DUNE far and near detectors are modular, so ``module'' must be understood in context.  It may refer to one of the four far detector modules,  distinct portions of a subdetector as in a ``field cage module,'' a software or electronics module,  e.g., a separate framework plug-in,  and so on}

\newduneword{detunit}{detector unit}{A portion of a \gls{detmodule} may be further partitioned into a number of similar parts.   For example, the \gls{sphd} \gls{tpc} is made up of \gls{apa}  units (and other elements)}

\newduneword{diskbuffer}{secondary DAQ buffer}{A secondary
  \gls{daq} buffer holds a small subset of the full rate as
  selected by a \gls{trigcommand}. 
  This buffer also marks the interface with the DUNE Offline}

\newduneabbrev{om}{OM}{online monitoring}{Processes that run inside
  the \gls{daq} on data ``in flight,'' specifically before landing on the
  offline disk buffer, and that provide feedback on the operation of
  the \gls{daq} itself and the general health of the data it is marshalling}

\newduneabbrev{dqm}{DQM}{data quality monitoring}{Analysis of the raw
  data to monitor the integrity of the data and the performance of the
  detectors and their electronics. This type of monitoring may be
  performed in real time, within the \gls{daq} system, or in later
  stages of processing, using disk files as input}

\newduneword{dumpbuffer}{DAQ dump buffer}{This \gls{daq} buffer
  accepts a high-rate data stream, in aggregate, from an associated
  portion of a \gls{detmodule} sufficient to collect all data likely relevant to
  a potential \gls{snb}}
\newduneabbrev{etf}{ETF}{Experiment Test Framework}{\gls{wlcg} testing middleware that runs grid jobs that actively test distributed sites' services and capabilities,  and reports back to monitoring services}

\newduneabbrev{etl}{ETL}{external trigger logic}{Trigger processing
  that consumes \gls{detmodule} level \gls{trignote} information
  and other global sources of trigger input and emits
  \gls{trigcommand} information back to the \glspl{mtl}}
\newduneabbrev{daqeti}{ETI}{external trigger interface}{Interface between \glspl{mtl} and external source and sinks of relevant trigger information}

\newduneword{trignote}{trigger notification}{Information provided by
  \gls{mtl} to \gls{etl} about \gls{trigdecision} 
  processing}

\newduneword{trigprimitive}{trigger primitive}{Information derived by
  the \gls{daq} \gls{fe} hardware that describes a region of space (e.g.,
  one or several neighboring channels) and time (e.g., a contiguous set
  of \gls{adc} sample ticks) associated with some activity}

\newduneword{externtrigger}{external trigger candidate}{Information
  provided to the \gls{mtl} about events external to a
  \gls{detmodule} so that it may be considered in forming
  \glspl{trigcommand}}

\newduneabbrev{daqoob}{OOB dispatcher}{out-of-band trigger command
  dispatcher}{This component is responsible for dispatching a \gls{snb} dump
  command to all \glspl{daqfer} in the \gls{detmodule}}

\newduneabbrev{mtl}{MTL}{module trigger logic}{Trigger processing
  that consumes \gls{detunit} level \gls{trigcommand} information
  and emits \glspl{trigcommand}. 
  It provides the \gls{etl} with \glspl{trignote} and receives back any
  \glspl{externtrigger}}

\newduneword{octant}{octant}{Any of the eight parts into which 4$\pi$
  is divided by three mutually perpendicular axes. 
  In particular in referencing the value for the mixing angle
  $\theta_{23}$}

\newduneword{trigcandidate}{trigger candidate}{Summary information derived
  from the full data stream and representing a contribution toward
  forming a \gls{trigdecision}}

\newduneword{trigcommand}{trigger command}{Information derived from
  one or more \glspl{trigcandidate}  that directs elements of a
  \gls{detmodule} to read out a portion of the data stream}

\newduneabbrev{tcm}{TCM}{trigger command message}{A message flowing
  down the trigger hierarchy from global to local context.  Also see \gls{tpm}}

\newduneabbrev{mlt}{MLT}{module level trigger}{The \gls{daq} component responsible for producing a \gls{trigdecision} that will be used to command the readout of a detector module}

\newduneword{trigdecision}{trigger decision}{The process by which
  \glspl{trigcandidate} are converted into \glspl{trigcommand}}

\newduneabbrev{tpm}{TPM}{trigger primitive message}{A message flowing
  up the trigger hierarchy from local to global context.  Also see \gls{tcm}}

\newduneabbrev{ipc}{IPC}{inter-process communication}{A system for software elements to exchange information between threads, local processes or across a data network.  An IPC system is typically specified in terms of protocols  composed of message types and their associated data schema}

\newduneword{daqdispre}{discovery and presence}{As used in the context of the \gls{ipc}, a system that provides mechanisms for a node on a communication network to learn of the existence of peers and their identity (discovery) as well as determine if they are currently operational or have become unresponsive (presence)}

\newduneabbrev{pubsub}{PUB/SUB}{publish-subscribe communication pattern}{An \gls{ipc} communication pattern where one element, the publisher, sends data to all connected elements, the subscribers.  Each subscriber may connect to multiple publishers.  A variant is PUB/SUB with topics where a subscriber may register an identifier, the topic, to limit the information received to just an associated subset}

\newduneabbrev{eb}{EB}{event builder}{A software agent that executes \glspl{trigcommand}  for one  \gls{detmodule} by reading out the requested data}

\newduneabbrev{daqdfo}{DFO}{data flow orchestrator}{The process by which trigger commands are executed in parallel and asynchronous manner by the back-end output subsystem of the \gls{daq}}

\newduneabbrev{daqubi}{UBI}{upstream DAQ buffer interface}{The process which provides read-only access to data residing in the upstream \gls{daq} buffers to processes on the network}

\newduneabbrev{cob}{COB}{cluster on board}{An ATCA motherboard housing four RCEs}

\newduneabbrev{rce}{RCE}{reconfigurable computing element}{Data processor located outside of the cryostat on a \gls{cob} that contains \gls{fpga}, RAM and \gls{ssd} resources, responsible for buffering data, producing trigger primitives, responding to triggered requests for data and synching \gls{snb} dumps}

\newduneabbrev{bow}{BOW}{Bump On Wire}{A working name for the front-end readout computing elements used in the nominal \gls{daq} design to interface the \gls{dp}  crates to the \gls{daq} front-end computers} 

\newduneabbrev{atca}{ATCA}{Advanced Telecommunications Computing
  Architecture}{An advanced computer architecture specification developed for the telecommunications, military, and aerospace industries that incorporates the latest trends in  high-speed interconnect technologies, next-generation processors, and improved reliability, availability and serviceability} 

\newduneabbrev{utca}{$\mu$TCA}{Micro Telecommunications Computing Architecture}{The computer architecture specification followed by the crates that house charge and light readout electronics; used in the \gls{spvd} and \gls{dp} technologies} 

\newduneabbrev{udp}{UDP}{user datagram protocol}{A simple,
  connectionless Internet protocol that supports data integrity
  checksums, requires no handshaking, and does not guarantee packet delivery}

\newduneabbrev{amc}{AMC}{advanced mezzanine card}{Holds digitizing
  electronics and lives in \gls{utca} crates}

\newduneabbrev{rf}{RF}{radio frequency}{Electromagnetic emissions
  that are within the (radio) frequency band of sensitivity of the detector
  electronics}

\newduneabbrev{fpga}{FPGA}{field programmable gate array}{An
integrated circuit technology that allows the hardware to be reconfigured to
execute different algorithms after its manufacture and deployment}

\newduneabbrev{fmc}{FMC}{FPGA mezzanine card}{Boards holding \glspl{fpga} and other integrated circuitry that attach to a motherboard}

\newduneabbrev{felix}{FELIX}{Front-End Link eXchange}{A
  high-throughput interface between \gls{fe} and trigger electronics
  and the standard PCIe computer bus}

\newduneword{daqpart}{DAQ partition}{A cohesive and
 coherent collection of \gls{daq} hardware and software working together to trigger and read out some portion of one detector module; it consists of an integral number of \glspl{daqfrag}. 
 Multiple \gls{daq} partitions may operate simultaneously, but each instance operates independently}

\newduneabbrev{fec}{DAQ FEC}{DAQ front-end computer}{The portion of one
  \gls{daqpart} that hosts the \gls{daqdr}, \gls{daqbuf} and
  \gls{daqds}.  It hosts the \gls{daqfer} and corresponding portion of the \gls{daqbuf}}

\newduneword{daqfrag}{DAQ front-end fragment}{The portion of one
  \gls{daqpart} relating to a single \gls{fec} and corresponding to an
  integral number of \glspl{detunit}.  See also \gls{datafrag}}

\newduneword{datafrag}{data fragment}{A block of data read out from a single \gls{daqfrag} that
span a contiguous period of time as requested by a \gls{trigcommand}}

\newduneabbrev{daqfer}{FER}{DAQ front-end readout}{The portion of a
  \gls{daqfrag} that accepts data from the detector electronics and
  provides it to the \gls{fec}}

\newduneabbrev{daqdr}{DDR}{DAQ data receiver}{The portion of the
  \gls{daqfrag} that accepts data from the \gls{daqfer}, emits
  trigger candidates produced from the input trigger primitives, and
  forwards the full data stream to the \gls{daqbuf}}

\newduneword{daqbuf}{DAQ primary buffer}{The portion
  of the \gls{daqfrag} that accepts full data stream from the
  corresponding \gls{detunit} and retains it sufficiently long for it
  to be available to produce a \gls{datafrag}}

\newduneword{daqds}{data selector}{The portion of the \gls{daqfrag}
  that accepts \glspl{trigcommand} and returns the corresponding
  \gls{datafrag}.  Not to be confused with \gls{daqdsn}}

\newduneword{daqdsn}{data selection}{The process of forming a trigger decision for selecting a subset of detector data for output by the \gls{daq} from the content of the detector data itself.  Not to be confused with \gls{daqds}}

\newduneabbrev{daqros}{DAQ RO}{DAQ readout subsystem}{The subsystem of the \gls{daq} for accepting and buffering data input from detector electronics}

\newduneabbrev{daqdss}{DAQ DS}{DAQ data selection subsystem}{The subsystem of the \gls{daq} responsible for forming a trigger decision based on a portion of the input data stream.  The majority subset of the \gls{daqtrs}}

\newduneabbrev{daqtrs}{DAQ TS}{DAQ trigger subsystem}{The subsystem of the \gls{daq} responsible for forming a trigger decision}

\newduneabbrev{daqbes}{DAQ BE}{DAQ back-end subsystem}{The portion of the \gls{daq} that is generally toward its output end.  It is responsible for accepting and executing trigger commands and marshaling the data they address to output storage buffers}

\newduneabbrev{daqtss}{DAQ TSS}{DAQ timing and synchronization subsystem}{The portion of the \gls{daq} that provides for timing and synchronization to various components}

\newduneabbrev{femb}{FEMB}{front-end mother board}{Refers to a unit of
  the \gls{sp} \gls{ce} that contains the \gls{fe} amplifier
  and \gls{adc} \glspl{asic} covering 128 channels}

\newduneword{asic}{ASIC}{application-specific integrated circuit}

\newduneword{lv}{LV}{low voltage}

\newduneabbrev{iceberg}{ICEBERG}{ICEBERG R\&D cryostat and electronics}{Integrated Cryostat and Electronics Built for Experimental Research Goals: a double-walled cryostat built and installed at \gls{fnal}  
for liquid argon detector R\&D and for testing of DUNE detector components}

\newduneword{coldadc}{ColdADC}{A newly developed 16-channels \gls{asic} providing analog to digital conversion}

\newduneword{coldata}{COLDATA}{A 64-channel control and communications \gls{asic}}

\newduneword{cryo}{CRYO}{(1) Integrated ASIC including \gls{fe} circuitry providing signal amplification and pulse shaping, analog to digital conversion, and control and communication functionalities for 64 channels; (2) acryonym for cryogenic systems and cryostat work scopes in \gls{lbnf}}

\newduneword{larasic}{LArASIC}{A 16-channel \gls{fe} \gls{asic} that provides signal amplification and pulse shaping}

\newduneword{cmos}{CMOS}{Complementary metal-oxide-semiconductor}

\newduneabbrev{enc}{ENC}{equivalent noise charge}{The equivalent noise charge is the input charge that corresponds to a \gls{s/n}$=1$}

\newduneword{sar}{SAR}{successive approximation register}

\newduneword{protodune}{ProtoDUNE}{Either of the two initial DUNE prototype detectors constructed at \gls{cern}. 
  One prototype implemented \gls{sp} technology and the other \gls{dp}}
  
\newduneword{protodune2}{ProtoDUNE-II}{The second run of a ProtoDUNE detector}  

\newduneword{pdsp}{ProtoDUNE-SP}{The \gls{sphd} \gls{protodune} detector constructed at \gls{cern}  in \gls{np04}}

\newduneword{pddp}{ProtoDUNE-DP}{The \gls{dp} \gls{protodune} detector constructed at \gls{cern} in \gls{np02}}

\newduneword{wa105}{WA105 DP demonstrator}{The 
3m$\times$1m$\times$1m WA105 \gls{dp} prototype detector at \gls{cern}}

\newduneword{rawevent}{DAQ event block}{The unit of data output by the
  \gls{daq}.  
  It contains trigger and detector data spanning a unique, contiguous
  time period and a subset of the detector channels}

\newduneabbrev{ssd}{SSD}{solid-state disk}{Any storage device that
  may provide sufficient write throughput to receive, both collectively and
  distributed, the sustained full rate of data from a \gls{detmodule}
  for many seconds}
\newduneabbrev{nvme}{NVMe}{Non-volatile memory express}{A specification for an interface to storage media attached via PCIe}

\newduneabbrev{hlt}{HLT}{high-level trigger}{This is actually a filter applied to data that has been triggered and aggregated in order to further reduce or characterize it}

\newduneabbrev{pid}{PID}{particle ID}{Particle identification}

\newduneword{readout window}{readout window}{A fixed, atomic and
  continuous period of time over which data from a \gls{detmodule}, in
  whole or in part, is recorded. 
  This period may differ based on the trigger that initiated the
  readout}

\newduneabbrev{zs}{ZS}{zero-suppression}{Used to delete some portion of a
  data stream that does not significantly deviate from zero or
  intrinsic noise levels. 
  It may be applied at different granularity from per-channel to per
  \gls{detunit}}

\newduneword{rc}{RC}{Depending on context, one of (1) resistive-capacitive (circuit), (2) run control, the system for configuring, starting and terminating the \gls{daq}, or (3) resource coordinator, a member of the \gls{dune} management team responsible for coordinating the financial resources of the project} 

\newduneabbrev{daqccm}{CCM}{DAQ control, configuration and monitoring subsystem}{A system for controlling, configuring and monitoring other systems in particular those that make up the \gls{daq} where the CCM encompasses \gls{rc}}

\newduneword{daqrun}{DAQ run}{A period of time over which relevant data taking conditions and \gls{daq} configuration are asserted to be unchanged. 
  Multiple \gls{daq} runs may occur simultaneously when multiple \glspl{daqpart} are active. 
  This term should not be confused with DUNE experiment or beam ``runs'' that typically span many \gls{daq} runs}
\newduneword{daqrunnum}{DAQ run number}{A monotonically increasing count that uniquely and globally identifies a \gls{daqrun}}

\newduneabbrev{snb}{SNB}{supernova neutrino burst}{A prompt 
  increase in the flux of low-energy neutrinos emitted in the first few seconds of a core-collapse supernova.  It can also refer to a trigger command type that may be due to this phenomenon,
  or detector conditions that mimic its interaction signature}

\newduneabbrev{snble}{SNB/LE}{supernova neutrino burst and low
  energy}{Supernova neutrino burst and low-energy physics program}

\newduneabbrev{snews}{SNEWS}{SuperNova Early Warning System}{A global
  supernova neutrino burst trigger formed by a coincidence of \gls{snb} 
  triggers collected from participating experiments}

\newduneabbrev{pps}{1PPS signal}{one-pulse-per-second signal}{An
  electrical signal with a fast rise time and that arrives in real
  time with a precise period of one second}

\newduneabbrev{sls}{SLS}{spill location system}{A system residing at
  the DUNE far detector site that provides information, possibly
  predictive, indicating periods of time when neutrinos are being
  produced by the \gls{fnal} Main Injector beam spills}

\newduneabbrev{wib}{WIB}{warm interface board}{Digital electronics
  situated just outside a FD cryostat that receives digital data
  from the \glspl{femb} (part of \gls{ce}) over cold copper connections and sends it to the \gls{rce}
  \gls{fe} readout hardware}

\newduneabbrev{gps}{GPS}{Global Positioning System}{A satellite-based system that provides a highly accurate \gls{pps} that may be used to synchronize clocks and determine location}

\newduneabbrev{ntp}{NTP}{Network Time Protocol}{A networking protocol that allows synchronizing of clocks to within a few \si{\milli\second} of a time standard on a local network and within a few tens of \si{\milli\second} over the Internet} 

\newduneword{ptp}{PTP}{Depending on context, either p-terphenyl, a \gls{wls} material, or Precision Time Protocol, a networking protocol that allows synchronizing of clocks to within a few \si{\micro\second} of a time standard on a local network}  

\newduneabbrev{irig}{IRIG}{inter-range instrumentation group}{A standards body that defined a time-code standard for transferring timing information}

\newduneabbrev{nic}{NIC}{network interface controller}{Hardware for controlling the interface to a communication network.  Typically, one that obeys the Ethernet protocol}

\newduneabbrev{wiec}{WIEC}{warm interface electronics crate}{Crates mounted on the signal flanges that contain the \glspl{wib}}

\newduneabbrev{ptc}{PTC}{power and timing card}{Cards that provide further processing and distribution of the signals entering and exiting the \gls{sp} cryostat}

\newduneabbrev{ptb}{PTB}{power and timing backplane}{Backplane used to connect the \glspl{wib} and the \glspl{ptc} on the \gls{wiec}. Also connects the \gls{ce} flange on the cryostat penetration}

\newduneabbrev{sipm}{SiPM}{silicon photomultiplier}{A solid-state
  avalanche photodiode sensitive to single \phel signals}

\newduneabbrev{cisc}{CISC}{cryogenic instrumentation and slow controls}{Includes equipment to monitor all detector  components and  \gls{lar} quality and behavior, and provides a control system for many of the detector components}

\newduneword{fte}{FTE}{full-time equivalent. A unit of labor
  for the project. One year of work from one person}

\newduneword{art}{art}{A software framework implementing an
  event-based execution paradigm} 

\newduneabbrev{sam}{SAM}{sequential
  access via metadata}{A data-handling system to store and retrieve
  files and associated metadata, including a complete record of the
  processing that has used the files}

\newduneword{artdaq}{artdaq}{A data acquisition toolkit for data transfer, aggregation and processing}

\newduneword{beamline}{beamline}{A sequence of control and monitoring devices used for the formation of a directed collection of particles; also subproject within \gls{usproj}} 

\newduneword{cdr}{CDR}{Depending on context, either ``conceptual design report,'' a formal project  document  that describes the experiment at a conceptual level, or ``conceptual design review,'' a formal review of the conceptual design of the experiment or of a component}  

\newduneabbrev{cf}{CF}{conventional facilities}{Pertaining to
  construction and operation of buildings and conventional infrastructure, and includes  cavern excavation}

\newduneabbrev{cp}{CP}{charge conjugation and parity}{Product of charge conjugation and parity
  transformations} 

\newduneabbrev{cpt}{CPT}{charge, parity, and time reversal symmetry}{product of charge, parity
  and time-reversal transformations}

\newduneabbrev{cpv}{CPV}{charge-parity symmetry violation}{Lack of
  symmetry in a system before and after charge and parity
  transformations are applied. 
  For \gls{cp} symmetry to hold,  a particle turns into its
 corresponding antiparticle under a charge transformation,  and a parity
transformation inverts its space coordinates,  i.e. produces the mirror image}

\newduneword{doe}{DOE}{U.S. Department of Energy}

\newduneabbrev{fra}{FRA}{Fermi Research Alliance}{A joint partnership of the University of Chicago and the Universities Research Association (URA) that manages and operates \gls{fnal} on behalf of the \gls{doe}}

\newduneabbrev{dune}{DUNE}{Deep Underground Neutrino Experiment}{A leading-edge, international experiment for neutrino science and proton decay studies; refers to the entire international experiment and collaboration} 

\newduneabbrev{esh}{ES\&H}{environment, safety and health}{A discipline and specialty that studies and implements practical aspects of environmental protection and safety at work} 

\newduneabbrev{ppe}{PPE}{personnel protective equipment}{Equipment worn to minimize exposure to hazards that cause serious workplace injuries and illnesses}

\newduneabbrev{odh}{ODH}{oxygen deficiency hazard}{a hazard that occurs when inert gases such as nitrogen, helium, or argon displace room air and thus reduce the percentage of oxygen below the level required for human life}

\newduneabbrev{feshm}{FESHM}{Fermilab Environment, Safety and Health Manual}{The document that contains \gls{fnal} 's policies and procedures designed to manage environment, safety, and health in all its programs}

\newduneabbrev{fscf}{FSCF}{Far Site Conventional Facilities}{The \gls{cf} at the DUNE far detector site, \gls{surf}, including all detector caverns and support infrastructure}
  
\newduneabbrev{nscf}{NSCF}{Near Site Conventional Facilities}{The \gls{cf} at the DUNE near detector site, \gls{fnal}}

\newduneabbrev{fd1c}{FD1+C}{far detector module 1 + cryogenics}{The first far detector module to be built at \gls{surf}, including integration and installation, and all cryogenics infrastructure to support FD1 and \gls{fd2}; also a subproject of \gls{usproj}} 

\newduneabbrev{fd2}{FD2}{far detector module 2}{The second DUNE far detector module to be built at \gls{surf}} 

\newduneabbrevs{gut}{GUT}{grand unified theory}{grand unified theories}{A class of theories that unifies the electroweak and strong forces}

\newduneabbrev{lar}{LAr}{liquid argon}{Argon in its liquid phase; it is a cryogenic liquid with a boiling point of \SI{87}{K} and density of \SI{1.4}{g/ml}}

\newduneabbrev{lbl}{LBL}{long-baseline}{Refers to the distance between the 
  neutrino source  and the \gls{fd}.  It can also refer to the distance between the near and far detectors. 
  The ``long'' designation is an approximate and relative distinction. For DUNE, this distance  (between \gls{fnal} and \gls{surf}) is approximately \SI{1300}{km}}

\newduneabbrev{lbnf}{LBNF}{Long-Baseline Neutrino Facility}{Long-Baseline Neutrino Facility; refers to the facilities that support the experiment including in-kind contributions under the line-item project. The portion of \gls{usproj} responsible for developing the neutrino beam, the far site cryostats,  and far and near site cryogenics systems, and the conventional facilities, including the excavations } 
  
\newduneabbrev{lbnf-dune}{LBNF/DUNE}{LBNF and DUNE enterprise}{Long-Baseline Neutrino Facility/Deep Underground Neutrino Experiment; refers to the overall enterprise or program including \gls{usproj}, participating international projects, and the \gls{dune} experiment and collaboration} 

\newduneword{usproj}{LBNF/DUNE-US}{Long-Baseline Neutrino Facility/Deep Underground Neutrino Experiment - United States; project to design and build the conventional and beamline facilities and the \gls{doe} contributions to the detectors. It is organized as a \gls{doe}/\gls{fnal} project and incorporates contributions to the facilities from international partners. It also acts as host for the installation and integration of the DUNE detectors} 

\newduneword{duneus}{DUNE-US}{Deep Underground Neutrino Experiment - United States; refers to the U.S. contribution to DUNE under the line-item \gls{usproj} project} 

\newduneabbrev{lbnc}{LBNC}{Long-Baseline Neutrino Committee}{The committee, composed of internationally prominent scientists with relevant expertise, charged by the \gls{fnal} director to review the scientific, technical, and managerial progress, plans and decisions associated with \gls{dune}}

\newduneabbrev{ncg}{NCG}{Neutrino Cost Group}{A group of internationally prominent scientists with relevant experience that is charged by the \gls{fnal} director to review the cost, schedule, and associated risks for the \gls{dune} experiment}

\newduneabbrev{mh}{MH}{mass hierarchy}{Describes the separation
  between the mass squared differences related to the solar and
  atmospheric neutrino problems (also written as \gls{mo})}

\newduneabbrev{mo}{MO}{mass ordering}{See \gls{mh}}

\newduneabbrev{mi}{MI}{Fermilab Main Injector}{An accelerator at
  \gls{fnal} that provides a beam of high-energy protons to the the \gls{usproj}  beamline} 

\newduneabbrev{pot}{POT}{protons on target}{Typically used as a unit
  of normalization for the number of protons striking the neutrino
  production target}

\newduneabbrev{qa}{QA}{quality assurance}{The process of ensuring that 
the quality of each element meets requirements during design and development, and to detect and correct poor results prior to production} 

\newduneabbrev{qc}{QC}{quality control}{The process (e.g., inspection, testing, measurements) 
of ensuring that each manufactured element meets its quality requirements prior to assembly or installation} 

\newduneabbrev{sm}{SM}{Standard Model}{Refers to a theory describing
  the interaction of elementary particles}

\newduneword{tdr}{TDR}{Depending on context, either ``technical design report,'' a formal project  document  that describes the experiment at a technical level, or ``technical design review,'' a formal review of the technical design of the experiment or of a component}  

\newduneabbrev{tp}{IDR}{interim design report}{An intermediate
milestone on the path to a full \gls{tdr}} 

\newduneabbrev{ckm}{CKM matrix}{Cabibbo-Kobayashi-Maskawa
  matrix}{Refers to the matrix describing the mixing between mass and
  weak eigenstates of quarks}

\newduneabbrev{cl}{CL}{confidence level}{Refers to a probability
  used to determine the value of a random variable given its
  distribution}

\newduneabbrev{pmns}{PMNS}{Pontecorvo-Maki-Nakagawa-Sakata}{A type of matrix that describes the mixing between mass and weak eigenstates of
  the neutrino}

\newduneword{hnl}{HNL}{heavy neutral lepton} 

\newduneabbrevs{cpa}{CPA}{cathode plane assembly}{cathode plane assemblies}{The component of the \gls{sphd} detector module that provides the drift HV cathode}

\newduneword{fc}{field cage}{The component of a \gls{lartpc} that contains and shapes the applied \efield}

\newduneword{cpafc}{CPA/FC}{A pair of \gls{cpa} panels and the top and bottom \gls{fc} portions that attach to the pair; an intermediate assembly for installation into the \gls{spmod} }

\newduneabbrev{topfc}{top FC}{top field cage}{The horizontal portions of the \gls{sphd} \gls{fc}   on the top of the \gls{tpc}}

\newduneabbrev{botfc}{bottom FC}{bottom field cage}{The horizontal portions of the \gls{sphd} \gls{fc} on the bottom of the \gls{tpc}}

\newduneabbrev{ewfc}{endwall FC}{endwall field cage}{The vertical portions of the \gls{fc} near the end walls}

\newduneword{gp}{ground plane}{An electrode held electrically neutral relative to Earth ground voltage; it is mounted on the \gls{fc} to protect the cryostat wall}

\newduneword{gg}{ground grid}{An electrode held electrically neutral relative to Earth ground voltage; it is installed between the cathode and the \glspl{pd} in a \gls{dpmod} to protect the \glspl{pmt}, maintaining high transparency to light} 

\newduneabbrev{alara}{ALARA}{as low as reasonably
  achievable}{Typically used with regard management of radiation
  exposure but may be used more generally. It means making every
  reasonable effort to maintain e.g., exposures, to as far below the
  limits as practical, consistent with the purpose for that the
  activity is undertaken}

\newduneabbrev{ecal}{ECAL}{electromagnetic calorimeter}{A detector
  component that measures energy deposition of traversing particles (in the \gls{dune} near detector design)}

\newduneabbrev{hv}{HV}{high voltage}{Generally describes a voltage
  applied to drive the motion of free electrons through some media, e.g., LAr}

\newduneword{spmod}{SP module}{single-phase DUNE \gls{fd} module}  
\newduneword{vdmod}{vertical drift module}{vertical drift DUNE \gls{fd} module} 
\newduneword{dpmod}{DP module}{dual-phase DUNE \gls{fd} module} 
\newduneword{dsp}{DUNE-SP}{a single-phase DUNE far detector module} 

\newduneabbrev{tcoord}{TC}{technical coordinator}{A member of the \gls{dune} management team responsible for organizing the technical aspects of the project effort; is head of \gls{tc}}

\newduneabbrev{tc}{TCN}{technical coordination}{The DUNE organization responsible for overall integration of the detector elements and successful execution of the detector construction project; areas of responsibility include general project oversight, systems engineering, \gls{qa} and safety}  

\newduneabbrev{exb}{EB}{executive board}{The highest level DUNE
  decision-making body for the collaboration}

\newduneabbrev{tb}{TB}{technical board}{The DUNE organization responsible for
  evaluating technical decisions}

\newduneabbrev{rrb}{RRB}{Resources Review Board}{A part of \gls{dune}'s international project governance structure, composed of representatives of all funding agencies that sponsor the project, and of  \gls{fnal} management, established to provide coordination among funding partners and oversight of \gls{dune}}

\newduneabbrev{inc}{INC}{International Neutrino Council}{A highest-level international advisory body to the U.S. \gls{doe} and the  \gls{fnal} directorate on matters related to the  \gls{lbnf} and the  \gls{pip2} projects. This council is composed of representatives from the international funding agencies and  \gls{cern} that make major contributions the infrastructure}

\newduneabbrev{cc}{CC}{charged current}{Refers to an interaction
  between elementary particles where a charged weak force carrier
  ($W^+$ or $W^-$) is exchanged}

\newduneabbrev{dis}{DIS}{deep inelastic scattering}{Refers to interaction between
  elementary particles and a nucleus in an energy range where the
  interaction can be modeled as occurring between constituent quarks
  of one nucleon and resulting in no bulk recoil of the resulting
  nucleus} 

\newduneabbrev{fsi}{FSI}{final-state interactions}{Refers to
  interactions between elementary or composite particles subsequent to
  the initial, fundamental particle interaction, such as may occur as
  the products exit a nucleus}
  
\newduneabbrev{fsint}{FSI}{far site integration}{The scope of work at the \gls{fs} for the \gls{integoff}} 

\newduneword{geant4}{Geant4}{A
  software toolkit for the simulation of the passage of particles
  through matter using \gls{mc} methods}

\newduneabbrev{genie}{GENIE}{Generates Events for Neutrino Interaction
  Experiments}{Software providing an object-oriented neutrino
  interaction simulation resulting in kinematics of the products of
  the interaction}

\newduneabbrev{mc}{MC}{Monte Carlo}{Refers to a method of numerical
  integration that entails the statistical sampling of the integrand
  function. 
  Forms the basis for some types of detector and physics simulations}

\newduneabbrev{qe}{QE}{quasi-elastic}{Refers to the 
  interaction of an elementary charged particle with a nucleus in an
  energy range where the interaction can be modeled as taking place with
  individual nucleons} 

\newduneabbrev{mou}{MoU}{memorandum of understanding}{A project management methodology that  \gls{usproj} uses to document agreement,s,  e.g., between \gls{fnal} and the Project, for how \gls{fnal}  will support the Project. More generally, a document
  summarizing an agreement between two or more parties} 

\newduneabbrev{pip2}{PIP-II}{Proton Improvement Plan II}{A \gls{fnal} project for
  improving the protons on target delivered delivered by the \gls{lbnf} neutrino production beam. 
  This is version two of this plan and it is planned to be followed by a PIP-III}
  
\newduneabbrev{sdsta}{SDSTA}{South Dakota Science and Technology
  Authority}{The legal entity that manages \gls{surf}, in Lead, S.D}
  
\newduneabbrev{sdsd}{SDSD}{Fermilab South Dakota Services Division}{A \gls{fnal}  division responsible providing host laboratory functions at SURF in South Dakota}

\newduneabbrev{firus}{FIRUS}{Facility Information Reporting Utility System}
 {Facility incident reporting systems, one at \gls{fnal} and at \gls{surf}, that monitors and reports the status of various fire, security and utility sensors} 

\newduneabbrev{bsi}{BSI}{building and site infrastructure}
 {The work package for outfitting of the \gls{lbnf} underground infrastructure}

\newduneabbrev{wbs}{WBS}{work breakdown structure}{An organizational
  project management tool by which the tasks to be performed are
  partitioned in a hierarchical manner}

\newduneabbrev{br}{BR}{branching ratio}{A fractional probability for a
  decay of a composite particle to occur into some specified set or
  sets of products}
\newduneword{bsm}{BSM}{beyond the Standard Model}

\newduneabbrev{dm}{DM}{dark matter}{The term given to the unknown
  matter or force that explains measurements of galaxy motion 
  that are otherwise inconsistent with the amount of mass associated
  with the observed amount of photon production}
  
\newduneabbrev{bdm}{BDM}{boosted dark matter}{A new model that describes a relativistic dark matter particle boosted by the annihilation of heavier dark matter particles in the galactic center or the sun}

\newduneabbrev{cern}{CERN}{European Laboratory for Particle Physics}{The leading particle physics laboratory in Europe and home to the \glspl{protodune} and other prototypes and demonstrators, including the \glspl{mod0}} 

\newduneabbrev{dsnb}{DSNB}{diffuse supernova neutrino background}{The
  term describing the pervasive, constant flux of neutrinos due to all
  past supernova neutrino bursts}

\newduneabbrev{espp}{ESPP}{European Strategy for Particle Physics}{The
cornerstone of Europe's
decision-making process for the long-term future of the
field. Mandated by the \gls{cern} Council, it is formed through a broad
consultation of the grass-roots particle physics community, it
actively solicits the opinions of physicists from around the world,
and it is developed in close coordination with similar processes in
the USA and Japan in order to ensure coordination between regions and
optimal use of resources globally}

\newduneabbrev{gar}{GAr}{gaseous argon}{argon in its gas phase}
\newduneabbrev{gartpc}{GArTPC}{gaseous argon time-projection chamber}{A \gls{tpc} filled with gaseous argon} 

\newduneabbrev{globes}{GLoBES}{General Long-Baseline Experiment
  Simulator}{A software package for simulating energy spectra of
  neutrino flux, interactions, and energy spectra measured after application of some
  model of a detector response)}

\newduneabbrev{snowglobes}{SNOwGLoBES}{SuperNova
Observatories with GLoBES} {From the official description: SNOwGLoBES is public software for computing interaction rates and distributions of observed quantities for \gls{snb} neutrinos in common detector materials} 

\newduneword{l/e}{L/E}{length-to-energy ratio}
\newduneword{lri}{LRI}{long-range interactions}

\newduneabbrev{nc}{NC}{neutral current}{Refers to an interaction
  between elementary particles where a neutrally charged weak force carrier
  ($Z^0$) is exchanged}

\newduneabbrev{nh}{NH}{normal hierarchy}{Refers to the neutrino mass
  eigenstate ordering whereby the sign of the mass squared difference
  associated with the atmospheric neutrino problem is positive}

\newduneabbrev{ih}{IH}{inverted hierarchy}{Refers to the neutrino mass
  eigenstate ordering whereby the sign of the mass squared difference
  associated with the atmospheric neutrino problem is negative}

\newduneabbrev{no}{NO}{normal ordering}{Refers to the neutrino mass
  eigenstate ordering whereby the sign of the mass squared difference
  associated with the atmospheric neutrino problem is positive}

\newduneabbrev{io}{IO}{inverted ordering}{Refers to the neutrino mass
  eigenstate ordering whereby the sign of the mass squared difference
  associated with the atmospheric neutrino problem is negative}

\newduneabbrev{msw}{MSW}{Mikheyev-Smirnov-Wolfenstein effect}{Explains
  the oscillatory behavior of neutrinos produced inside the sun as
  they traverse the solar matter}

\newduneabbrev{nsi}{NSI}{nonstandard interaction}{A general class of
  theory of elementary particles other than the Standard Model}

\newduneabbrev{pfive}{P5}{Particle Physics Project Prioritization
Panel}{The Particle Physics Project Prioritization Panel (P5) was a
subpanel of the High Energy Physics Advisory Panel (HEPAP). It completed
its Report, a ten-year strategic plan for high energy physics in the
U.S., in 2014. This report included a recommendation that ``host a world-leading neutrino
program that will have an optimized set of short- and long-baseline neutrino oscillation experiments, and its long-term focus
is a reformulated venture referred to here as the Long Baseline
Neutrino Facility (LBNF)''}

\newduneabbrev{sme}{SME}{standard-model extension}{an effective field theory that contains the \gls{sm}, general relativity, and all possible operators that break Lorentz symmetry (Wikipedia)}

\newduneabbrev{susy}{SUSY}{supersymmetry}{Theoretical symmetry between a fermion and a boson}

\newduneabbrev{wimp}{WIMP}{weakly-interacting massive particle}{A
  hypothesized particle that may be a component of dark matter}

\newduneabbrev{ce}{CE}{cold electronics}{Analog and digital readout electronics that operate at cryogenic temperatures}

\newduneabbrev{crp}{CRP}{charge-readout plane}{An anode technology using a stack of perforated \glspl{pcb} with etched electrode strips to provide \gls{cro} in 3D; it has two induction layers and one collection layer; it is used in the \gls{spvd} \gls{fd} and \gls{dp} designs} 

\newduneabbrev{dram}{DRAM}{dynamic random access memory}{A computer memory technology}

\newduneabbrev{fnal}{Fermilab}
{Fermi National Accelerator Laboratory}{U.S. national laboratory in Batavia, IL.  It is the laboratory that hosts \gls{lbnf} and \gls{dune}, and serves as the experiment's near site}

\newduneabbrev{bnl}{BNL}{Brookhaven National Laboratory}{US national laboratory in Upton, NY}

\newduneabbrev{slac}{SLAC}{SLAC National Accelerator Laboratory}{US national laboratory in Menlo Park, CA}

\newduneabbrev{lbnl}{LBNL}{Lawrence Berkeley National Laboratory}{US national laboratory in Berkeley, CA}

\newduneabbrev{anl}{ANL}{Argonne National Laboratory}{US national laboratory in Lemont, IL}

\newduneabbrev{lanl}{LANL}{Los Alamos National Laboratory}{US national laboratory in Los Alamos, NM}

\newduneword{fs}{FS}{Depending on context, one of (1) the far site, \gls{surf}, where the DUNE far detector is located; (2) ``Full Stream'' relates to a data stream that has not undergone selection, compression or other form of reduction} 

\newduneabbrev{lem}{LEM}{large electron multiplier}{A micro-pattern detector suitable for use in ultra-pure argon vapor; LEMs consist of copper-clad PCB boards with sub-millimeter-size holes through which electrons undergo amplification}

\newduneabbrev{lng}{LNG}{liquefied natural gas}{Pertaining to natural gas in its liquid phase}

\newduneabbrev{mip}{MIP}{minimum ionizing particle}{Refers to a
  particle traversing some medium such that the particle's mean energy loss is  
  near the minimum}

\newduneabbrev{pd}{PD}{photon detector}{The detector
  elements involved in measurement of the number and arrival times of
  optical photons produced in a detector module} 

\newduneabbrev{pmt}{PMT}{photomultiplier tube}{A device that makes use
  of the photoelectric effect to produce an electrical signal from the
  arrival of optical photons}

\newduneabbrev{ppm}{ppm}{parts per million}{A concentration equal to one part in $10^{6}$}
\newduneabbrev{ppb}{ppb}{parts per billion}{A concentration equal to one part in $10^{9}$}
\newduneabbrev{ppt}{ppt}{parts per trillion}{A concentration equal to one part in $10^{12}$}

\newduneword{rio}{RIO}{reconfigurable input output}

\newduneabbrev{s/n}{S/N}{signal-to-noise}{signal-to-noise ratio}

\newduneword{ssp}{SSP}{\gls{sipm} signal processor}

\newduneabbrev{sbn}{SBN}{Short-Baseline Neutrino}{A \gls{fnal} program consisting of three collaborations, \gls{microboone}, \gls{sbnd}, and \gls{icarus}, to perform sensitive searches for $\nue$ appearance and $\numu$ disappearance in the Booster Neutrino Beam}

\newduneword{stt}{STT}{straw tube tracker}

\newduneword{wire board}{wire board}{At the head end of the APA in the \gls{sphd} \gls{tpc}, stacks of electronics boards referred to as ``wire boards'' are arrayed to anchor the wires.  They also provide the connection between the wires and the cold electronics} 

\newduneabbrev{wls}{WLS}{wavelength-shifting}{A material or process by
  which incident photons are absorbed by a material and photons are
  emitted at a different, typically longer, wavelength}
  
\newduneabbrev{tpb}{TPB}{tetra-phenyl butadiene}{A \gls{wls} material}

\newduneabbrev{sft}{SFT}{signal feedthrough}{A cryostat penetration allowing for the passage of cables or other extended parts}

\newduneabbrev{sftchimney}{SFT chimney}{signal feedthrough chimney}{A volume above the cryostat penetration used for a signal feedthrough} 

\newduneabbrev{catiroc}{CATIROC}{charge and time integrated readout chip}{A complete read-out chip manufactured in AustriaMicroSystem designed to read arrays of 16 photomultipliers}

\newduneabbrev{wr}{WR}{White Rabbit}{A component of the timing system that forwards clock signal and time-of-day reference data to the master timing unit}

\newduneabbrev{mch}{MCH}{MicroTCA Carrier Hub}{A network switching device}

\newduneabbrev{wrmch}{WR-MCH}{White Rabbit \gls{utca} Carrier Hub}{A card mounted in \gls{utca} crate that recieves time syncronization information and trigger data packets over \gls{wr} network and disributes them to the \gls{amc} over \gls{utca} backplane} 

\newduneabbrev{wrtsn}{WR-TSN}{White Rabbit TimeStamping Node}{A unit on the \gls{wr} network that timestamps the trigger signals and sends out trigger data packets to \gls{wrmch}}

\newduneword{qap}{QAP}{quality assurance plan} 
\newduneword{ieshp}{IESHP}{integrated environmental, safety and health plan}
\newduneword{dmp}{DMP}{data management plan} 
\newduneword{qam}{QAM}{quality assurance manager} 

\newduneabbrev{dss}{DSS}{detector support system}{The system of rails suspended from the cryostat ceiling in a \gls{spmod} used to support the \gls{apa}s, \gls{cpa}s, and the \glspl{ewfc}} 

\newduneabbrev{ddss}{DDSS}{DUNE detector safety system}{The hardware system responsible for the safety of the detector, implemented either via a \gls{plc} or via custom hardware protections} 

\newduneabbrev{tco}{TCO}{temporary construction opening}{An opening in the side of a cryostat through which detector elements are brought into the cryostat; utilized during construction and installation}

\newduneabbrev{surf}{SURF}{Sanford Underground Research Facility}{The laboratory in Lead, SD, USA where the \gls{dune} \gls{fd} will be installed and operated; also where the \gls{usproj} \gls{fscf} and the \gls{fs} cryostat and cryogenics systems will be constructed} 

\newduneabbrev{uit}{UIT}{underground installation team}{An organizational unit responsible for installation in the underground area at the \gls{surf} site}

\newduneabbrev{cmgc}{CMGC}{construction manager/general contractor}{The contracted company hired to manage overall construction, used by \gls{lbnf} at the \gls{surf} site for the \gls{fscf} construction} 

\newduneword{pdr}{PDR}{Depending on context, either ``preliminary design report,'' a formal project document  that describes the experiment at a preliminary level, or ``preliminary design review,'' a formal review of the preliminary design of the experiment or of a component} 

\newduneword{fdr}{FDR}{Depending on context, either ``final design report,'' a formal project document  that describes the experiment at a final level, or ``final design review,'' a formal review of the final design of the experiment or of a component} 

\newduneabbrev{prr}{PRR}{production readiness review}{A project management mechanism by which the production readiness is reviewed}  

\newduneabbrev{irr}{IRR}{installation readiness review}{A project management mechanism by which the plan for installation is reviewed}  
\newduneabbrev{orr}{ORR}{operational readiness review}{A project management mechanism by which the operational readiness is reviewed}  

\newduneabbrev{ppr}{PPR}{production progress review}{A project management mechanism by which the progress of production is reviewed}  

\newduneabbrev{edms}{EDMS}{engineering document management system}{A computerized document management system developed and supported at the \gls{cern} in which some \gls{dune} project and collaboration documents, drawings and engineering models are managed} 
\newduneabbrev{docdb}{DocDB}{Document DataBase}{A computerized document management system developed and supported at \gls{fnal} in which most \gls{lbnf-dune} documents are managed (docs.dunescience.org); some documents are maintained in \gls{edms}}

\newduneword{wrgm}{WR grandmaster}{White Rabbit grandmaster}

\newduneabbrev{larsoft}{LArSoft}{Liquid Argon Software}{A shared base of physics software across \gls{lartpc} experiments}
\newduneword{nova}{NOvA}{The \gls{nova} off-axis neutrino oscillation experiment at \gls{fnal}}
\newduneword{minerva}{MINERvA}{Neutrino cross sections experiment at \gls{fnal},  shut down in 2019}
\newduneword{microboone}{MicroBooNE}{A \gls{lartpc} neutrino oscillation experiment at \gls{fnal}}
\newduneword{sbnd}{SBND}{The Short-Baseline Near Detector experiment at  \gls{fnal}}
\newduneabbrev{nexo}{nEXO}{Enriched Xenon Observatory}{Experiment at Lawrence Livermore National Laboratory (U.S. national lab in Livermore, CA) searching for new physics with neutrinoless double-beta decay}
\newduneword{argoneut}{ArgoNeuT}{The ArgoNeuT test-beam experiment and \gls{lartpc} prototype at  \gls{fnal}}
\newduneword{icarus}{ICARUS}{A neutrino experiment that was located at the Laboratori Nazionali del Gran Sasso (LNGS) in Italy, then refurbished at \gls{cern} for re-use in the same neutrino beam from \gls{fnal} used by the \gls{miniboone} , \gls{microboone} and \gls{sbnd} experiments at \gls{fnal}}
\newduneword{atlas}{ATLAS}{One of two general-purpose detectors at the \gls{lhc}. It investigates a wide range of physics, from the measurements of the Higgs boson properties to searches for extra dimensions and particles that could make up \gls{dm}}

\newduneword{lbne}{LBNE}{Long Baseline Neutrino Experiment; (1) a terminated U.S. experiment that was reformulated in 2014 under the auspices of the new \gls{dune} collaboration, an internationally coordinated and internationally funded program, with \gls{fnal} as host; and (2) the former incarnation of \gls{usproj} } 

\newduneabbrev{lbno}{LBNO}{Long Baseline Neutrino Observatory} {A terminated European project that, during its six-year duration, assessed the feasibility of a next-generation deep underground neutrino observatory in Europe)}

\newduneword{wirecell}{Wire-Cell}{A tomographic automated \threed neutrino event reconstruction method for \glspl{lartpc}}
\newduneabbrev{wct}{WCT}{Wire-Cell Toolkit}{A software toolkit with data flow processing components for \gls{lartpc} noise and signal simulation, noise filtering, signal processing, and tomographic \threed ionization activity imaging}
\newduneword{ftslite}{F-FTS-lite}{Light-weight version of the \gls{fnal} File Transfer system used for rapid data transfers out of the online systems}
\newduneabbrev{fts}{FTS}{File Transfer System}{A file transfer system developed at \gls{fnal} to catalog and move data to permanent storage}

\newduneword{35t}{35 ton prototype}{A prototype cryostat and \gls{sp} detector built at \gls{fnal} before the \gls{protodune} detectors}

\newduneabbrev{mcr}{MCR}{main communications room}{Space at the \gls{fd} site for cyber infrastructure}

\newduneabbrev{cuc}{CUC}{central utility cavern}{The utility cavern at the 4850L of \gls{surf} located between the two detector caverns. It contains utilities such as central cryogenics and other systems, and the underground data center and control room}

\newduneabbrev{cfd}{CFD}{computational fluid dynamics}{High performance computer-assisted modeling of fluid dynamical systems}
\newduneword{vuv}{VUV}{vacuum ultra-violet}
\newduneword{tallbo}{TallBo}{A cylindrical cryostat at \gls{fnal} primarily used for developing scintillation light collection technologies for \gls{lartpc} detectors}

\newduneword{root}{ROOT}{A modular scientific software toolkit. It provides all the functionalities needed to deal with big data processing,  statistical analysis, visualization, and storage.  It is mainly written in C++ but integrated with other languages such as Python and R}

\newduneabbrev{eos}{EOS}{EOS}{The XRootD-based distributed file system developed by CERN}
\newduneabbrev{ehn1}{EHN1}{Experiment Hall North One}{Location at \gls{cern} of the \gls{np02} and \gls{np04} areas used for the \glspl{protodune} and for other test and prototyping activities for DUNE} 

\newduneword{led}{LED}{Light-emitting diode}
\newduneabbrev{rtd}{RTD}{resistance temperature detector}{A temperature sensor consisting of a material with an accurate and reproducible resistance/temperature relationship}
\newduneword{swc}{SWC}{Software \& Computing}
\newduneabbrev{las}{LAS}{LEM-anode Sandwich}{In the \gls{dp} technology, a \gls{lem} and its corresponding anode are mounted together in a module called a LEM-anode sandwich}

\newduneword{roi}{ROI}{region of interest}
\newduneabbrev{hpc}{HPC}{high-performance computing}{high-performance computing facilities; generally computing facilities emphasizing parallel computing with aggregate power of more than a teraflop}

\newduneword{comfund}{common fund}{The shared resources of the collaboration}
\newduneabbrev{ims}{IMS}{integrated master schedule}{A project management device consisting of linked tasks and milestones}

\newduneword{hvdb}{HVDB}{HV divider board}

\newduneword{hvft}{HVFT}{HV feedthrough}  

\newduneword{sas}{SAS}{Another term for the materials airlock; a pass-through chamber used to ensure safe transfer of materials into a clean room, avoiding contamination in both directions}

\newduneabbrev{fea}{FEA}{finite element analysis}{Simulation of a physical phenomenon using the numerical technique called Finite Element Method (FEM), a numerical method for solving problems of engineering and mathematical physics}

\newduneword{fss}{FSS}{field shaping strips}
\newduneword{lvds}{LVDS}{low-voltage differential signaling}

\newduneword{esd}{ESD}{electrostatic discharge}

\newduneabbrev{rp}{RP}{resistive panel}{Resistive panels form the constant potential surfaces for a \gls{spmod} \gls{cpa}; they are composed of a thin layer of carbon-impregnated Kapton and laminated to both sides of a \frfour sheet}

\newduneword{uhmwpe}{UHMWPE}{ultra-high molecular weight polyethylene}

\newduneword{cts}{CTS}{Cryogenic Test System}

\newduneabbrev{plc}{PLC}{programmable logic controller}{An industrial digital computer that has been ruggedized and adapted for the control of manufacturing or other processes that require high reliability, ease of programming, and process fault diagnosis} 

\newduneword{mppc}{MPPC}{\SI{6}{mm}$\times$\SI{6}{mm} Multi-Pixel Photon Counters produced by Hamamatsu\texttrademark{} Photonics K.K}

\newduneabbrev{sfp}{SFP}{small form-factor pluggable}{a particular standard for optical transceivers}

\newduneabbrev{minipod}{MiniPOD}{miniature parallel optical device}{a family of types of multi-channel optical transceivers}

\newduneword{ccc}{CCC}{configuration change command}
\newduneword{ccondc}{CCC}{code of conduct committee}

\newduneword{act}{ACT}{activation time stamp}
\newduneword{lcm}{LCM}{light calibration module}
\newduneword{lpm}{LPM}{light pulser module}
\newduneword{dac}{DAC}{digital-to-analog converter}
\newduneword{arapuca}{ARAPUCA}{A \gls{pds} design that consists of a light trap that captures wavelength-shifted photons inside boxes with highly reflective internal surfaces until they are eventually detected by \gls{sipm} detectors or are lost}
\newduneword{sarapu}{S-ARAPUCA}{Standard \gls{arapuca} design with different \gls{wls} coatings on both faces of the dichroic filter window(s) of the cell}
\newduneword{xarapu}{X-ARAPUCA}{Extended \gls{arapuca} design with \gls{wls} coating on only the external face of the dichroic filter window(s) but with a \gls{wls} doped plate inside the cell}
\newduneword{feb}{FEB}{front-end board}

\newduneabbrev{lsnd}{LSND}{Liquid Scintilator Neutrino Detector}{A scintillation detector and associated experiment located at Los Alamos National Laboratory}

\newduneabbrev{cvn}{CVN}{convolutional visual network}{An algorithm for identifying neutrino interactions based on their topology and without the need for detailed reconstruction algorithms}

\newduneword{pandora}{Pandora}{The Pandora multi-algorithm approach to pattern recognition} 

\newduneabbrev{pma}{PMA}{Projection Matching Algorithm}{A reconstruction algorithm that combines \twod reconstructed objects to form a \threed representation}
\newduneabbrev{bdt}{BDT}{boosted decision tree}{A method of multivariate analysis}
\newduneabbrev{cnn}{CNN}{convolutional neural network}{A deep learning technique most commonly applied to analyzing visual imagery}
\newduneword{pdg}{PDG}{Particle Data Group}

\newduneword{pci}{PCI}{Peripheral Component Interconnect}

\newduneword{labview}{LabVIEW}{Laboratory Virtual Instrument Engineering Workbench is a system-design platform and development environment for a visual programming language from National Instruments}

\newduneword{pcb}{PCB}{printed circuit board}

\newduneword{crio}{cRIO}{Compact Reconfigurable Input Output}

\newduneword{dcs}{DCS}{Distributed Communications System}

\newduneword{opc-ua}{OPC-UA}{OPC  Unified Architecture is a machine to machine communication protocol for industrial automation developed by the OPC Foundation. OPC stands for Object Linking and Embedding for Process Control}

\newduneword{cabangle}{Cabibbo angle}{A quark mixing parameter that governs the coupling of up quarks to strange quarks}
\newduneword{valor}{VALOR}{A neutrino oscillation fitting framework that is used by \gls{t2k}; the name stands for VALencia-Oxford-Rutherford, the original three institutions that developed it}
\newduneword{cafana}{CAFAna}{Common Analysis File Analysis}
\newduneabbrev{pca}{PCA}{principal component analysis}{A statistical procedure that uses an orthogonal transformation to convert a set of observations of possibly correlated variables into a set of values of linearly uncorrelated variables called principal components (Wikipedia)}
\newduneword{numi}{NuMI}{a set of facilities at \gls{fnal}, collectively called ``Neutrinos at the Main Injector.''  The NuMI neutrino beamline target system converts an intense proton beam into a focused neutrino beam}
\newduneword{gibuu}{GiBUU}{Giessen Boltzmann-Uehling-Uhlenback Project; a unified theory and transport framework in the MeV and GeV energy regimes for elementary reactions on nuclei }
\newduneabbrev{rpa}{RPA}{random phase approximation} {an approximation method commonly used for describing the dynamic linear electronic response of electron systems (Wikipedia)}
\newduneword{t2k}{T2K}{T2K (Tokai to Kamioka) is a long-baseline neutrino experiment in Japan studying neutrino oscillations}
\newduneword{mptdet}{MPT detector}{multipurpose tracking detector}

\newduneword{lariat}{LArIAT}{The repurposed ArgoNeuT \gls{lartpc}, modified for use in a charged particle beam, dedicated to the calibration and precise characterization of the output response of these detectors}

\newduneword{captain}{CAPTAIN}{Experimental program sited at \gls{lanl} that is designed to make measurements of scientific importance to \gls{lbl} neutrino physics and physics topics that will be explored by large underground detectors}

\newduneword{dayabay}{Daya Bay}{a neutrino-oscillation experiment in Daya Bay, China, designed to measure the mixing angle $\Theta_{13}$  using antineutrinos produced by the reactors of the Daya Bay and Ling Ao nuclear power plants}

\newduneword{nuwro}{NuWro}{neutrino interaction generator}

\newduneabbrev{neut}{NEUT}{neutrino interaction generator}{A neutrino interaction simulation program library for the studies of atmospheric and accelerator neutrinos}

\newduneword{minos}{MINOS}{A long-baseline neutrino experiment, with a near detector at \gls{fnal} and a far detector in the Soudan mine in Minnesota,  designed to observe the phenomena of neutrino oscillations (ended data runs in 2012)}

\newduneabbrev{efig}{EFIG}{Experimental Facilities Interface Group}{The body that provides a means to coordinate and discuss issues related to the interfaces within and between the facility and the DUNE detectors at both the \gls{fnal} and \gls{surf} sites} 

\newduneword{ashriver}{Ash River}{The Ash River, Minnesota, USA \gls{nova} experiment far site, used as an assembly test site for \gls{dune}} 

\newduneword{ipd}{project integration director}{Responsible for integration and installation of DUNE detector deliverables. Manages the integration project} 

\newduneabbrev{jpo}{JPO}{Joint Project Office}{The framework used to facilitate close and coherent coordination across all elements with many shared management resources; 
JPO functions include systems engineering, procurement, \gls{esh}, \gls{qa}, finance, project controls, risk management, compliance, internal reviews, partner agreement management, document management, and administrative support} 

\newduneword{ifbeam}{IFbeam}{Database that stores beamline information indexed by timestamp}

\newduneabbrev{marley}{MARLEY}{Model of Argon Reaction Low Energy Yields}{Developed at UC Davis, MARLEY is the first realistic model of neutrino electron interactions on argon for enegies less than \SI{50}{MeV}. This includes the energy range important for \gls{snb} neutrinos and also solar 8--boron neutrinos}

\newduneabbrev{es}{ES}{elastic scattering}{Events in which a neutrino
elastically scatters off of another particle}

\newduneabbrev{cno}{CNO}{carbon nitrogen oxygen}{The CNO cycle (for carbon-nitrogen-oxygen) is one of the two known sets of fusion reactions by which stars convert hydrogen to helium, the other being the proton-proton chain reaction (pp-chain reaction). In the CNO cycle, four protons fuse, using carbon, nitrogen, and oxygen isotopes as catalysts, to produce one alpha particle, two positrons and two electron neutrinos}

\newduneabbrev{sdwf}{SDWF}{South Dakota Warehouse Facility}{Warehousing operations in South Dakota responsible for receiving LBNF and DUNE goods and coordinating shipments to the access shaft (Ross Shaft) at \gls{surf}}

\newduneabbrev{wms}{WMS}{warehouse management system}{Commercial software package used to track shipments and interface to freight forwarders. This includes a database for shipping}

\newduneabbrev{dcdb}{DCDB}{DUNE construction database}{Database used by DUNE to track the history and testing of all parts of each \gls{detmodule}}

\newduneabbrev{aup}{AUP}{acceptance for use and possession}{Required for beneficial occupancy of the underground areas at \gls{surf} for \gls{lbnf} and \gls{dune}}

\newduneabbrev{bms}{BMS}{building management system}{A system provided by the \gls{cf} to manage the utility (cooling, ventilation, power, etc.) and fire/life safety systems. Separate systems are provided at \gls{surf} and at \gls{fnal}} 

\newduneabbrev{fls}{FLS}{fire and life safety system}{Fire and life safety; systems designed with \gls{cf} to meet building/safety code compliance for safe facilities at \gls{surf} and at \gls{fnal}} 

\newduneabbrev{sno}{SNO}{Sudbury Neutrino Observatory}{The Sudbury Neutrino Observatory was a detector built 6800 feet under ground, in INCO's Creighton mine near Sudbury, Ontario, Canada. SNO was a heavy-water Cherenkov detector designed to detect neutrinos produced by fusion reactions in the sun}

\newduneword{sk}{Super-Kamiokande}{Experiment sited in the Kamioka-mine, Hida-city, Gifu, Japan that uses a large water Cherenkov detector to study neutrino properties through the observation of solar neutrinos, atmospheric neutrinos and man-made neutrinos}

\newduneabbrev{id}{ID}{inner diameter}{Inner diameter of a tube}

\newduneabbrev{od}{OD}{outer diameter}{Outer diameter of a tube}

\newduneabbrev{rms}{RMS}{root mean square}{The square root of the arithmetic mean of the squares of a set of values, used as a measure of the typical magnitude of a set of numbers, regardless of their sign}

\newduneabbrev{orc}{ORC}{operational readiness clearance}{Final safety approval prior to the start of operation}

\newduneabbrev{gsc}{GSC group}{global safety coordination group}{DUNE group that evaluates applicable codes and standards, including international code equivalency, for the design, assembly, and installation of the \gls{fd}}

\newduneabbrev{ha}{HA}{hazard analysis}{A first step in a process to assess risk; the result of hazard analysis is the identification of the hazards present for a task or process}
\newduneword{har}{HAR}{hazard analysis report}

\newduneabbrev{tap}{TAP}{trip action plan}{A document required for any trip by a worker to the underground area at \gls{surf}, per that site's access control program; it describes the work to be accomplished during the trip} 

\newduneword{em}{EM}{emergency management}
\newduneword{ert}{ERT}{emergency response team}

\newduneabbrev{ndk}{NDK}{nucleon decay}{The hypothetical, baryon number violating decay of a proton or a bound neutron into lighter particles}

\newduneabbrev{emi}{EMI}{electromagnetic interference}{Disturbance generated by an external source that affects an electrical circuit by electromagnetic induction, electrostatic coupling, or conduction}

\newduneabbrev{pe}{PE}{photoelectron}{An electron ejected from the surface of a material by the photoelectric effect}

\newduneabbrev{spe}{SPE}{single photoelectron}{A single photoelectron}

\newduneabbrev{fwhm}{FWHM}{full width at half maximum}{Width of a distribution measured between those points at which the distribution is equal to half of its maximum amplitude}

\newduneabbrev{gdml}{GDML}{geometry description markup language}{An application-independent, geometry-description format based on XML}

\newduneabbrev{xml}{XML}{extensible markup language}{A markup language that defines a set of rules for encoding documents in a format that is both human-readable and machine-readable}

\newduneabbrev{crt}{CRT}{cosmic-ray tagger}{Detector external to the TPC designed to tag TPC-traversing cosmic ray particles}

\newduneabbrev{sn}{SN}{supernova}{Event that occurs upon the death of certain types of stars}

\newduneabbrev{wg}{WG}{working group}{A group of persons working together to achieve specified goals}

\newduneabbrev{ctsf}{CTSF}{coating, testing and storage facility}{A facility where the the \gls{dp} photon detectors will be coated, tested, and stored} 

\newduneword{rucio}{Rucio}{Data management system originally developed
by \gls{atlas} but now open-source and shared across \gls{hep}}
\newduneabbrev{doma}{DOMA}{data organization, management, and access}{data organization, management, and access efforts through the \gls{hsf}}

\newduneabbrev{hsf}{HSF}{HEP Software Foundation}{A foundation that facilitates cooperation and common efforts in high energy physics software and computing internationally} 

\newduneabbrev{wlcg}{WLCG}{Worldwide LHC Computing Grid}{Worldwide LHC
Computing Grid}
\newduneabbrev{osg}{OSG}{Open Science Grid}{Open Science Grid}
\newduneabbrev{sci}{SCI}{Scientific Computing Infrastructure}{Proposed
extension of the infrastructure component of \gls{wlcg} to other
experiments}
\newduneabbrev{csc}{CSC}{computing and software consortium}{DUNE
computing and software consortium}

\newduneword{dirac}{DIRAC}{Computing workflow management designed for LHCb and now used by many HEP experiments}

\newduneword{frp}{FRP}{fiber-reinforced plastic}
\newduneabbrev{hdpe}{HDPE}{high-density polyethylene}{A thermoplastic polymer made from petroleum commonly used to make plastic bottles}
\newduneword{hvps}{HVPS}{\gls{hv} power supply}
\newduneword{aisi}{AISI}{American Iron and Steel Institute}
\newduneword{ific}{IFIC}{Instituto de Fisica Corpuscular (in Valencia, Spain)}
\newduneabbrev{rsds}{RSDS}{radioactive source deployment system}{Proposed calibration system based on the deployment of
radioactive sources inside the \gls{dune} cryostat}
\newduneword{2p2h}{2p2h}{two particle, two hole}
\newduneabbrev{duneprism}{DUNE-PRISM}{\gls{dune} Precision Reaction-Independent Spectrum Measurement}{a mobile near detector that can perform measurements over a range of angles off-axis from the neutrino beam direction in order to sample many different neutrino energy distributions}
\newduneword{arcube}{ArgonCube}{The name of the core part of the \gls{dune} \gls{nd}, a \gls{lartpc}}

\newduneabbrev{citf}{CITF}{cryogenic instrumentation test facility}{A facility at \gls{fnal} with small ($<$\SI{1}{ton}) to intermediate ($\sim$\SI{1}{ton}) volumes of instrumented, purified TPC-grade \gls{lar}, used for testing devices intended for use in \gls{dune}}

\newduneabbrev{3dst}{3DST}{3D scintillator tracker}{The core part of the \threed projection scintillator tracker spectrometer in the near detector conceptual design}
\newduneabbrev{3dsts}{3DST-S}{3D scintillator tracker spectrometer}{The \threed projection scintillator tracker spectrometer  in the near detector conceptual design}
\newduneabbrev{mpd}{MPD}{multi-purpose detector}{A component of the near detector conceptual design; it is a magnetized system consisting of a \gls{hpgtpc} and a surrounding \gls{ecal}}
\newduneabbrev{hpg}{HPG}{high-pressure gas}{gas at high pressure to be used in a \gls{hpgtpc}} 
\newduneabbrev{hpgtpc}{HPgTPC}{high-pressure gaseous argon TPC}{A \gls{tpc} filled with gaseous argon; a possible component of the \gls{dune} \gls{nd}}

\newduneword{src}{SRC}{short-range correlated nucleon-nucleon interactions}
\newduneword{larpix}{LArPix}{ \gls{asic} pixelated charge readout for a \gls{tpc} }
\newduneword{arclt}{ArCLight}{a light detector for the \gls{arcube} effort}
\newduneword{fhc}{FHC}{forward horn current ($\numu$ mode)}
\newduneword{rhc}{RHC}{reverse horn current (\numubartonumubar mode)}
\newduneword{mwpc}{MWPC}{multi-wire proportional chamber}
\newduneword{na61}{NA61}{CERN hadron production experiment}
\newduneword{pdnd}{ProtoDUNE-ND}{a prototype \gls{dune} \gls{nd}}
\newduneabbrev{roc}{ROC}{readout chamber}{readout chamber for gaseous argon \gls{tpc}}
\newduneabbrev{iroc}{IROC}{inner readout chamber}{inner (radial) readout chamber for gaseous argon \gls{tpc}}
\newduneabbrev{oroc}{OROC}{outer readout chamber}{outer (radial) readout chamber for gaseous argon \gls{tpc}}

\newduneword{lux}{LUX}{Large Underground Xenon (LUX) dark matter detector at \gls{surf} }

\newduneword{mjdemo}{Majorana Demonstrator}{Experiment sited at \gls{surf} that  seeks to determine whether neutrinos are their own antiparticles}

\newduneword{lz}{LZ}{Experiment sited at \gls{surf} that  seeks to detect faint interactions between galactic dark matter and regular matter}

\newduneword{mu2e}{Mu2e}{An experiment sited at \gls{fnal} that searches for charged-lepton flavor violation and seeks to discover physics beyond the \gls{sm}}

\newduneword{pdsp2}{ProtoDUNE-SP-II}{A second test run in the single-phase ProtoDUNE test stand at CERN, acting as a validation of the final
single-phase detector design}  

\newduneword{osha}{OSHA}{Occupational Safety and Health Administration (USA Department of Labor) formed by the Occupational Safety and Health Act of 1970}
\newduneabbrev{pns}{PNS}{pulsed neutron source}{Calibration system based
on neutron capture gamma showers spread out in the whole detector}

\newduneabbrev{fv}{FV}{fiducial volume}{The detector volume within the \gls{tpc} that is selected for physics analysis through cuts on reconstructed event position}

\newduneword{p6}{P6}{software framework used to plan and status the resource-loaded schedule of activities associated with \gls{usproj}}

\newduneabbrev{evms}{EVMS}{earned value management system}{Earned Value Management is a systematic approach to the integration and measurement of cost, schedule, and technical (scope) accomplishments on a project or task. It provides both the government and contractors the ability to examine detailed schedule information, critical program and technical milestones, and cost data (text from the US DOE); the EVMS is a system that implements this approach}

\newduneword{core}{CORE}{CORE contributions are in either monetary units or labor hours. They can be technical components for the facility or experiment and the effort of the staff needed to produce, install, and test them;  major facilities for the experiment; or other products and services relevant for the completion of the facility or experiment} 

\newduneabbrev{ahj}{AHJ}{Authority Having Jurisdiction}{An organization, office, or individual responsible for enforcing the requirements of a code or standard, or for approving equipment, materials, an installation, or a procedure (\gls{osha})}
\newduneword{cte}{CTE}{coefficient of thermal expansion}

\newduneabbrev{opc}{OPC}{open platform communications}{Open platform communications is a series of standards and specifications for industrial telecommunication} 
\newduneword{scada}{SCADA}{supervisory control and data acquisition}
\newduneword{ln}{LN$_2$}{liquid nitrogen}
\newduneabbrev{lapd}{LAPD}{Liquid Argon Purity Demonstrator}{Cryostat at \gls{fnal}  for long-term studies requiring a large volume of argon}

\newduneabbrev{pab}{PAB}{Proton Assembly Building}{Home of several \gls{lar} facilities at \gls{fnal} }
\newduneword{hep}{HEP}{high energy physics}
\newduneword{cms}{CMS}{Compact Muon Solenoid experiment; one of two general-purpose detectors at the \gls{lhc}. }
\newduneword{alice}{ALICE}{A Large Ion Collider Experiment, at CERN}
\newduneword{gpib}{GPIB}{general purpose interface bus}

\newduneabbrev{pfparticle}{PFParticle}{particle flow particle}{Each of the individual reconstructed particles in the hierarchy (or particle flow) describing the reconstructed event interaction}

\newduneabbrev{mcparticle}{MCParticle}{Monte Carlo Particle}{Individual true simulated particle}
\newduneword{au}{AU}{astronomical unit}
\newduneword{nufit}{NuFIT 4.0}{The NuFIT 4.0 global fit to neutrino oscillation data}

\newduneword{uhv}{UHV}{ultra high vacuum}
\newduneword{lps}{LPS}{laser positioning system}

\newduneword{unicamp}{UNICAMP}{University of Campinas, Sao Paulo, Brazil}
 
\newduneabbrev{fbk}{FBK}{Fondazione Bruno Kessler}{FBK is a research non-profit entity in Trento, Italy that partners in the development of technology with applications in various fields including High Energy Physics}

\newduneabbrev{enob}{ENOB}{effective number of bits}{The effective number of bits is a measure of the dynamic range of an \gls{adc} and its associated circuitry. The resolution of an \gls{adc} is specified by the number of bits used to represent the analog value, in principle giving 2N signal levels for an N-bit signal. However, all real \gls{adc} circuits introduce noise and distortion. ENOB specifies the resolution of an ideal \gls{adc} circuit that would have the same resolution as the circuit under consideration}
\newduneabbrev{sndr}{SNDR}{signal to noise and distortion ratio}{Also known as SINAD. Ratio of the \gls{rms} signal amplitude to the mean value of the root-sum-square of all other spectral components, including harmonics, but excluding \gls{dc} levels. It is a good indication of the overall dynamic performance of an \gls{adc} because it includes all components which make up noise and distortion}
\newduneabbrev{sfdr}{SFDR}{spurious free dynamic range}{Spurious free dynamic range is the ratio of the \gls{rms} value of the signal to the \gls{rms} value of the worst spurious signal regardless of where it falls in the frequency spectrum. The worst spur may or may not be a harmonic of the original signal}
\newduneabbrev{thd}{THD}{total harmonic distortion}{Total harmonic distortion is the ratio of the \gls{rms} value of the fundamental signal to the mean value of the root-sum-square of its harmonics} 
\newduneword{tvs}{TVS}{transient voltage suppression}

\newduneword{riskprob}{risk probabilities}{The risk probability, after taking into account the planned mitigation activities, is ranked as 
 L (low $<\,$\SI{10}{\%}), 
M (medium \SIrange{10}{25}{\%}), or 
H (high $>\,$\SI{25}{\%}). 
The cost and schedule impacts are ranked as 
L (cost increase $<\,$\SI{5}{\%}, schedule delay $<\,$2 months), 
M (\SIrange{5}{25}{\%} and 2--6 months, respectively) and 
H ($>\,$\SI{20}{\%} and $>\,$2 months, respectively)}

\newduneabbrev{lbls}{LBLS}{laser beam location system}
{Auxiliary calibration system providing an independent location measurement of the ionization laser beams direction}

\newduneabbrev{lsst}{LSST}{Large Synoptic Survey Telescope}{8.4 m telescope with 3.2G-pixel camera that will start taking data in 2023}
\newduneabbrev{ska}{SKA}{Square Kilometer Array}{International radio telescope array planned to start data-taking in 2027}
\newduneabbrev{hyperk}{HyperK}{Hyper Kamiokande}{260 kt water Cerenkov neutrino detector to begin construction at Kamiokande in 2020}
\newduneword{lhcb}{LHCb}{LHC experiment dedicated to forward physics}
\newduneword{belleii}{Belle II}{B-factory experiment now running at KEK}

\newduneabbrev{ldm}{LDM}{light-mass dark matter}{Refers to dark matter particles with mass values much lower than the electroweak scale, specifically below the 1~GeV level}
 
\newduneabbrev{bnv}{BNV}{baryon-number violating}{Describing an interaction where \gls{baryonnumber} is not conserved}

\newduneword{bugey}{Bugey}{Neutrino experiment that operated at the Bugey nuclear power plant in France}

\newduneword{minosplus}{MINOS$+$}{The successor to the \gls{minos} experiment, utilizing the same detectors and beam line, but operating at higher beam energy tune than \gls{minos}, parasitic with \gls{nova}}

\newduneword{baryonnumber}{baryon number}{A quantity expressing the total number of baryons in a system minus the number of antibaryons}

\newduneword{np04}{NP04}{
The CERN North Area in \gls{ehn1} intersected by the \gls{h4} hadron beamline,  the location of  \gls{pdsp} and \gls{pdsp2}; also used to refer to the 800\,t cryostat in this area}

\newduneword{np02}{NP02}{
The CERN North Area in \gls{ehn1} intersected by the \gls{h2} hadron beamline, the location of  the 800\,t cryostat used for \gls{pddp} and for \gls{spvd} tests and prototypes; also used to refer to the 800\,t cryostat in this area}

\newduneword{h4}{H4}{CERN North Area hadron beamline used for \gls{pdsp} and \gls{pdsp2}}  

\newduneword{h2}{H2}{CERN North Area hadron beamline used for \gls{pddp} and \gls{spvd} prototypes and demonstrators}  

\newduneword{ua1}{UA1}{UA1 (Underground Area 1) was a particle detector at \gls{cern}'s  Super Proton Synchrotron (SPS). It ran from 1981 until 1990, when the SPS was used as a proton-antiproton collider, searching for traces of W and Z particles in collisions. (CERN) The UA1 dipole magnet was reused in the NOMAD experiment and currently provides the magnetic field for the \gls{t2k} ND280 detector}

\newduneword{ssc}{SSC}{The Superconducting Super Collider was to be a huge underground ring complex beneath the area near Waxahachie, Texas, USA, that would have been the world’s most energetic particle accelerator. It was begun in 1990, but canceled by the U.S. Congress in 1993 (scientificamerican.com Oct 2013)}

\newduneword{daphne}{DAPHNE}{Detector electronics for Acquiring PHotons from NEutrinos is a custom-developed warm front-end waveform digitizing electronics module derived from the readout system developed at \gls{fnal}  for the Mu2e experiment}
 
\newduneword{nersc}{NERSC}{National Energy Research Computing Facility at \gls{lbnl}}

\newduneword{integoff}{integration project}{The \gls{doe} project element that organizes the onsite teams responsible for coordinating far detector installation and detector-facility integration activities at \gls{surf} as well as near detector installation activities at \gls{fnal}.  The integration project office includes overall LBNF/DUNE systems engineering, compliance and review offices} 

\newduneabbrev{sma}{SMA}{SubMiniature version A}{Connector interface for coaxial cables
with a screw-type coupling mechanism}

\newduneword{kloe}{KLOE}{KLOE is a $e^+ e^-$ collider detector spectrometer operated at DAFNE,  the $\phi$-meson factory at Frascati, Rome.  In DUNE it will consist of a \SI{26}{cm} Pb+scintillating fiber \gls{ecal} surrounding a cylindrical open detector region that is  \SI{4.00}{m} in diameter and \SI{4.30}{m} long.  The \gls{ecal} and detector region are embedded in a \SI{0.6}{T} magnetic field created by a \SI{4.86}{m} diameter superconducting coil and a \SI{475}{tonne} iron yoke}

\newduneabbrev{ro}{RO}{review office}{An office within the \gls{integoff} that organizes reviews} 

\newduneabbrev{doecd}{CD}{critical decision}{The U.S. DOE's Order 413.3B outlines a series of staged project approvals, each of which is referred to as a critical decision (CD)}

\newduneabbrev{lbnfspac}{LBNF/DUNE-US SPAC}{LBNF / DUNE-US Strategic Project Advisory Committee}{A committee charged by the host laboratory director to provide expert, independent advice on significant issues and strategies related to \gls{usproj} project organization, management, and risks} 

\newduneabbrev{sand}{SAND}{System for on-Axis Neutrino Detection}{The beam monitor component of the near detector that remains on-axis at all times and serves as a dedicated neutrino spectrum monitor}

\newduneword{4850l}{4850L}{The depth in feet (1480 m) of the access level for the DUNE underground area at SURF; called the ``4850 level''} 

\newduneword{apb}{APB}{authorship and publications board}
\newduneword{crb}{CRB}{collaboration resources board}
\newduneword{cube}{CuBe}{beryllium copper, used to make \gls{sphd} \gls{apa} readout planes}
\newduneword{drift}{drift}{(1) refers to electron drift under the influence of an electric field in a \gls{tpc}; (2) an excavated horizontal corridors (tunnels) in the underground areas at \gls{surf}}
\newduneword{exposure}{exposure}{The integrated detector fiducial mass times beam intensity; it is proportional to the number of interactions and is used to normalize cross sections in a data sample}
\newduneword{fr4}{FR-4}{Flame-retardant fiberglass-reinforced epoxy resin laminate used in making PCBs and other detector components}
\newduneword{g10}{G-10}{Non-flame-retardant fiberglass-reinforced epoxy resin laminate used in making PCBs}
\newduneword{kapton}{Kapton}{A polyimide plastic film that is stable over a broad range of temperatures and is resistant to radiation damage}
\newduneword{shaft}{shaft}{A vertical excavation at \gls{surf} connecting with the surface}
\newduneword{winze}{winze}{A vertical excavation at \gls{surf} connecting two drifts, not connecting to the surface}
\newduneword{ib}{IB}{institutional board; all institutions participating in DUNE are represented on this board}
\newduneword{irb}{IBR}{institutional board representative}
\newduneword{sc}{SC}{depending on context, either speakers committee or scientific computing} 
\newduneword{sac}{SAC}{spokespersons advisory committee}
\newduneword{eoc}{EOC}{education and outreach committee}
\newduneword{indico}{Indico}{Web-based meeting organization tool}
\newduneabbrev{htc}{HTC}{High Throughput Computing}{Computing facilities typically consisting of large numbers of commodity servers as opposed to a single large machine. Best suited for running large numbers of independent jobs in parallel, these facilities are what is usually meant by ``grid computing''}
\newduneword{dcache}{dCache}{A distributed, highly scalable (multi-PB) storage system, usable as both a standalone system and as a high-speed frontend to a tape storage system (such as \gls{pnfs} at \gls{fnal} )}
\newduneabbrev{ifdhc}{IFDHC}{Intensity Frontier Data Handling Client}{A multi-protocol tool for data transfer and file delivery in jobs. It is able to automatically select transfer protocols based on source and destination characteristics}
\newduneabbrev{ifdh}{ifdh}{Intensity Frontier Data Handling}{The actual command invoked when using \gls{ifdhc}, on the command line, e.g. ifdh cp source\_file dest\_file}
\newduneabbrev{pnfs}{PNFS}{Pseudo Network File System}{A file system often used in large storage systems. Typically interaction is very similar to a regular NFS volume, but there can be some subtle and important differences}

\newduneword{nde}{NDE}{non-destructive evaluation} 
\newduneword{psv}{PSV}{pressure safety valve}
\newduneword{pickling}{pickling}{steel pickling and oiling is a metal surface treatment finishing process used to remove surface impurities such as rust and carbon scale from hot rolled carbon steel}

\newduneabbrev{tof}{ToF}{time of flight}{The time a particle takes to fly between two visible interactions observed in the detector. If combined with the distance traveled by the particle, for example a neutron, it can be used for energy reconstruction}

\newduneword{pep4}{PEP-4}{TPC for the Positron Electron Project 4 Collider at Stanford}

\newduneword{ndlar}{ND-LAr}{\gls{lartpc} component of the near detector based on \gls{arcube} technology}

\newduneword{ndgar}{ND-GAr}{component of the near detector with a core gaseous argon \gls{tpc} surrounded by an \gls{ecal} and a magnet}

\newduneword{sfgd}{SuperFGD}{Super Fine-Grained Detector (SuperFGD) is a 3D granular plastic scintillator detector that adopts the same technology as \dword{3dst}. It will be installed in the \dword{t2k} \dword{nd280} system . The \dword{3dst} design will inherit in large part from the SuperFGD detector} 

\newduneword{nd280}{ND280}{Near Detector 280, is the \dword{t2k} magnetized near detector} 

\newduneword{fee}{FEE}{front-end electronics}

\newduneword{mpgd}{MPGD}{MicroPattern Gas Detectors}

\newduneword{rmm}{RMM}{Resistive MicroMegas}

\newduneabbrev{stv}{STV}{Single Transverse Variables}{Kinematical variables obtained by projecting the neutrino interaction onto the transverse plane}

\newduneabbrev{ccqe}{CCQE}{charged current quasielastic interaction} {An interaction where a neutrino scatters from a nucleon, producing a charged lepton and converting a neutron to a proton or vice versa}

\newduneword{sis}{SIS}{shallow inelastic scattering}

\newduneabbrev{res}{RES}{resonant scattering}{The mode of scattering where the target nucleon is excited to a resonant state and decays, typically producing one or more pions}

\newduneabbrev{hadw}{W}{invariant mass of the hadronic system}{Refers to the invariant mass of the hadronic system formed during the neutrino scatter}

\newduneabbrev{agky}{AGKY}{Andreopoulos-Gallagher-Kehayias-Yang}{A model for hadronization of non-resonant inelastic neutrino reactions used in \gls{genie}. At low invariant hadronic masses, typically less than 2.3\,GeV/c$^2$, it is a KNO-inspired empirical model anchored on several bubble chamber measurements of neutrino-induced shower characteristics. For invariant hadronic masses between 2.3 and 3.0\,GeV/c$^2$, the model transitions linearly to a \gls{genie}-tuned version of PYTHIA, which is also used for the simulation of events at higher invariant masses} 

\newduneabbrev{clas}{CLAS}{CEBAF Large Acceptance Spectrometer}{A nuclear and particle physics detector located in the experimental Hall B at Jefferson Laboratory in Newport News, Virginia, United States. It is used to study the properties of the nuclear matter by the collaboration of over 200 physicists. Of particular relevance is its study of electron interactions with nuclei, including argon} 

\newduneabbrev{e4nu}{e4nu}{Electrons for Neutrinos}{A collaboration dedicated to using JLab's electron-scattering data to deliver improved neutrino-nucleus cross sections} 

\newduneabbrev{ldmx}{LDMX}{Light Dark Matter eXperiment}{The LDMX detector concept consists of a small precision tracker, and electromagnetic and hadronic calorimeters, all with near $2\pi$ azimuthal acceptance from the forward beam axis out to $\sim40^\circ$ angle. This detector would be capable of measuring correlations among electrons, pions, protons, and neutrons in electron-nucleus scattering at exactly the energies relevant for DUNE physics} 

\newduneabbrev{mec}{MEC}{meson-exchange currents} {An nuclear effect wherein pairs or larger groups of nucleons within a nucleus are bound together through the exchange of pions or other mesons. Neutrinos and other particles can scatter from these correlated pairs}

\newduneword{imt}{IMT}{Intranuclear momentum transfer}

\newduneword{miniboone}{MiniBooNE}{The Mini Booster Neutrino Experiment,  at \gls{fnal} , was designed to fully explore the LSND result}

\newduneabbrev{tms}{TMS}{The Muon Spectrometer}{A muon spectrometer for the Near Detector that will be installed for the initial running period of DUNE, before the \gls{mpd} detector component is ready} 

\newduneabbrev{cvmfs}{CVMFS}{CERN VM File System}{A distributed file system designed for scalable, high-performance distribution of software to interactive and batch computers} 

\newduneword{kerberos}{Kerberos}{A strong authentication system used by the computing resources at \gls{fnal} and \gls{cern}} 

\newduneabbrev{mrb}{MRB}{Multi Repository Build System}{A \gls{fnal}-developed build system based on \dword{cmake} that allows development and builds of code from multiple repositories} 

\newduneword{cmake}{Cmake}{CMake is an open-source, cross-platform family of tools designed to build, test and package software}

\newduneabbrev{ups}{UPS}{UNIX product support}{A software tool that sets up a consistent environment of versions of pre-installed products and their dependencies on UNIX-like platforms} 

\newduneabbrev{upd}{UPD}{UNIX product distribution}{A tool for uploading and downloading pre-built software products between local systems and centralized software distribution servers.  UPD is not frequently used on DUNE because newer tools are more convenient} 

\newduneabbrev{sso}{SSO}{single sign-on}{Used at \gls{fnal} to indicate that a group of services,  such as DocDB or the DUNE Wiki share common sign-in credentials and active sessions.  \gls{fnal}  services that say "Sign in with SSO username and password" mean to use your \gls{fnal} Services or federated username and password} 

\newduneabbrev{vo}{VO}{virtual organization}{A database containing a list of member names, certificate distinguishing information, and a list of permissions members have to access computing grid and data resources} 

\newduneabbrev{nas}{NAS}{network attached storage}{Disk storage that is available on computers but shared between them.  Relies on \gls{nfs} mounts rather than authenticated file transfer protocols.  Usually found on interactive servers to provide space for home directories, app and data storage} 

\newduneabbrev{nfs}{NFS}{network file system}{Industry-standard mechanism for mounting disks over a network.  Provides regular UNIX file and directory access} 

\newduneword{recombination}{recombination}{Electrons freed from Argon atoms will sometimes recombine with the positive argon ions, either the same ones from which they came or nearby ones.  Sometimes called ``quenching"} 

\newduneword{elife}{electron lifetime}{The attachment of electrons drifting through liquid argon to impurity molecules such as oxygen or water is parameterized by an exponential as a function of time with a time constant called the electron lifetime} 

\newduneword{gplane}{grid plane}{The uninstrumented plane of wires or electrodes on an anode plane 
facing the drift volume.   It shapes the signals and provides \gls{esd} protection} 

\newduneword{gmesh}{grounding mesh}{A metal mesh attached to the SP \gls{apa} frame between the collection-plane wires and the space inside the frame where the \gls{pd} modules are installed.  It provides electric field uniformity so the collection-plane wires all have similar fields around them} 

\newduneabbrev{bcr}{BCR}{baseline change request}{A DOE project change, part of the change management system process}

\newduneword{ncav}{North Cavern}{the location of two of the planned four DUNE far detector modules at \gls{surf}}

\newduneword{scav}{South Cavern}{the location of two of the planned four DUNE far detector modules at \gls{surf}}

\newduneword{semp}{SEMP}{systems engineering management plan}  

\newduneabbrev{tpcost}{TPC}{total project cost}{The DOE terminology for the total budget and contingency for the entire \gls{usproj} project from CD-0 to CD-4}  
\newduneabbrev{nsint}{NSI}{near site integration}{The scope of work at the near site for the \gls{integoff}} 

\newduneabbrev{opcost}{OPC}{other project costs}{The DOE project costs that support conceptual design, pre-operations commissioning, technical coordination, and power}

\newduneabbrev{moa}{MOA}{memorandum of agreement}{A project management methodology that documents an agreement between \gls{fnal} and the \gls{usproj}  Project for how \gls{fnal}  will support the project} 

\newduneabbrev{croc}{CROC}{central readout chamber}{central (radial) readout chamber for the ND \gls{gartpc}} 

\newduneabbrev{tki}{TKI}{transverse kinematic imbalance}{The imbalance among final-state particle momenta in the transverse plane to the neutrino direction; different aspects of the imbalance are sensitive to the detail of the nuclear effects in neutrino-nucleus interactions} 

\newduneabbrev{iandi}{I\&I}{Integration and Installation}{One of the three project areas in the LBNF/DUNE-US \gls{doe} project, along with LBNF and DUNE-US} 

\newduneword{hmi}{HMI}{human-machine interface}

\newduneword{lhe}{LHe}{liquid helium}

\newduneword{echain}{energy chain}{mechanical machine elements used to carry and guide power to moving parts of machines or structures, as required for \gls{duneprism} to carry power, data, and utilities to and from each movable near detector component at any arbitrary position along its travel path}

\newduneabbrev{sphd}{FD1-HD}{horizontal drift detector module}{LArTPC design used in FD1 in which electrons drift horizontally to wire plane anodes (\glspl{apa}) that along with the front-end electronics are immersed in LAr} 

\newduneabbrev{spvd}{FD2-VD}{vertical drift detector module}{LArTPC design used in FD2 in which electrons drift vertically to PCB-based anodes at the top and bottom of the LAr volume,  with a cathode in the middle} 

\newduneabbrev{cru}{CRU}{charge-readout unit}{In the \gls{spvd} design an assembly of the \glspl{pcbp} plus adapter boards; two to a \gls{crp}} 

\newduneword{mpv}{MPV}{most probable value}

\newduneword{pcbp}{PCB panel}{In the \gls{spvd} design, one of four \glspl{pcb} of size  
1.5 $\times$ 1.7\.,m assembled into a \gls{cru}}

\newduneword{anodepln}{anode plane}{a planar array of charge readout devices covering an entire face of a detector module}

\newduneword{msps}{MSPS}{megasamples per second}

\newduneword{gpsdo}{GPSDO}{\gls{gps} disciplined oscillator}

\newduneword{tdaq}{TDAQ}{trigger and DAQ system} 

\newduneword{nios}{NIOS}{network identity operating system} 

\newduneabbrev{corsika}{CORSIKA}{COsmic Ray SImulations for KAscade}{a program for detailed simulation of extensive air showers initiated by high-energy cosmic ray particles}

\newduneabbrev{scd}{SCD}{scientific computing division}{\gls{fnal}'s Scientific Computing Division}

\newduneabbrev{garsoft}{GArSoft}{Gaseous Argon Software}{A software toolkit similar to \gls{larsoft}, but targeted at the gaseous argon time projection chamber and calorimeter of \gls{ndgar}}

\newduneword{ndgarlite}{ND-GAr-Lite}{a temporary muon spectrometer consisting of the magnet and steel flux return of \gls{ndgar}, but with a simplified tracking chamber made with scintillating bars}

\newduneword{github}{GitHub}{a commercial web service providing code version management, storage,  and browsing via \gls{git}}

\newduneword{git}{git}{a distributed version-control system, commonly used to manage software}

\newduneword{xrootd}{xrootd}{a high-performance data system widely used in \gls{hep} to store and to distribute data to jobs.  It allows streaming of data}

\newduneabbrev{gpuaas}{GPUAAS}{GPU As A Service}{a technique that allows many non-GPU-enabled compute nodes to share a GPU resource by sending it work over the network and waiting for results to be returned}
\newduneword{sce}{SCE}{space charge effect}

\newduneabbrev{nest}{NEST}{noble element simulation technique}{Comprehensive simulation code modeling the excitation, ionization, and corresponding scintillation and electroluminescence processes in liquid noble elements}

\newduneword{bgr}{BGR}{A bandgap voltage reference is a circuit block in ASIC for providing stable reference voltages} 

\newduneword{comsol}{COMSOL}{General-purpose simulation software based on advanced numerical methods (comsol.com)}

\newduneword{greyrm}{grey room}{ISO-8 clean room with a cleanliness level of 3.5M particles of 0.5 micron or less per cubic meter volume. ISO-8 clean rooms are referred to as grey rooms because at this level of cleanliness most standard clean room attire is not required.}

\newduneabbrev{pof}{PoF}{power-over-fiber}{a technology in which a fiber optic cable carries optical power, which is used as an energy source rather than, or as well as, carrying data; this allows a device to be remotely powered, while providing electrical isolation between the device and the power supply} 

\newduneword{ppc}{PPC}{photovoltaic power converter}
\newduneword{eelaser}{edge-emitting laser}{a laser in which  light is emitted from the edge of the substrate}
\newduneword{peek}{PEEK}{Polyether ether ketone, a colorless organic thermoplastic polymer}

\newduneword{fsm}{Finite State Machine}{a mathematical model of computation; it is an abstract machine that can be in exactly one of a finite number of states at any given time} 
\newduneword{icecube}{IceCube}{South Pole Neutrino Observatory}
\newduneword{pde}{PDE}{photo detection efficiency} 
\newduneword{pmos}{PMOS}{(see \gls{cmos}; PMOS is constructed with the p-type source and drain and an n-type substrate}
\newduneabbrev{poe}{PoE}{power-over-Ethernet}{systems that pass electric power along with data on twisted-pair Ethernet cabling} 

\newduneabbrev{daqdts}{DTS}{DUNE timing and synchronization subsystem}{The portion of the \gls{daq} that provides for timing and synchronization to various detector systems}

\newduneword{tai}{TAI}{International Atomic Time}

\newduneword{larzic}{LARZIC}{The cryogenic amplifier \dword{asic} that is the principal component of the \gls{spvd} top drift \dword{fe} analog cards}

\newduneword{dpdfd}{DPDFD}{Deputy Project Director for far detectors}

\newduneword{bsws}{BSWS}{bearing sensor wire compression seal}

\newduneabbrev{ly}{LY}{light yield}{detected photons per unit deposited energy}

\newduneword{mtbf}{MTBF}{mean time between failures}

\newduneword{ingaas}{InGaAs}{Indium gallium arsenide is a room-temperature semiconductor commonly used as a high-speed, high-sensitivity photodetector for optical fiber telecommunications}

\newduneword{intlproj}{LBNF/DUNE Construction Project}{The international project to design and build the facilities and detectors for the \gls{lbnf-dune}; it includes the \gls{usproj} and projects at multiple international partners to manage the contributions from non-US institutions and funding agencies to design, build, and install the detector components}

\newduneword{ddmp}{DUNE Management Plan}{DUNE Management Plan}

\newduneword{pd2hd}{ProtoDUNE-II-HD}{The second ProtoDUNE for the HD design, using NP04 to test installation, integration, and detector component performance}

\newduneword{mod0}{Module 0}{The final pre-production instance of a detector; for the DUNE \glspl{detmodule}, the \glspl{protodune2} in the 800\,t cryostats in \gls{np02} and \gls{np04} serve this purpose}

\newduneword{vdmod0}{FD2-VD Module 0}{The final pre-production \gls{protodune} instance for the DUNE \gls{spvd} \gls{detmodule},  it will use the 800\,t cryostat in \gls{np02} }

\newduneword{hdmod0}{FD1-HD Module 0}{The final pre-production \gls{protodune} instance for the DUNE \gls{sphd} \gls{detmodule},  it will use the 800\,t cryostat in \gls{np04} }

\newduneword{fsii}{FSII}{far site integration and installation}
\newduneword{fdc}{FDC}{Far Detector and Cryogenics Subproject}

\newduneabbrev{co}{CO}{compliance office}{a team of engineers from multiple partners that provides clear direction for designing and constructing components that will be used during integration}

\newduneword{fscfbsi}{FSCF-BSI}{\gls{usproj} subproject for far site conventional facilities, building and site infrastructure}

\newduneabbrev{poms}{POMS}{Production Operations Management System}{A workflow management system available for all DUNE users to submit and monitor grid jobs as well as view job log files}

\newduneabbrev{fifemon}{FIFEMON}{FabrIc for Frontier Experiments MONitoring}{Comprehensive suite of job and storage monitoring information available for most \gls{fnal} experiments, including DUNE}

\newduneword{larg4}{LArG4}{LArG4 is the replacement for LArsim/LArG4. LArG4 is based on \gls{artg4tk}}

\newduneword{artg4tk}{artg4tk}{artg4tk provides a general interface between \gls{geant4} and \gls{art}}

\newduneword{tde}{TDE}{top detector electronics}

\newduneword{bde}{BDE}{bottom detector electronics}

\newduneword{sigproc}{Signal Processing}{The goal of TPC signal processing is to reconstruct the distribution of ionization electrons arriving at wire planes from the digitized TPC waveform}

\newduneword{mcnd}{MCND}{More Capable Near Detector} 

\newduneabbrev{compass}{COMPASS}{
Common Muon and Proton Apparatus for Structure and Spectroscopy }{a multipurpose experiment at \gls{cern}’s Super Proton Synchrotron (SPS)} %

\newduneword{vd}{vertical drift}{single-phase, vertical drift \gls{lartpc} technology}

\newduneword{hd}{horizontal drift}{single-phase,  horizontal drift \gls{lartpc} technology}

\newduneabbrev{esnet}{ESnet}{Energy Sciences Network}{The \gls{doe}'s dedicated science network} %

\newduneword{project.py}{project.py}{XML-based job configuration system developed by the \gls{microboone} collaboration} %

\newduneabbrev{ppfx}{PPFX}{Package to Predict the FluX}{\gls{fnal}-supported package that implements hadron production corrections to geant4 simulations and propagates uncertainties for the NuMI  and LBNF beam lines} %

\newduneabbrev{ml}{ML}{Machine Learning}{Machine Learning}

\newduneword{datalake}{data lake}{The not-yet-realized concept of a storage service with multiple levels of quality of service in which the end user can access data without knowing the data's source location} 

\newduneabbrev{metacat}{MetaCat}{MetaCat}{Metadata Catalog, a modern replacement for the file description portion of the sam metadata catalog} %

\newduneword{g4lbnf}{g4lbnf}{LBNF neutrino beamline simulation program} 

\newduneword{Geant4Reweight}{Geant4Reweight}{Framework for evaluating and propagating hadronic interaction uncertainties in Geant4} 

\newduneword{geantnet}{G\'EANT}{G\'EANT interconnects Europe's national research and education networking (NREN) organizations with the high bandwidth, high speed and highly resilient pan-European backbone.
}

\newduneabbrev{nren}{NREN}{National Research and Education Network}{National level research computing network infrastructure} %

\newduneabbrev{vrf}{VRF}{Virtual Routing and Forwarding}{Networking overlays that provide  a logical routing infrastructure  that allows flexible traffic engineering} %

\newduneword{samvalue}{value}{A generic quantity describing a file in the \dword{sam} data catalog} %

\newduneword{samparameter}{parameter}{A user or experiment described quantity describing a file in the \dword{sam} data catalog} %

\newduneword{samproject}{SAM-project}{A server process running centrally that maintains a predefined list of files and delivers information about  their locations when asked by distributed processes. The project tracks success and failure of file processing} %

\newduneword{samconsumer}{SAM-consumer}{A client process that requests information about file locations from a \dword{samproject}, process the file and reports success or failure} %

\newduneword{samdataset}{SAM-dataset}{ A dynamic collection of files defined by queries to the \dword{sam} data catalog} %
\newduneword{metacatdataset}{MetaCat-dataset}{ A fixed but mutable collection of files defined by queries to the \dword{metacat} data catalog} %

\newduneword{samsnapshot}{SAM-snapshot}{A fixed collection of files corresponding to a \dword{sam} \dword{samdataset} at a particular point in time} %

\newduneabbrev{mql}{MQL}{\dword{metacat} Query Language} {A query language which supports queries of the \dword{metacat} data catalog, including parentage and logical functions such as union, join and subtraction} %

\newduneword{pd2}{ProtoDUNE-II}{\gls{protodune} test runs at CERN in 2022-2023; also called \gls{mod0}}

\newduneabbrev{ucondb}{uconDB}{Unstructured Conditions Database}{Unstructured conditions database developed for \gls{fnal} fixed target experiments} 

\newduneabbrev{json}{JSON}{JavaScript Object Notation}{Open standard data interchange format that uses pair-value pairs and maps well onto python data formats such as tuples and lists} 

\newduneabbrev{dqmdb}{DQMDB}{Data Quality and Monitoring Database}{Database storing the results of data-quality monitoring} 

\newduneword{db}{DB}{database}

\newduneabbrev{vm}{VM}{virtual machine}{Emulator of a physical computer that allows multiple users to configure different operating systems  while sharing physical hardware}

\newduneword{enstore}{Enstore}{A mass storage system developed by \gls{fnal} that provides distributed access and management of data stored on tapes} 

\newduneword{stashcache}{StashCache}{A distributed caching federation that enables opportunistic users to utilize nearby opportunistic storage} 

\newduneabbrev{hdf5}{HDF5}{Hierarchical Data Format}{Data format widely used in \gls{ml}} 

\newduneword{glideinwms}{GlideinWMS}{A system of submitting pilot jobs to grid computing sites, inside of which user jobs run, presenting a uniform setup across many different sites} 

\newduneword{mars}{MARS}{The MARS code system is a set of \gls{mc} programs for detailed simulation of coupled hadronic and electromagnetic cascades, with heavy ion, muon and neutrino production and interactions} 

\newduneabbrev{fife}{FIFE}{Fabric for Intensity Frontier Experiments}{\gls{fnal} computing infrastructure for \gls{if} Experiments} 

\newduneword{tr}{trigger record}{A data record produced by the \gls{dune} \gls{daq} system.  A Trigger Record can contain multiple interaction ``events'' or none} 

\newduneword{postgres}{Postgres}{also known as PostgreSQL, Postgres is a free and open-source relational database management system used extensively for databases in \gls{hep}}

\newduneword{fhicl}{FHICL}{Fermilab Hierarchical Configuration Language; a standard configuration language for the storage, communication, and manipulation of scientific parameter sets}

\newduneword{api}{API}{Application Programming Interface}

\newduneword{hwdb}{HWDB}{hardware database}   

\newduneword{s3}{S3}{The Amazon cloud-based commercial storage service} 

\newduneword{openstack}{OpenStack}{An open source cloud software used to deploy instances of containers}

\newduneabbrev{cric}{CRIC}{Computing Resource Information System}{a framework providing a centralized (and flexible) way to describe which resources are being used by the experiment and how}

\newduneabbrev{ccb}{CCB}{Computing Contributions Board}{a board made up of institutional representatives for larger countries and laboratories. It meets annually to negotiate collaboration contributions to computing infrastructure} 

\newduneword{garfield}{Garfield}{A simulation program developed at \gls{cern} for gaseous detectors } 

\newduneword{datatier}{data tier}{Differing data types produced in a processing sequence, for example, raw data, reconstructed, derived analysis sample, histograms} 

\newduneabbrev{gpu}{GPU}{Graphical Processing Unit}{Specialized computing hardware optimized for image processing} 

\newduneabbrev{gpvm}{dunegpvm}{DUNE General Purpose Virtual Machine}{Centrally managed virtual Linux systems at \gls{fnal} with access to network attached and \gls{pnfs} storage. Used for small-scale data analysis and algorithm development} 

\newduneword{madx}{MAD-X}{framework that provides the de facto standard scripting language to describe particle accelerators, simulate beam dynamics, and optimize beam optics at \gls{cern}}

\newduneabbrev{cpu}{CPU}{Central Processing Unit}{A computing processor, when used as a unit of processing; generally refers to a single core} 

\newduneabbrev{fft}{FFT}{Fast Fourier Transform}{An algorithm that calculates the frequency components of a time-domain waveform in a computationally efficient manner} 

\newduneword{grain}{GRAIN}{In the \gls{sand} detector, a small cryostat containing \gls{lar} installed upstream of the straw-tube tracker inside the \gls{ecal}}

\newduneword{voms}{VOMS}{Virtual Organization Membership Service (in grid computing)}

\newduneword{cry}{CRY}{A cosmic ray shower library and software tool}

\newduneword{jwt}{JWT}{\gls{json} web token}

\newduneabbrev{ferry}{FERRY}{Frontier Experiments RegistRY}{a central database that keeps track of all scientific computing users at \gls{fnal}, the experiments and groups of which they are members, and the various capabilities they are allowed}

\newduneabbrev{perfsonar}{perfSONAR}{performance Service-Oriented Network monitoring ARchitecture}{a network measurement toolkit designed to provide federated coverage of paths and to help establish end-to-end usage expectations}

\newduneword{ipv6}{IPv6}{the most recent version of the Internet communications protocol that provides an identification and location system for computers on networks and routes traffic across the Internet}

\newduneword{monit}{MONIT}{a suite of tools using open source technology used in the monitoring of the \gls{cern} IT data center and \gls{wlcg} infrastructure. }

\newduneword{garana}{GArAna}{provides a facility for making an analysis ntuple from information stored in \gls{garsoft} data products}

\newduneword{ci}{CI}{continuous integration}

\newduneword{datadis}{Data Dispatcher}{communicates between running processing jobs and the data delivery systems.  It provides file location information and basic bookkeeping on file access and transfers} 

\newduneabbrev{castor}{CASTOR}{CERN Advanced STORage manager}{a hierarchical storage system with tape and disk developed at \gls{cern}. It is being replaced by \gls{cta}} 

\newduneabbrev{cta}{CTA}{CERN Tape Archive}{a hierarchical storage system with tape and disk developed at \gls{cern}. It is replacing \gls{castor}} 

\newduneabbrev{sonic}{SONIC}{Services for Optimized
Network Inference on Coprocessors}{Framework for implementing machine learning algorithms on co-processors developed by \gls{fnal}}

\newduneword{jobsub}{JobSub}{The \gls{fnal} \gls{if} job submission client that supports user submission of complex workflows to HT-Condor} 

\newduneword{hepcloud}{HEPCloud}{routes jobs to local or remote computing resources based on the policy for a particular experiment, workflow requirements, and cost and efficiency of accessing the various resources. It expands the resources available to include \gls{hpc} centers and commercial cloud resources}

\newduneword{wmsmon}{WMS monitoring}{Workflow Management System Monitoring}

\newduneword{redmine}{Redmine}{an open source code repository and issue tracking tool that was historically used by many of the \gls{fnal} general computing projects and \gls{if} experiments} 

\newduneabbrev{if}{IF}{Intensity Frontier}{refers to \gls{hep} experiments, particularly in the U.S., that rely on high luminosity instead of high energy for discovery.  Includes B-physics, neutrino, and muon experiments}

\newduneword{slack}{Slack}{a commercial business communication platform} 

\newduneword{servicenow}{ServiceNow}{a commercial enterprise workflow system used by \gls{fnal} for formal issue tracking and IT workflows such as user account preparation} 

\newduneabbrev{rse}{RSE}{Rucio Storage Element}{A storage element that is known to the DUNE \gls{rucio} instance} 

\newduneabbrev{coffea}{COFFEA}{Columnar Object Framework For Effective Analysis}{Columnar data analysis frame\-work developed at \gls{fnal}} 

\newduneword{nuisance}{Nuisance}{An open source C++ framework for studying neutrino interaction cross sections} 

\newduneabbrev{highland}{HighLAND}{High Level Analysis at a Neutrino Detector}{Analysis framework developed by the \gls{t2k} collaboration} 
 
\newduneabbrev{raii}{RAII}{Resource Acquisition Is Initialization}{a C++ programming technique that binds the lifecycle of a resource that must be acquired before use (e.g., allocated heap memory, thread of execution, open socket, open file, locked mutex, disk space, database connection—anything that exists in limited supply) to the lifetime of an object}  
 
\newduneword{fts3}{FTS3}{File Transfer Service version 3, developed by \gls{cern}, and distinct from, the \gls{fnal} File Transfer Service} 
 
\newduneword{wfms}{WFMS}{Workflow Management System}
 
\newduneword{condmeta}{conditions metadata}{defined as the information necessary to understand the context of physics data, e.g., beam data or calibrations} 
 
\newduneword{cdb}{conditions DB}{The Conditions Database stores the \gls{condmeta} needed for data processing and analysis}

\newduneword{rntuple}{RNtuple}{The next generation of the \gls{root} I/O system} 
 
\newduneabbrev{lan}{LAN}{Local Area Network}{Computing network confined to a relatively small geographic area} 
 
\newduneabbrev{wan}{WAN}{Wide Area Network}{Computing network that interconnects \glspl{lan} across relatively broad geographic areas} 

\newduneword{layer3}{Layer-3}{Networking protocol  {https://en.wikipedia.org/wiki/Network\_layer}}
 
\newduneabbrev{rest}{REST}{Representational State Transfer}{Standard for web interfaces} 
 
\newduneabbrev{faq}{FAQ}{Frequently Asked Questions}{A software system for collecting and answering the most common questions about an activity} 
 
\newduneword{spack}{Spack}{Software package manager for UNIX and mac OS systems} 

\newduneabbrev{lxplus}{LXPLUS}{Linux Public Login User Service)}{The interactive logon service to Linux for all \gls{cern} users} 

\newduneword{dunesw}{dunesw}{The base DUNE software release} 

\newduneword{posix}{POSIX}{Portable Operating System Interface - Unix standard for operating system interfaces} 

\newduneword{nutools}{NuTools}{shared code for \gls{larsoft} + NOvA + other neutrino experiments that use \gls{art}.  Includes beam and event generators and re-weighting packages}

\newduneword{singlecube}{SingleCube}{A cubical time projection chamber 30 cm on a side with a single ND-LAr pixel readout tile, used for prototyping and tests}

\newduneword{hllhc}{HL-LHC}{High-luminosity \gls{lhc}} 

\newduneabbrev{sof}{SoF}{signal-over-fiber}{a technology in which a fiber optic cable carries detector output that has been converted from an electrical to an optical pulse}

\newduneword{icd}{interface document}{Interface document}

\newduneword{ipmi}{IPMI}{Intelligent Platform Management Interface}

\newduneword{grafana}{Grafana}{add def}

\newduneword{trigact}{trigger activitation}{a correlation (or cluster) in time and space within a stream of \glspl{trigprimitive}}

\newduneword{cad}{CAD}{computer aided design}

\newduneword{gkvm}{GKVM}{Gava, Kneller, Volpe, McLaughlin Supernova model}

\newduneword{daqrou}{DAQ readout unit}{The basic component of the \gls{daqros}}

\newduneword{kpp}{key performance parameter}{(KPP) Defined by \gls{doecd}-2 (U.S.) as characteristic,  function, requirement or design basis that if changed would have a major impact on the system or facility performance, schedule,  cost and/or risk}
\newduneword{wust}{WUST}{Wroclaw University of Science and Technology in Poland}
\newduneword{seri}{SERI}{State Secretariat for Education, Research and Innovation in Switzerland}
\newduneword{feed}{FEED}{front-end engineering design}
\newduneword{prv}{PRV}{pressure-relief valve}
\newduneword{vfd}{VFD}{variable frequency drive}
\newduneword{hart}{HART}{highway addressable remote transducer}

\hypersetup{
    pdftitle={\expshort TDR \thedocsubtitle},
    pdfauthor={\expshort Collaboration},
    final=true,
    colorlinks=false,
    linktocpage=true,
    linkbordercolor=blue,
    citebordercolor=green,
    urlbordercolor=magenta,
    filecolor=black,
    pdfpagemode=UseOutlines,
    pdfborderstyle={/S/U},  
}

\begin{document}

\pagestyle{titlepage}

\begin{center}
   {\Huge LBNF/DUNE }

  \vspace{5mm}

  {\Huge Cryostats and Cryogenics Infrastructure \\ for the DUNE Far Detector}  

  \vspace{10mm}

  {
  \LARGE Design Report

  \vspace{5mm}
  
    }

  \vspace{15mm}

\includegraphics[width=0.8\textwidth]{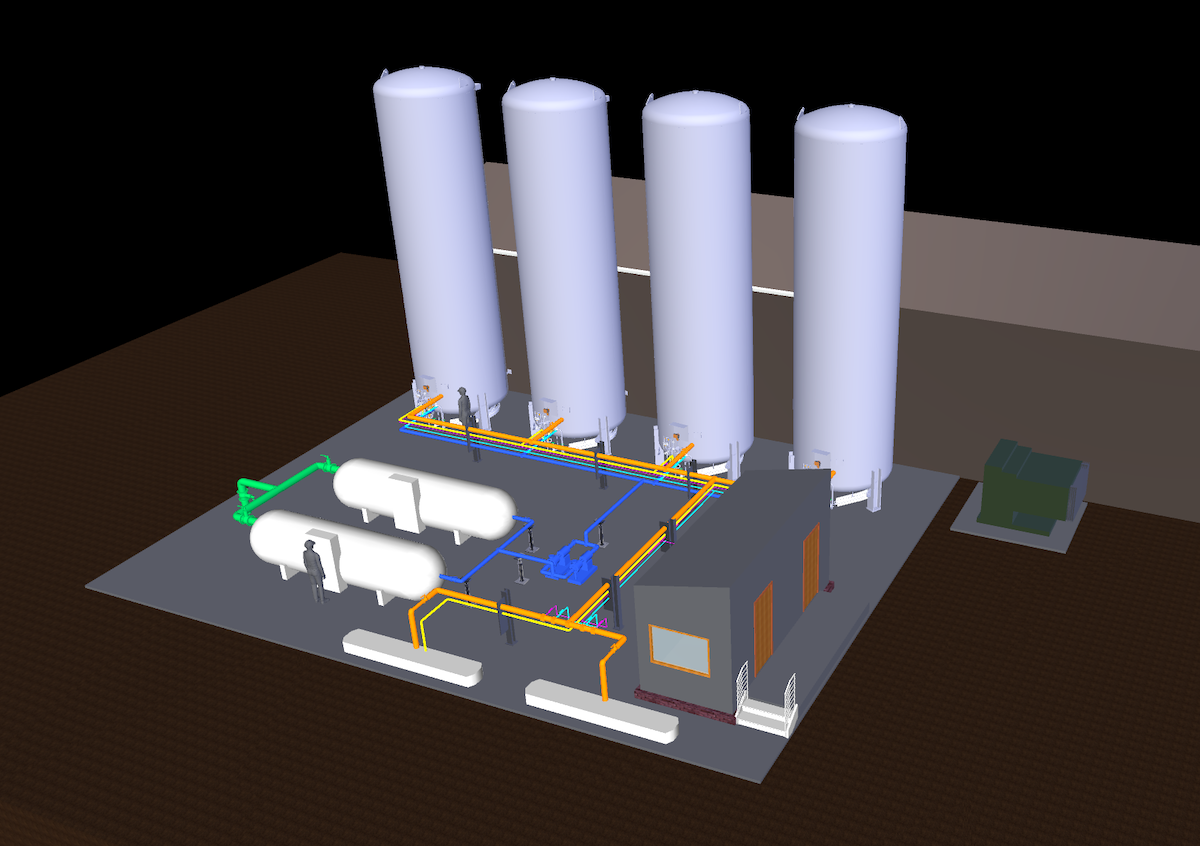}

  \vspace{10mm}
  \today
    \vspace{15mm}
    
    {\large{The LBNF/DUNE-US Project}}
\end{center}

\cleardoublepage
\vspace*{16cm} 
  {\small  This document was prepared by the LBNF/DUNE Project and the DUNE Collaboration using the resources of the Fermi National Accelerator Laboratory (Fermilab), a U.S. Department of Energy, Office of Science, HEP User Facility. Fermilab is managed by Fermi Research Alliance, LLC (FRA), acting under Contract No. DE-AC02-07CH11359.
  
LBNF/DUNE also acknowledges the international, national, and regional funding agencies supporting the institutions who have contributed to completing this Design Report.  
  }


\cleardoublepage

\title{Authors}
\author[1]{M.~Adamowski}
\author[1]{M.~Delaney}
\author[1]{R.~Doubnik}
\author[1]{D.~Montanari}
\author[1]{T.~Nichols}
\author[1]{I.~Young}
\affil[1]{Fermi National Accelerator Laboratory, Batavia, IL 60510, USA}
\author[2]{J.~Bremer}
\author[2]{D.~Mladenov}
\author[2]{A.~Parchet}
\author[2]{F.~Resnati}
\affil[2]{CERN, The European Organization for Nuclear Research, 1211 Meyrin, Switzerlan}
    
    \begin{center}
    \large{Authors}
    
M.~Adamowski, M.~Delaney, R.~Doubnik, D.~Montanari, T.~Nichols, and I.~Young \\
\vspace{1mm}
\textit{Fermi National Accelerator Laboratory, \\ Batavia, IL 60510, USA}
  
J.~Bremer, D.~Mladenov, A.~Parchet, and F.~Resnati \\
\vspace{1mm}
\textit{CERN, The European Organization for Nuclear Research, \\ 1211 Meyrin, Switzerland}

on behalf of the Long-Baseline Neutrino Facility and \\
the Deep Underground Neutrino Experiment (LBNF/DUNE)
    \end{center}

\cleardoublepage

\renewcommand{\familydefault}{\sfdefault}
\renewcommand{\thepage}{\roman{page}}
\setcounter{page}{0}

\pagestyle{plain} 


\textsf{\tableofcontents}

\textsf{\listoffigures}

\textsf{\listoftables}


\iffinal\else
\textsf{\listoftodos}
\clearpage
\fi

\renewcommand{\thepage}{\arabic{page}}
\setcounter{page}{1}

\pagestyle{fancy}

\renewcommand{\chaptermark}[1]{%
\markboth{Chapter \thechapter:\ #1}{}}
\fancyhead{}
\fancyhead[RO]{\textsf{\footnotesize \thechapter--\thepage}}
\fancyhead[LO]{\textsf{\footnotesize \leftmark}}

\fancyfoot{}
\fancyfoot[RO]{\textsf{\footnotesize Design Report}}
\fancyfoot[LO]{\textsf{\footnotesize \thedoctitle}}
\fancypagestyle{plain}{}

\renewcommand{\headrule}{\vspace{-4mm}\color[gray]{0.5}{\rule{\headwidth}{0.5pt}}}



\chapter{Introduction}
\label{ch:cryo-intro}

\section{Overview of \dshort{lbnf-dune}} 
\label{sec:cryo-intro-ovvw}

The \dword{lbnf-dune} represents an international collaborative effort in neutrino physics. The \dword{dune} will be a world-class neutrino observatory and nucleon decay detector designed to answer fundamental questions about elementary particles and their role in the universe. 
The international \dword{dune} experiment, hosted by the U.S. \dword{doe}’s \dword{fnal} in Illinois, will consist of a cryogenic \dword{fd} located about 1.5\,km underground (this level is referred to as the 4850 foot level underground, or the ``4850L'') at the \dword{surf}  in South Dakota, U.S.  and a \dword{nd}  located at \dword{fnal}. \dword{surf} is \SI{1300}{\km} (\SI{800}{miles}) from \dword{fnal}).

The \dword{dune} \dword{fd} will be a very large, four-module \dword{lartpc}, each detector module in its own cryostat containing approximately \larmass (metric kilotons)\footnote{All tonnage listed in this document (with the unit ``t'' or ``kt'') is metric.} of ultra-pure \dword{lar}. 
The \dword{lartpc} technology has the unique capability to reconstruct neutrino interactions with image-like precision and unprecedented resolution. \dword{dune} will enable the study of neutrino oscillations from a new 
beamline originating from \dword{fnal}, as well as from neutrino bursts from any core-collapse supernova that occurs in our galaxy during the experiment's lifetime. 
\dword{dune} will also enable a search for proton decay.

The \dword{lbnf-dune} Project includes only two of the planned \dwords{detmodule}. They will implement different drift geometries and readout designs, and are designated \dword{sphd} and \dword{spvd}. A future project that is currently in the early planning stages, referred to informally as DUNE's ``Phase II,'' will complete the four-module \dword{fd}, double the beamline power, and add a \dword{ndgar} to the \dword{nd}~\cite{EDMS2838447}.

The \dword{lbnf} refers to the facilities and infrastructure that will support this complex system of detectors at the Illinois and South Dakota sites. Also hosted by \dword{fnal}, it is the portion of \dword{lbnf-dune} responsible for developing the neutrino beam, excavating and outfitting the underground caverns, and providing the \dword{fd} cryostats, the far and near site cryogenics systems and cryogens, and the conventional facilities. 

The cryostats and cryogenics infrastructure for the far site is part of the Project's Far Detector and Cryogenics (\dword{fdc}) subproject. The Project structure is described in Section~\ref{sec:intro-org}. 

\section{The Far Site Cryogenics Infrastructure Overview}
\label{sec:cryo:intro:ovvw}

This design report describes the \dword{lbnf} \dword{fd} cryogenics infrastructure, which includes cryostats for the first two \dword{dune} \dwords{detmodule} and the cryogenics infrastructure for them.  These facilities are largely in the \dword{doe} scope (\dword{usproj}), but include critical, committed, in-kind contributions  from the \dword{cern}, Brazil (\dword{unicamp}), Switzerland (\dword{seri}), and Poland (\dword{wust}). 
This document presents the reference design of the various systems and provides references to the documents in which they are defined.

The cryogenics infrastructure includes systems to receive, transfer, store, purify, and maintain the almost 40,000\,t  of \dword{lar} required for these first two detector modules, and is designed to support expansion to up to four detector modules. Specifically, the scope includes: 
\begin{itemize}
\item two membrane cryostats,
\item space for the remaining membrane cryostats (for a total of four), 
\item receiving facilities for \dword{lar} tanker trucks, 
\item facilities to vaporize the \dword{lar} prior to transfer to the \dword{4850l}, 
\item 
a transfer system to deliver argon gas from the  surface to the underground cavern area, 
\item a closed-loop \dword{ln} refrigeration system for condensing the \dword{lar},
\item all required piping for the argon and nitrogen in both gas and liquid phases,
\item boil-off \dword{gar} reliquefaction equipment,
\item \dword{lar} purification facilities, 
\item LAr circulation systems, 
\item regeneration systems, 
\item equipment to control the pressure inside the cryostat,
\item process controls for all equipment, and
\item the capability for emptying the cryostats at the end of the experimental run.
\end{itemize}

A strong, successful prototyping effort to qualify the cryogenics system and the cryostat technologies has been in progress for over ten years.   
A series of detectors of increasing size, starting with a \dword{35t} (metric tons of \dword{lar}), outfitted with associated
cryostats and cryogenics systems have been built and tested at \dword{fnal} and   \dword{cern} as part of the \dword{sbnd} and \dword{protodune} programs, respectively. 

\section{Principal Parameters for the Detector's Cryogenic Environment}
\label{sec:cryo-intro-princ-reqs}

\dword{lbnf-dune} has compiled a comprehensive set of requirements on the cryostats and the cryogenics system~\cite{lar-fd-req}, ensuring that these deliverables satisfy the needs and constraints of both the \dword{dune} far detector and the \dword{usproj}  \dword{fscf}, and take into account regulatory and safety codes and standards, as well as logistical concerns. This section presents a subset of the parameters that derive from these requirements, concentrating on those that affect the physical design of these systems most directly.

The overarching requirements are to provide a high-purity, 
stable \dword{lar} environment and mechanical support for the \dword{fd} \dword{lartpc} modules.

\begin{dunetable}[Parameters for the cryogenics system]{p{.3\textwidth}p{.3\textwidth}p{.3\textwidth}}
{tab:param-1-csys}
{Parameters for the cryogenics system} 
{Parameter} &  \textbf{Value} &  Note \\ \toprowrule

GAr purge flow rate &  \SI{1123}{m$^3$/hr} & From 1.2 m/hr  \\ \colhline

LAr filling fill time (Phase I) & 243/400 days & 1st/2nd cryostats only (w/ three \dword{ln} refrigeration units) \\ \colhline

LAr filling fill time  (Phase II) & 362/540 days\footnotemark & 3rd/4th cryostats only (w/ four \dword{ln} refrigeration units) \\ \colhline

Cryostat static heat leak  & 48.7\,kW & Each cryostat \\ \colhline

Electronics heat leak  & 23.7 kW & Each cryostat \\ \colhline

Total estimated heat leak   & 87.1/98.1 kW & Each cryostat with two/four pumps in operation\\ \colhline

Available cooling power  & 4 $\times$ 100 kW = 400 kW & Three \dword{ln} units for cryostats 1 and 2, fourth unit for 3 and 4 \\ \colhline

Maximum LAr circulation speed
(assuming 5 days turnover per cryostat)&  \SI{1.73}{m$^3$/min} (40 kg/s) & All four LAr pumps in operation \\ \colhline

Nominal LAr circulation per cryostat & \SI{0.43}{m$^3$/min} (10 kg/s) & Only one LAr pump in operation \\ \colhline

LAr Purity (FD1/FD2) & 100\,\dword{ppt}/50\,\dshort{ppt} & Oxygen equivalent contamination (O$_2$, H$_2$O) \\

\end{dunetable}
\footnotetext{The cryostat 4 fill time includes supplemental \dword{lar} transfer, targeting an 18 month fill time.}

\section{Scope and Design Parameters}
\label{sec:cryo-intro-scope-param}

\subsection{The Cryostats}
\label{sec:cryo-intro-cstats}

The scope of the Phase I cryostat deliverables for the \dword{dune} \dfirst{fd} includes the design, 
procurement, fabrication, testing, delivery and installation
 of the cryostats for 
 the first two \dword{fd} modules. 

The 
cryostats 
are free-standing and  constructed using membrane cryostat technology, which is widely used for transportation and storage of liquified natural gas (LNG) (Figure~\ref{fig:memb-tank-int}). This technology implements a \SI{1.2}{mm} thick stainless steel membrane to  contain the liquid and transfers the load to the insulation and support structure.  The membrane 
liner is corrugated to provide strain relief resulting from 
temperature-related expansion and contraction  (Figure~\ref{fig:inner-membrane}). Table~\ref{tab:param-1-cstat} gives the cryostat dimensions and other parameters. 
Each cryostat is passively insulated by a \SI{0.8}{m} thick layer of polyurethane foam  on all sides and the roof.  The surrounding steel support structure for the cryostats
includes a \SI{12}{mm} thick stainless steel plate serving as a vapor barrier, as well as \SI{1.1}{m} tall I-beams bearing the
weight of the  cryostat, the enclosed detector, and the contained liquid and gaseous argon.

\begin{dunefigure}[Interior of a LNG tanker ship]
{fig:memb-tank-int}
{Interior of a LNG tanker ship. 
The tank shown is 24~m high by 35~m wide with interior grid-like 
corrugations on a 0.34~m pitch. By comparison, a single LBNF 
cryostat is 14.0~m high by 15.1~m wide.}
\includegraphics[width=.85\textwidth]{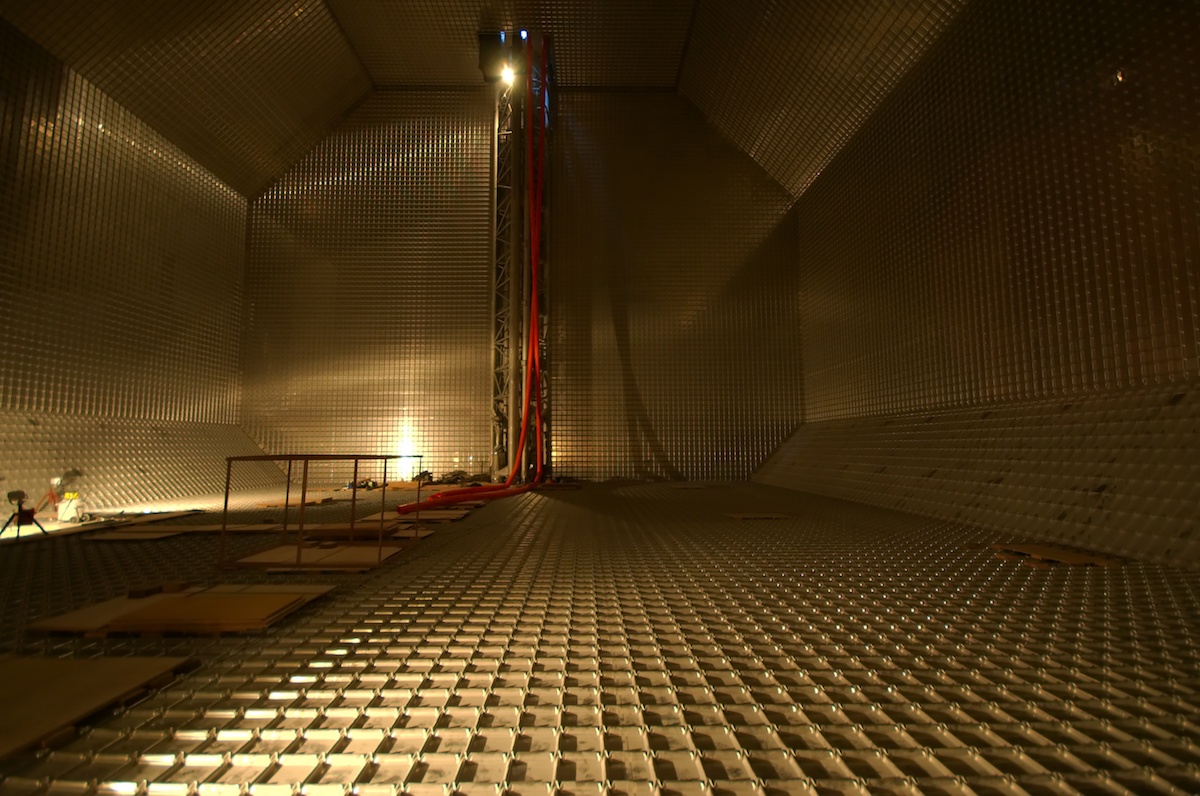} 
\end{dunefigure}

\begin{dunefigure}[Cryostat corrugated stainless steel primary barrier] 
{fig:inner-membrane}
{Corrugated stainless steel primary barrier of the membrane cryostat (this barrier is referred to as ``the membrane''). The corrugations range in height from $\sim\,$\SI{70}{mm} at the intersections (``knuckles'') down to $\sim\,$\SIrange{30}{50}{mm} in the straight sections.}
\includegraphics[width=.65\textwidth]{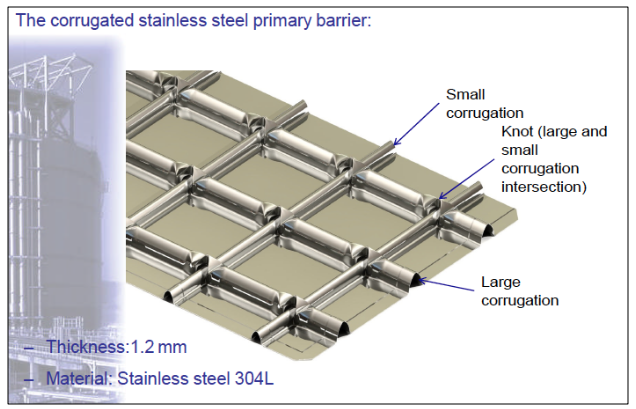}
\end{dunefigure}

\begin{dunetable}[Parameters for one \dshort{fd} cryostat]{p{.35\textwidth}p{.6\textwidth}}
{tab:param-1-cstat}
{Parameters for one \dword{fd} cryostat}
{Parameter} &  \textbf{Value} \\ \toprowrule

Cryostat Internal Volume &  13,107 m$^3$ \\
\colhline
Total LAr mass contained & \larmass \\
\colhline
Cryostat inside/outside height & 14.0 m / \cryostatht \\
\colhline
Cryostat inside/outside width & 15.1 m /  \cryostatwdth \\
\colhline
Cryostat inside/outside length & 62.0 m /  \cryostatlen \\
\colhline
Insulation &  Reinforced \SI{80}{cm} thick polyurethane; \\
           &  two (inner/outer) layers around secondary \\
           &  containment (\SI{40}{cm} per layer) \\ 
\colhline
Primary membrane (GTT Design) & \SI{1.2}{mm}  thick type 304L stainless steel \\
                              & with corrugations on \SI{340}{mm} $\times$ \SI{340}{mm} \\
                       & rectangular pitch\\
\colhline
Secondary containment (GTT Design) & $\approx$ 0.07 mm thick aluminum between 
fiberglass cloth; overall thickness is 
0.8 mm located between insulation layers \\
\colhline
External vapor barrier thickness & \SI{12}{mm} \\
(steel plates)                   &       \\
\colhline
External support structure thickness & \SI{1.1}{m}  (steel) \\
on sides, top, and bottom & \\
\colhline
LAr Temperature & \lartemp{} $\pm$ \SI{1}{K} \\
\colhline
Minimum LAr depth (liquid head) & \SI{13.37}{m} for \dword{sphd}; \SI{13.51}{m} for \dword{spvd} \\
\colhline
Ullage contents  & Ar gas (3 to 5\% of cryostat volume)  \\ 
\colhline
Ullage operating pressure & \SI{50}{mbarg} (range:\SIrange{50}{150}{mbarg}) \\
\colhline
Cryostat design pressure & 350 mbarg \\
\colhline
Personnel and equipment access to cavern & Ross Shaft\\
\colhline
Base, side walls, and roof & steel structures \\
\colhline
Moisture protection (outside cryostat) & combination of convection and continuous exhaust of cold air, replaced with heated fresh air \\
\colhline
Vapor barrier & steel plates  \\
\colhline
LAr containment system & stainless steel primary membrane; \\
                       & secondary barrier \\
\end{dunetable}

Two membrane cryostat vendors have been 
identified that are technically capable 
of delivering a membrane cryostat that meets the design requirements for 
\dword{lbnf}, GTT\footnote{GTT (Gaztransport \& Technigaz) \url{https://www.gtt.fr/}}  and 
IHI\footnote{IHI (Ishikawajima-Harima Heavy Industries) \url{https://www.ihi.co.jp/en/}}.  \dword{lbnf} has selected GTT. 
Chapter~\ref{ch:cryo-cryost} describes the GTT cryostat design. 

\FloatBarrier

\subsection{The Cryogenics System}
\label{sec:cryo-intro-sys}

The scope of the \dword{fdc} cryogenics system includes the design, procurement, on-site installation and testing of the cryogenics systems to support four \dword{dune} \dwords{detmodule}. 
The scope also includes coordination of construction,  \dword{qa}, and \dword{qc} of all the activities related to the cryogenics systems and cryostats, as well as process controls and procurement of the required \dword{lar}. This system is funded by both  \dword{doe} and non-DOE collaborative partners. 

 The overall cryogenics system is required to enable several modes of operation: 

\begin{itemize}
    \item gaseous argon (GAr) purge,
    \item cryostat and detector \cooldown{},
    \item cryostat fill with \dword{lar},
    \item steady-state operations with LAr circulation, and
    \item cryostat emptying.
\end{itemize}

The cryogenics subsystems are defined as follows:

\begin{itemize} 

\item The \textit{infrastructure cryogenics} supports the needs of the cryostat and the proximity cryogenics. This subsystem  provides the
equipment to receive \dword{lar}, vaporize the liquid and transfer the gas underground. 
It is responsible for the designs of the \dword{ln} refrigeration system and of the nitrogen system (composed of the refrigeration system and the \dword{ln} buffer tanks,  
the nitrogen generation, and \dword{ln}/GN$_2$ distribution). It also provides the argon distribution system. This subsystem, described in Section~\ref{ch:cryosys:subsys}, is funded by the \dword{doe}.

\item The \textit{proximity cryogenics} consists of all the systems that take the 
\dword{lar} from the infrastructure cryogenics and
deliver it to the cryogenics components inside the cryostat at the required temperature, pressure, purity, and mass flow rate. This subsystem circulates and
purifies the \dword{lar},  condenses  
the boil-off \dword{gar}, and condenses the \dword{gar} during the cryostat fill, returning it to the \dword{lar} circulation and purification stream. It comprises the \dword{gar} purification system, the argon condensers, the \dword{lar} filtration and associated regeneration systems, and the \dword{lar} pumps 
and monitoring instrumentation. See Section~\ref{sec:prox-cryogenics}.  Most of the proximity cryogenics is the responsibility of non-DOE partners.

\item The \textit{internal cryogenics} is located within the cryostats themselves and includes the design of all items needed to distribute
\dword{lar} and \dword{gar} throughout the volume, and all features needed for the commissioning, \cooldown, fill, and steady-state operations of the
cryostats and detectors. See Section~\ref{sec:internal-cryo}. The internal cryogenics is the responsibility of a non-\dword{doe} partner.
\end{itemize} 

The international engineering teams that will design, manufacture, commission, and qualify these subsystems 
benefit from the experience of the \dword{sbn} and \dword{protodune} programs. 

Table~\ref{table:cryog-params} lists parameters for the overall cryogenics system for one cryostat. 

\begin{dunetable}[Key parameters for cryogenics system]{p{0.05\textwidth}p{0.70\textwidth}p{0.20\textwidth}}{table:cryog-params}
{Key parameters for the cryogenics system for one detector module}
& {Parameter} &  \textbf{Value (per cryostat)} \\ \toprowrule
1& Piston purge \dshort{gar} vertical flow rate  &  \SI{1.2}{m/hr}\\ \colhline

2& Maximum \cooldown rate (\dword{sphd}) & \SI{60}{K/hr} \\ \colhline

3& Maximum temperature gradient between any two points in TPC during \cooldown & \SI{50}{K}\\ \colhline

4& Maximum available cooling power during fill for cryostats 1 and 2 &  226 and 139 kW, respectively \\ \colhline

5& Maximum available cooling power during fill for cryostats 3 and 4 &  151 and 64 kW, respectively  \\ \colhline
6& Maximum available cooling power (for all purposes except fill)   & \SI{100}{kW} \\ \colhline
7& Required \dshort{lar} purity (oxygen equivalent contamination) for \dword{sphd} normal operations  &  $<\,$\SI{100}{ppt} \\ \colhline
8& Required \dshort{lar} purity (oxygen equivalent contamination) for \dword{spvd} normal operations  &  $<\,$\SI{50}{ppt} \\ \colhline
9& Full volume turnover rate; normal operations  & \SI{5.5}{days} \\
\end{dunetable}

Each of the parameters outlined in Table~\ref{table:cryog-params} fulfills a specific need during the various operational modes of the cryogenics subsystem, namely:

The linear flow rate (row 1 of Table~\ref{table:cryog-params}) of \dword{gar} during the purge (Section~\ref{sec:cryosys-proc-purge-cool}) prevents back-diffusion of oxygen through the purge gas, which would inhibit the process. This flow rate has been experimentally verified in several other \dwords{lartpc}, including 
\dword{microboone}, the \dword{wa105}, and most recently, \dword{pdsp} and \dword{pddp}. At \dword{lbnf}, this corresponds to a volumetric flow rate of \SI{1123}{\cubic\meter\per\hour}.

The maximum \cooldown rate (row 2) of
the \dword{tpc} and the maximum temperature gradient between any two points (row 3) ensure mechanical stability of the \dword{tpc} during \cooldown (Section~\ref{sec:cryosys-proc-cool}). They have been selected in order to protect the \dword{tpc} components from unacceptable stresses arising from thermal gradients between these components and the surrounding structure.

The available cooling power during the fills (rows 4--5) is a function of the total available refrigeration, less the cooling required to offset the various sources of heat ingress into the cryogenics subsystems and the cooling required to maintain the liquid in previously filled cryostats. The \cooldown is discussed in Section~\ref{sec:cryostat-fill}. 

The maximum cooling power (row 6) is the sum of all estimated heat loads into the subsystem during peak operations with some additional margin. 

The \dword{lar} purity values (rows 7--8) ensure proper operation of the \dwords{tpc} by enabling an electron lifetime greater than \SI{3}{ms} as required by \dword{dune} for \dword{sphd}, and greater than \SI{6}{ms} for \dword{spvd}, which has a longer maximum drift length. These purity levels shall be achieved by recirculating each \dword{lar} volume through the purification system (discussed in Section~\ref{subsec:argon-pur}) at a nominal turnover rate (row 9) of 5.5 days, corresponding to a purification rate of \SI{40}{kg/s}. Previous experience with \dword{pdsp} validates this parameter, as the experiment demonstrated that the smaller detector was able to achieve and sustain an electron lifetime in excess of \SI{30}{ms} with a comparably slower 10.8 day turnover.

\section{Management and Organization}
\label{sec:intro-org}

Modeled after the project structure of traditional accelerator-based particle physics experiments, such as the \dword{atlas} and \dword{cms} experiments at the \dword{lhc}, the \dword{lbnf-dune} construction project (the ``Project'') includes

\begin{itemize}
\item The \dword{usproj} Project (the ``U.S. Project'') to design and build the conventional and beamline
facilities and the \dword{doe} contributions to the detectors; it is organized as a \dword{doe}/\dword{fnal} project
and incorporates contributions to the facilities from international partners. It also acts as host for
the installation and integration of the detectors.

\item Projects funded by multiple international partners for the detectors to design, build, and install
the detector components; the deliverables from these projects are collectively organized and
coordinated by the international \dword{dune} collaboration with oversight provided by \dword{fnal} on
behalf of all stakeholders.

\end{itemize}

\dword{usproj} is divided into five subprojects:
\begin{itemize}
\item Far Site Conventional Facilities Excavation (FSCF-EXC),
\item Far Site Conventional Facilities Buildings and Site Infrastructure (FSCF-BSI),
\item Near Site Conventional Facilities and Beamline (NSCF+B),
\item Far Detector modules 1 and 2 and Cryogenics (FDC), and
\item Near Detector (ND).
\end{itemize}

All work required for completion of the Project 
is included in a single, integrated \dword{wbs}. The \dword{doe}-funded portion of the work includes all work that began in FY10 (\dword{doecd}-0) and continues through to project completion at each subproject’s \dword{doecd}-4.

  The \dword{fscf} and \dword{fdc} subprojects all support the \dword{dune} Far Detector at \dword{surf}. The scope of the cryogenics part of \dword{fdc} is  to plan, design, and construct the cryostat and cryogenic systems.  The \dword{fdc} Deputy Project Manager for Cryogenics is responsible for leading the planning, design, and construction of the cryogenics systems and works in conjunction with \dword{cern} to deliver the cryostats and other cryogenics systems. 
  The \dword{cern} Manager for Cryogenics Infrastructure 
  is responsible for leading the planning, design, and construction of the cryostats and cryogenics infrastructure in conjunction with the rest of the LBNF team.
  These project managers will be assisted by other staff whom the managers will organize into a project team. This includes engineers, designers, and managers from \dword{fnal} and \dword{cern}, as well as contract engineers and designers, to execute the planning and design, oversee the technical aspects of procurements, and oversee the installation. This team will work closely with the two  \dword{fscf} Project Managers, \dword{fdc} Project Manager, 
  far site integration and installation (\dword{fsii}), and \dword{surf} staff.   The management structure is shown in Figure~\ref{fig:mgmt-struc}. 

\begin{dunefigure}[Management Structure] 
{fig:mgmt-struc}
{Cryogenics Infrastructure Project Management Structure}
\includegraphics[width=0.88\textwidth]{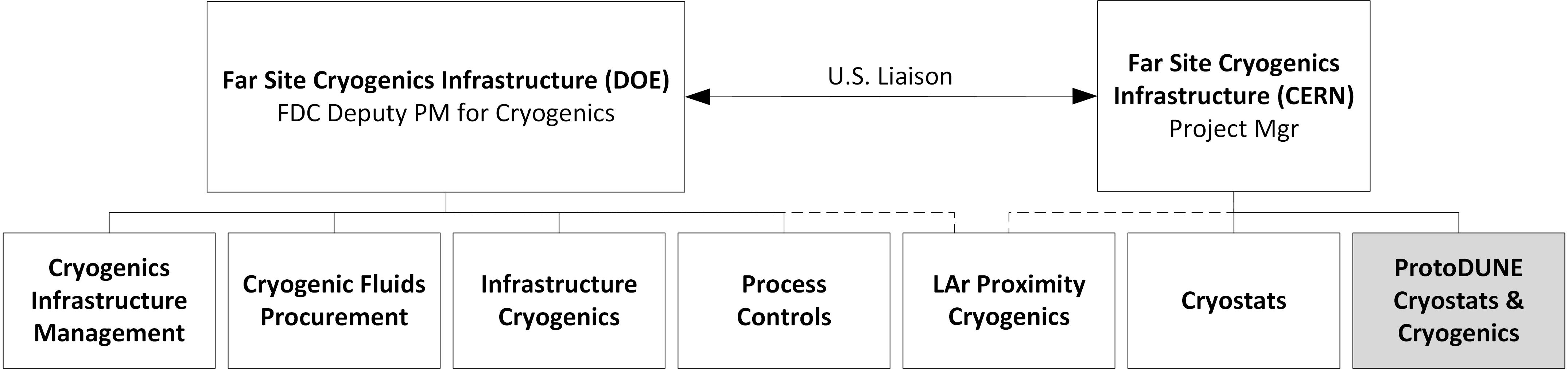}
\end{dunefigure}

\section{Participants and Scope Assignment}
\label{sec:intro-assign}

The \dword{fnal} Directorate and \dword{usproj} have been working with \dword{doe} leadership to
identify responsible parties for non-DOE scope. 
All scope to support the first two far detector modules is committed by either the \dword{doe} or partners. 
Table~\ref{table:cryo-scope-assgn} lists the projected scope assignments.

\begin{dunetable}[Scope assignments]{p{0.63\textwidth}p{0.15\textwidth}p{0.15\textwidth}}{table:cryo-scope-assgn}
{Scope assignments}
{Item} &  \textbf{FD1-HD} &  \textbf{FD2-VD} \\ \toprowrule

 Cryostat (part of Far Site integration) & CERN & CERN \\ \colhline
Cryogenics systems integration & DOE & DOE \\ \colhline
 Nitrogen System (Refrigeration + \dshort{ln} storage + distribution) -- engineering / manufacturing / installation/ commissioning & DOE & DOE\\ \colhline
 
\dshort{lar} receiving facilities (surface) & DOE + CERN  & DOE + CERN \\ \colhline
 Argon condensers system -- engineering / manufacturing & Switzerland & Brazil \\ \colhline
 Argon distribution & DOE & DOE \\ \colhline
 Argon purification and regeneration -- engineering / manufacturing & Brazil & Brazil \\ \colhline
 LAr circulation -- engineering / manufacturing & Brazil & Brazil \\ \colhline
 \dshort{gar} boil-off and pressure control & DOE & DOE \\ \colhline
 Process controls & DOE & DOE \\ \colhline
 Internal cryogenics -- engineering / manufacturing (part of Far Site Integration) & Poland & Poland \\ \colhline
Installation of in-kind contributions (IKC) & DOE & DOE \\  \colhline
Collect safety docs and obtain safety approval & DOE & DOE\\ \colhline
Purge and \cooldown & DOE & DOE \\  \colhline
\dshort{lar} procurement and fill ($\ast$ \dshort{lar} only, not labor) & DOE & DOE$\ast$ \\ 

\end{dunetable}

The organization and management is fully described in~\cite{bib:docdb117}.
\FloatBarrier

\section{Acquisition Strategy}
\label{sec:cryo-intro-acq}

The FDC Cryogenics team is using an acquisition strategy known as Engineering/Manufacturing/ Installation/Testing and/or Startup in which design documents and specifications are given to a subcontractor or partner to produce/assemble and deliver given systems or equipment. The subcontractor or partner then bears the performance risk for the deliverables. 

The team has 
developed conceptual designs for the cryogenics systems needed to support the DUNE far detector modules. 
The designs are mature enough 
to ensure 
that they meet the requirements and for the team to write performance specifications and interface documents that will be used by subcontractors or partners to produce/assemble and deliver the equipment. 

The scope of the FDC cryogenics systems acquisition includes engineering, manufacturing, installation, testing and startup as appropriate (e.g., for the nitrogen system and the argon receiving facilities).
The reference design allows partners to familiarize themselves with the systems (that they have to deliver) and write their own procurement documents. Or they can use the performance specifications and interfaces written by the FDC team and customize them for their specific needs.
This document presents the reference design of the various systems and references to the actual documents.

\section{Risks}

Figure~\ref{fig:risks} lists the open risks for the Far Site Cryogenics, along with their ranking as high, medium or low, and their probabilities and impacts. 

\begin{dunefigure}[Cryogenics infrastructure risks] 
{fig:risks}
{Cryogenics infrastructure risks}
\includegraphics[width=\textwidth]{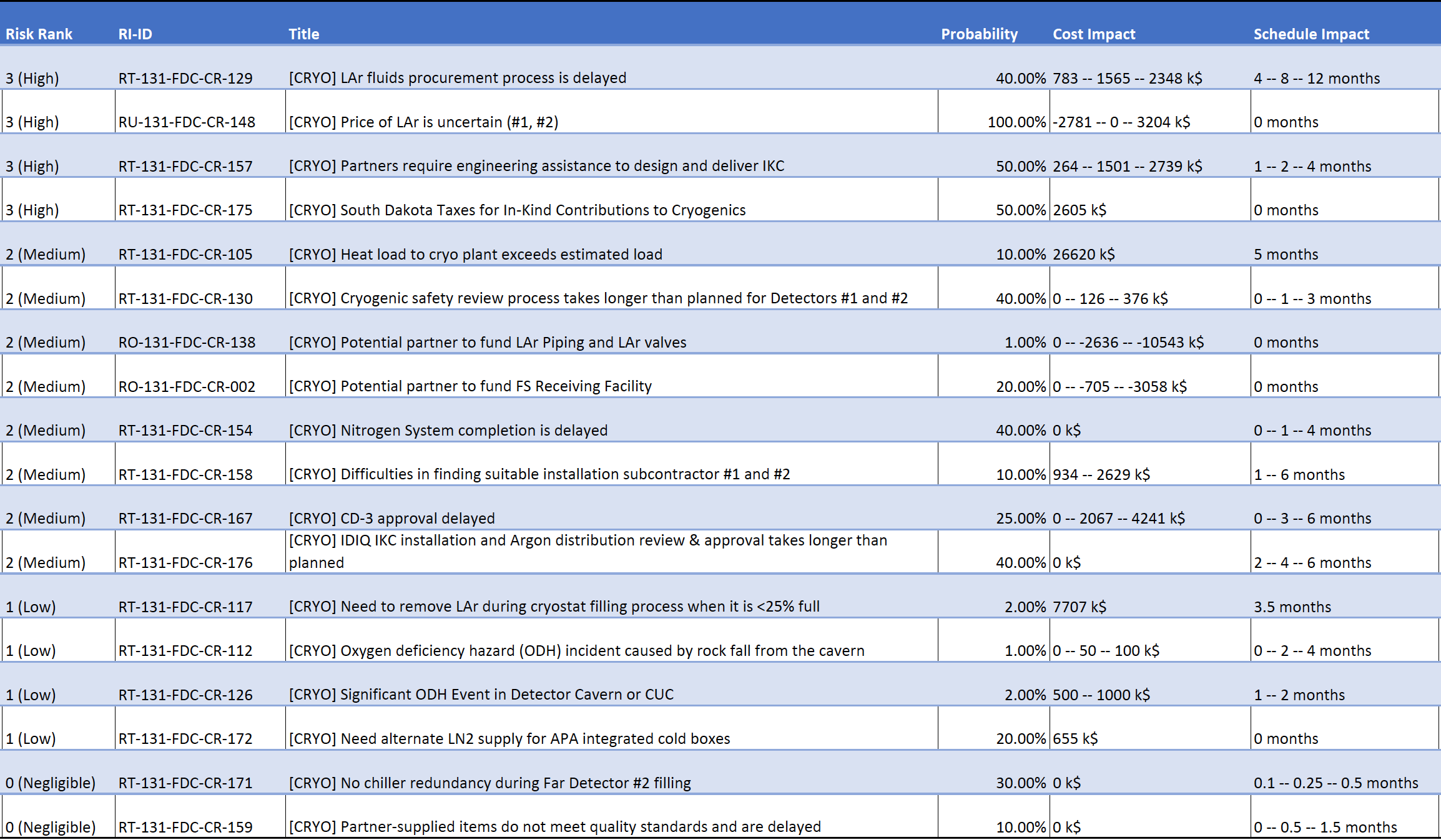}
\end{dunefigure}

\cleardoublepage


\chapter{Cryostat Configuration and Construction}
\label{ch:cryo-cryost}

The cryostats for the \dword{dune} \dword{fd} will be constructed using membrane cryostat technology. This technology, employing multiple layers of support and insulation, is commonly used for liquefied natural gas (LNG) storage and transport tanker ships, and has been proven to be an excellent option for \dword{lartpc} experiments~\cite{Adams_2020, adams2020protodunesp}. 

The cryostats are in-kind contributions from the \dword{cern}, and thus outside the \dword{usproj} scope. They are described here for completeness. 
Cryostat installation is included in the \dword{fdc} installation activity (\dword{fsii}) schedule.

\section{Cryostat Inner Structure}
\label{sec:cryostat:struct}

In the cryostat design, a \SI{1.2}{mm} thick, corrugated stainless steel membrane, Figure~\ref{fig:inner-membrane}, forms a sealed container for the \dword{lar}, with surrounding layers of thermal insulation and vapor barriers. Outside these layers, a free-standing steel frame forms the outer (warm) vessel, the bottom and sides of which support the  hydrostatic load. The roof of the cryostat supports most of the components and equipment within the cryostat, e.g., the \dword{tpc} and \dword{pds} components, electronics, and sensors. The majority of the cryogenic piping runs along the bottom and is supported by the floor. The cryostat dimensions and other parameters are given in Table~\ref{tab:param-1-cstat}.

Membrane tank vendors, including GTT, have a ``cryostat in a kit'' design that 
incorporates the insulation and secondary barriers into packaged 
units. Figure~\ref{fig:gst-composite}   
illustrates, from innermost to outermost layers, the composition of the bottom and side walls of the membrane cryostat, which consist of 
\begin{itemize}
 \item{a corrugated, stainless steel primary membrane in contact with the \dword{lar},} 
 \item{inner layer of insulation (polyurethane foam),}
 \item{a secondary barrier (a thin triplex secondary membrane that contains the 
       \dword{lar} in case of a leak in the primary membrane),}
 \item{outer layer of insulation,}
 \item{a steel barrier to prevent water vapor ingress, and}
 \item{the steel frame (the ``warm,'' i.e., room temperature, outer vessel, not pictured in Figure~\ref{fig:gst-composite}).}
\end{itemize}  

\begin{dunefigure}[GST composite system from GTT] 
{fig:gst-composite}
{GST composite system from GTT.}
\includegraphics[width=0.5\textwidth]{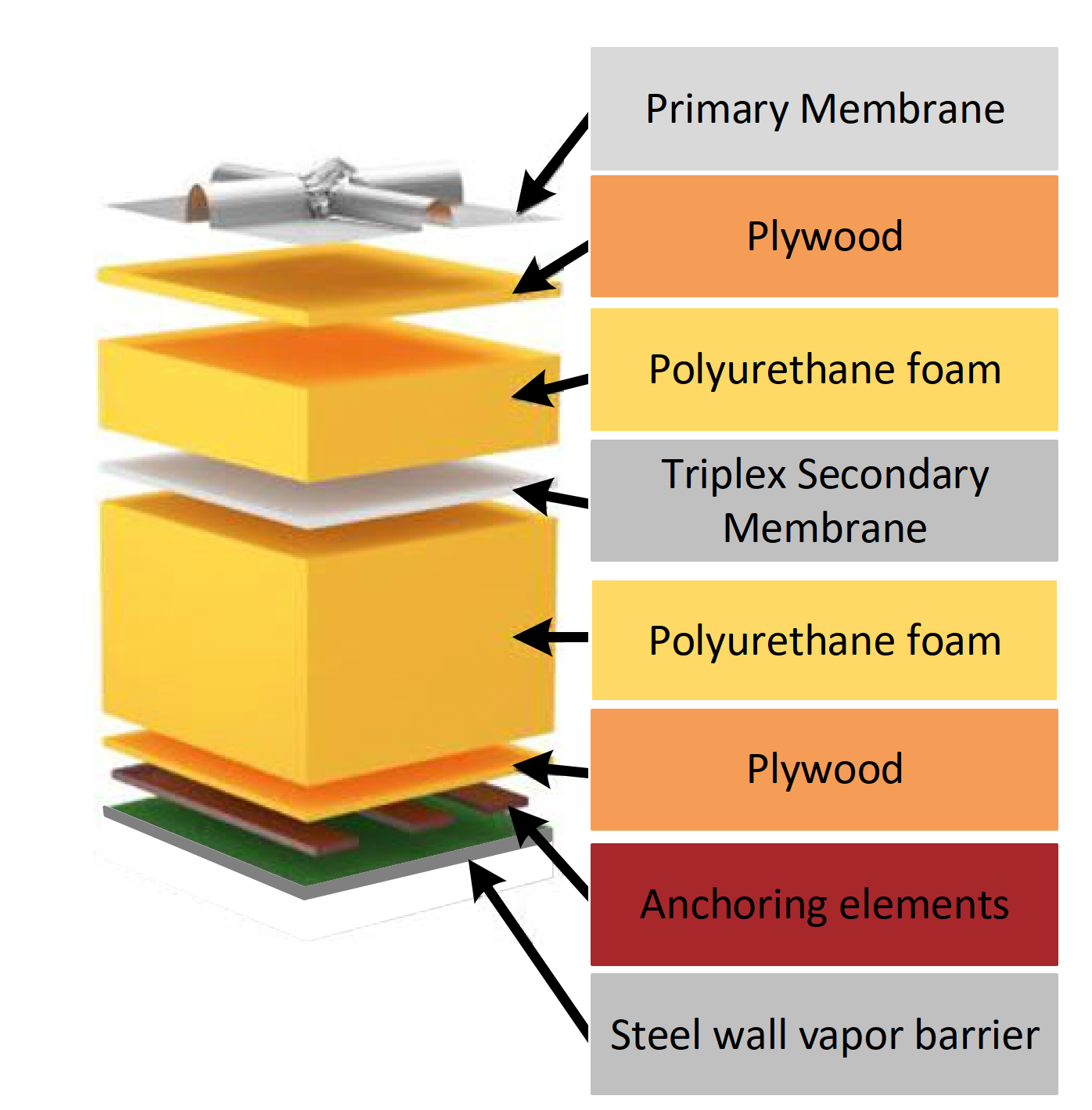}
\end{dunefigure}

The cryostat roof design, described in Section~\ref{sec:cryo-roof}, is somewhat different. 

Each cryostat is positioned at
one end of a detector cavern with enough air space reserved for a combination of convection
and forced air circulation, which maintains the outer steel structure temperatures above
freezing.


To minimize the heat ingress and the required 
refrigeration load, the membrane cryostat requires insulation between the 
primary stainless steel membrane and the exterior vapor barrier (steel plates). 
Our choice of insulation thickness of \SI{80}{cm}, tested in \dword{protodune}, limits the heat input to a cryostat to an acceptable \SI{48.7}{kW}, as described in Chapter~\ref{sec:cryo-cryosys-proto-plans}.

The insulation material, a solid polyurethane, is manufactured in 1\,m $\times$ 3\,m 
composite panels. The panels will be laid out in a grid with 
\SI{3}{cm} gaps between them (to be filled with loose fiberglass) and 
fixed onto anchor bolts embedded into the steel outer structure
at about $\sim$\SI{3}{m} intervals. 
The composite panels contain the outer insulation layer, the secondary 
membrane, and the inner insulation layer. After positioning adjacent 
composite panels and filling the \SI{3}{cm} gap, the secondary membrane 
is spliced together by epoxying. All seams are covered so that the secondary
membrane becomes a seamless liner. A corner detail is shown 
in Figure~\ref{fig:vessel-corner}.

\begin{dunefigure}[Membrane corner detail] 
{fig:vessel-corner}
{Membrane corner detail.}
\includegraphics[width=.65\textwidth]{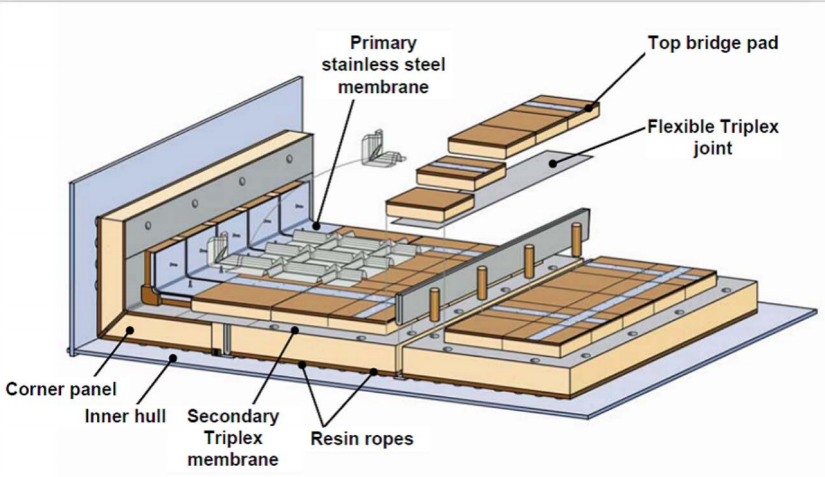}
\end{dunefigure}

The secondary membrane is composed of a thin aluminum sheet covered on both sides with 
fiberglass cloth, a roughly $\sim\,$\SI{1}{mm} thick composite called triplex that is 
very durable and flexible.  
This membrane is placed within the insulation space,
surrounding the internal tank on the bottom and sides, and  
separating the insulation space into two distinct, leak-tight, 
inner and outer volumes. 
It is connected 
to embedded metal plates in the vertical steel wall at the upper 
edge of the tank and separated from the steel frame by the outer insulation layer. In the unlikely event of an internal leak from 
the cryostat's primary membrane into the inner insulation space, the 
secondary membrane will contain 
the liquid cryogen.

\section{Exterior of Cryostat: Vapor Barrier and Steel Structure} 

The functions of the steel warm vessel are to (1) contain the membrane 
vessel and the intermediate layers, (2) provide mechanical support, and (3) provide a vapor barrier from the outside. 
The vapor barrier, welded to the inside surfaces of the 
outer steel structure, is required on all six faces of the cryostat  
to  prevent the ingress of water vapor into the insulation 
space. If water vapor were permitted to migrate into the 
insulation space, it could freeze and degrade the thermal 
performance of the insulation. The barrier must also 
reliably absorb the stresses and strains from all normal 
loading conditions. The selected vapor barrier material 
is \SI{12}{mm} thick steel plate. 

Each of the four free-standing cryostats will be positioned on a firm surface with 
no additional structural connections necessary for either 
the cavern floor or the cavern walls. It will be positioned with enough air space 
 for convection and forced air circulation, to maintain  
the steel temperature above the dew point temperature to prevent condensation and freezing. 
The distance from the outer surface of the structure to the cavern wall will
vary from \SIrange{20}{50}{cm}.
The internal and external dimensions of a cryostat are given in Table~\ref{tab:cstat-dim}. 

\begin{dunetable}[Cryostat inside and outside dimensions]{ll}{tab:cstat-dim}
{Cryostat inner and outer dimensions}
{Dimension} &  \textbf{Value} \\ \toprowrule

Cryostat inside/outside height & 14.0 m / \cryostatht \\
\colhline
Cryostat inside/outside width & 15.1 m /  \cryostatwdth \\
\colhline
Cryostat inside/outside length & 62.0 m /  \cryostatlen \\
\end{dunetable}

The outer vessel consists of outer supporting steel profiles, interconnected 
through a steel grid and a \SI{12}{mm} thick steel 
continuous plate on the inner side that is in contact with the membrane insulation.
The material used is S460ML structural carbon steel, with yield 
strength of \SI{430}{MPa} and tensile strength of \SI{510}{MPa}. The main 
profile used is HL 1100$\times$548 or its ASTM alternative 
W 44$\times$16$\times$368. Four profiles are bolted together, by 
four 
connections, forming a structural ``portal.'' Each 
bolting connection consists of 16 bolts (M42). The additional 
grid is made of the IPE300 profile. The total self-weight of 
the structure is approximately \SI{2000}{t} (metric).

The main advantage of this design is the fact that such a structure 
can be fully decoupled from the civil engineering work related to
the excavation and finishing of the four caverns. All components 
can be procured and prepared on the surface, ready to be lowered 
through the shaft. Underground installation will take 12   
months for each of the cryostats and can be done sequentially. 
The warm vessel will be fully accessible from outside and can be 
inspected at any time. Stairs and gangways are included 
in the design. 

The design was done mainly using ANSYS code, a commercial \dword{fea} method. 
The design adheres to the Eurocode III codes, which were verified using 
a safety equivalency review process with U.S. codes~\cite{EDMS2826471}. 

Approximately \SI{17.5}{kt} of \dword{lar} acts as load on the floor 
($\sim\,$\SI{20}{t\per\square\meter}). Approximately \SI{8}{kt} of hydrostatic force acts 
on each of the long walls, with triangular distribution over the 
height, and around \SI{2}{kt} of hydrostatic force acts on each 
of the short walls.  Additionally, a normal ullage operational pressure 
of \SI{50}{mbarg} (\SI{0.5}{t/\square\meter}) acts on every 
wall. The structure has been verified to accommodate all this plus an 
accidental overpressure, up to a total of 
of 350\,mbarg (3.5 t/m$^{2}$), the design pressure 
of the cryostat.  The calculations take into account the weight of the installed detector, as well as any potential 
seismic action.

The following \dword{fea} models and analysis methods have been developed and used\footnote{Solid, shell, and beam models refer to techniques used to model specific geometries. The latter two imply a large computational savings; a shell element is a 2D abstraction of a 3D shape (used when the thickness of the object is much smaller than the length), and a beam element is a 1D abstraction of a 3D shape (typically used for long shapes of constant cross section). }: 

\begin{itemize}
 \item{a 
 model for evaluation of the global behavior of the entire structure;}
 \item{analytical models of a single portal, i.e., four main beams  connected together (roof, floor, and the two side walls), adjacent portals 
 run the full length of the cryostat; }
 \item{an additional shell model of single cell (one portal and eight additional grid beams) 
 to study the main elements of the structure  in more detail, }
 \item{specific  analyses, i.e., linear (eigenvalue) and nonlinear buckling,  have been performed on the following parts of the structure to evaluate their stability:}
 \begin{itemize}
  \item{a single portal by using beam elements, and}
  \item{a single cell (one portal and two additional grid beams, one 
        on the left and the other on the right) using two different \dword{fea} models: 
 (1) beam and shell elements (using ANSYS Workbench), and (2) only shell elements (using ANSYS APDL).}
 \end{itemize}
 \item{very detailed models on the connections (bolting and/or welding) 
       on a single portal using solid and contact elements.}
\end{itemize}

The maximum stress levels on the main profiles at the location of 
the maximum moment are on the order of \SI{125}{MPa}, which allows a 
safety factor of four with respect to the tensile strength of the 
chosen material. Additional bracing of the main profiles increases 
the stability of the structure by a factor of 2.5, as verified by 
stability analyses.  

\section{Cryostat Roof}
\label{sec:cryo-roof}
The stainless steel primary membrane and all the intermediate layers 
except the secondary membrane, which is only needed to protect against liquid leaks,    
continue across the roof of the 
cryostat, providing a vapor-tight seal.  Recall that the cryostat roof must support 
detector components and equipment
inside the cryostat. 
Except for sidewall penetrations from
the external \dword{lar} recirculation pumps, all piping 
and electrical penetrations into the 
interior of the cryostat are made through 
the roof to minimize the potential for leaks. 

To construct the roof, studs are first welded to the underside of the steel plates of the roof to bolt the 
insulation panels to the steel plates. Insulation plugs are then inserted 
into the bolt-access holes.  The primary membrane panels 
are tack-welded to hold them in place, then fully 
welded to complete the inner cryostat volume. 

Feedthrough 
ports located at regular intervals along the corrugation pattern
of the primary membrane along the roof will accommodate \dword{tpc} hangers, electrical 
and fiber-optic cables, and piping. See Figure~\ref{fig:v5ch2-roof-nozzle}. 

All connections into the cryostat (again, except for the \dword{lar} 
recirculation pumps) will be made 
via nozzles or penetrations above the maximum liquid level, 
and located on the roof. See Figure 
\ref{fig:v5ch2-roof-nozzle} for a typical roof-port 
penetration.  

\begin{dunefigure}[Nozzles in the roof of membrane cryostat] 
{fig:v5ch2-roof-nozzle}
{Nozzles in the roof of membrane cryostat (Figure courtesy GTT).}
\includegraphics[width=0.48\textwidth]{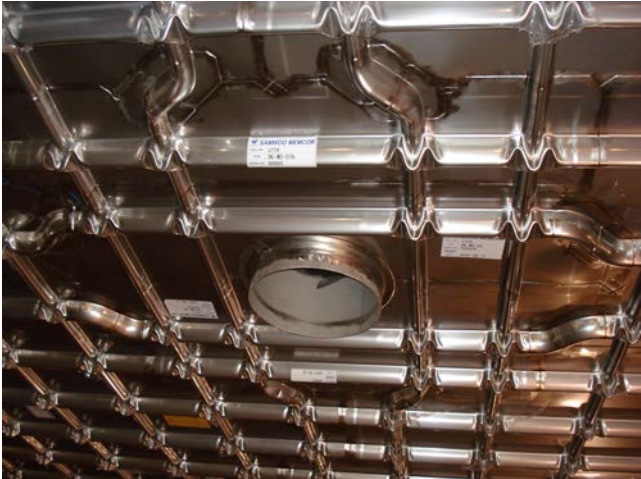} 
\end{dunefigure}

\section{Leak Prevention}
\label{sec:cryo-cryosys-leak}

To prevent infiltration of water-vapor or oxygen through 
microscopic membrane leaks (below detection level) the 
insulation spaces will be continuously purged to provide 
one volume exchange per day.  

The insulation space between the primary and 
secondary barriers will be maintained at \SI{15}{mbarg},  
slightly above atmospheric pressure. This space will
be monitored for changes that might indicate a leak 
from the primary membrane.  The outer insulation space 
will also be purged with nitrogen at a slightly different 
pressure. The pressure gradient across the membrane walls 
will be maintained in the outward direction. Pressure-control devices and relief valves will be installed on both insulation 
spaces to ensure that the pressures in those spaces do not 
exceed the operating pressure inside the cryostat. 

The purge gas will be provided by a nitrogen generator, separating nitrogen from compressed air. The purge system is not safety-critical, and an outage of the nitrogen generator would have no impact on operations.

All welds will be tested. The primary membrane will be subjected to several leak tests and to weld remediation, as necessary. 
\cleardoublepage

\chapter{Cryogenics Subsystems}
\label{ch:cryosys}
\section{Infrastructure Cryogenics}
\label{ch:cryosys:subsys}

The Infrastructure Cryogenics includes (1) the \dword{lar} receiving and vaporizing facilities (above-ground), (2) the systems to transfer the vaporized argon to the \dword{4850l}, (3) the nitrogen
system (below-ground), (4) argon and nitrogen distribution systems  (below-ground), and (5) the process controls (above-and below-ground). Piping and Instrumentation Diagrams (P\&ID)~\cite{EDMS2211532,EDMS2775007,
EDMS2775008,EDMS2775002} detail the cryogenics systems and associated process instrumentation.

\subsection{Argon Receiving and Vaporization Facilities at the Surface} 
\label{ch:cryosys:subsys:ic:receiv}

Figure~\ref{fig:cryog-surface-receiving-fac} illustrates 
the layout~\cite{EDMS2211569,EDMS2447338}  for the planned cryogen receiving station near the Ross Headframe at the \dword{surf}.  

\begin{dunefigure}[Argon receiving facilities near the Ross Headframe] 
{fig:cryog-surface-receiving-fac}
{View of \dword{lar} receiving facilities near the Ross Headframe. The vertical cylinders are \dword{lar} dewars providing a total of \SI{200}{\cubic\meter} storage. The horizontal cylinders are redundant \dshort{lar}  vaporizers; \dshort{lar}  must be vaporized before proceeding down the Ross Shaft. The gray stations serve to receive
\dshort{lar} from delivery trucks; received \dshort{lar} is tested for purity automatically before deliveries are allowed to proceed.} 
\includegraphics[width=0.88\textwidth]{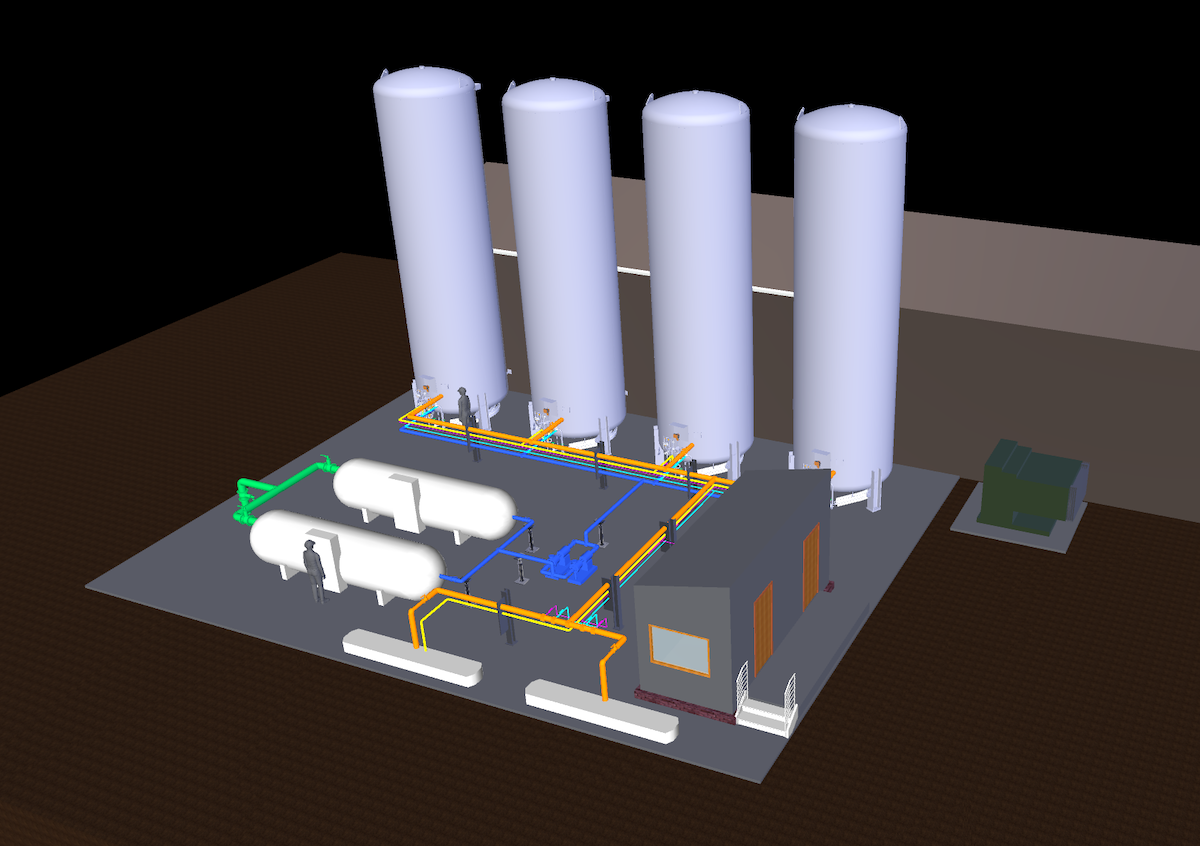}
\end{dunefigure}

To prepare for road tankers to arrive and efficiently offload \dword{lar} at the Ross Headframe, \dword{surf} must provide vehicle access and hard-surfaced driving areas 
adjacent to receiving facilities there.  

The receiving station will be available for \dword{lar} deliveries during the initial
filling period, with a set of four interconnected receiving tanks each with a capacity of \SI{50}{\cubic\meter}.  
The total receiving capacity is thus \SI{200}{\cubic\meter}. 

\subsubsection{Liquid Argon Testing and Vaporization}

To ensure that it meets the purity specification, each \dword{lar} road tanker load will be tested using an analyzer rack with instruments to check water, nitrogen, and oxygen content.   Only \dword{lar} that meets or exceeds the specification will be received into a tank. The tanks serve as a buffer volume 
enabling receipt of \dword{lar} at a pace of about four 20\,t-capacity \dword{lar} trucks  
per day during the fill period.  

The \dword{lar} is transferred from the storage tank to a \SI{300}{kW} vaporizer that vaporizes it and warms up the resulting gas to 
room temperature. The warm \dword{gar} is then transferred down the Ross Shaft (Section~\ref{ch:cryosys:subsys:ic:xfer-to-4850}).
The \dword{lar} transfer has been modelled with a process simulator to verify pipe sizing and operating parameters~\cite{EDMS2211567}. Pumps may be used to send the \dword{lar} through the vaporizers, if needed. They are accounted for in the space and utilities budget. The size of the vaporizer is driven by the size of the condensers available underground to recondense the \dword{gar} and fill the cryostats. See Sections~\ref{sec:comp:ln2} and~\ref{sec:prox-cryogenics-detcav} for details. 

\subsection{Systems to Transfer Vaporized Argon (GAr) to the 4850L}
\label{ch:cryosys:subsys:ic:xfer-to-4850}

\begin{figure}[htbp]
\centering
\includegraphics[width=.9\textwidth]{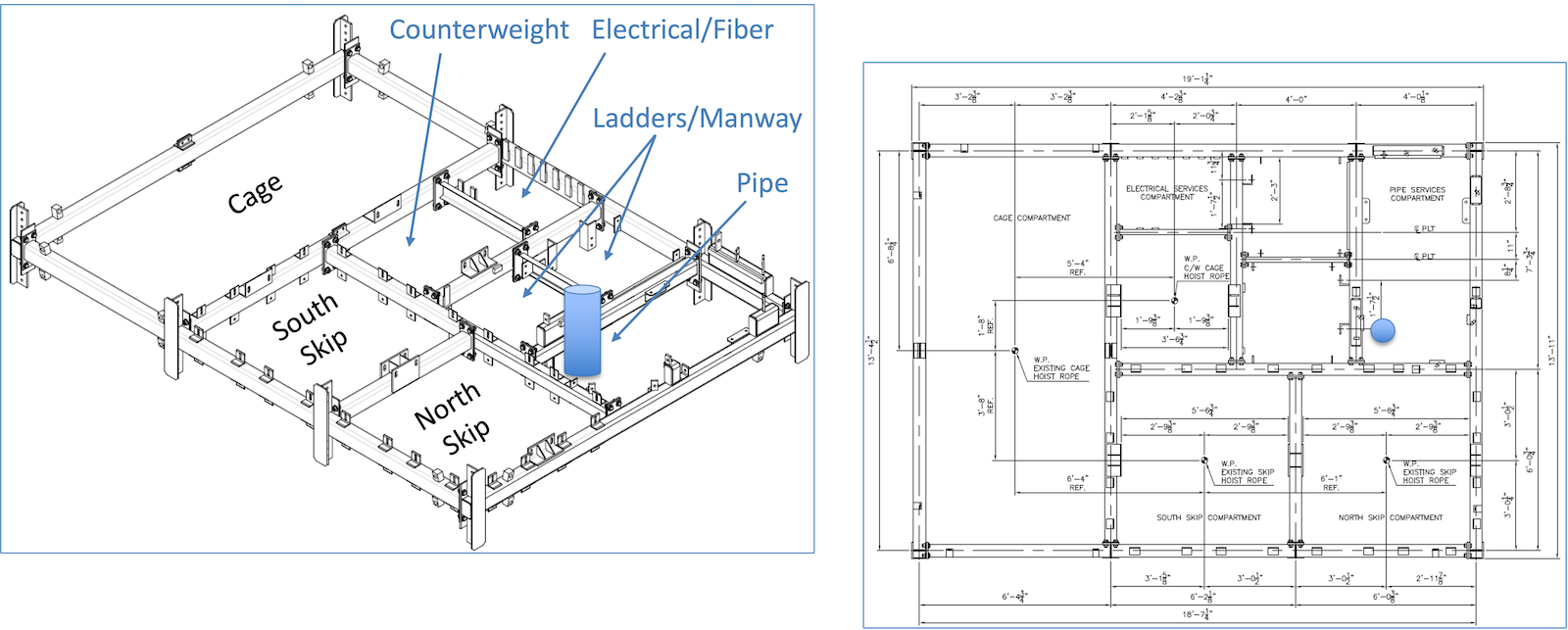} 
\caption[Ross Shaft frame with GAr pipe]{Ross Shaft frame with GAr pipe}
\label{fig:framing-at-ross-piping}
\end{figure}

The \dword{gar} is transferred to the \dword{4850l} at ambient 
temperature and a pressure of \SI{0.24}{MPa}, via a vertical \dword{gar} transfer line (one 6-inch SCH 40 stainless steel pipe) that passes down a utility chase in the Ross Shaft, as shown in Figures~\ref{fig:cryog-eqp-between} and~\ref{fig:framing-at-ross-piping}. The line is 8-inch SCH 40 above-ground; underground it is reduced to 6 inches inside the shaft to ease installation.

At the \dword{4850l}, the piping exits the 
shaft and runs along a drift to the \dword{cuc} where it deposits the \dword{gar} into a gas filtration system.

\begin{dunefigure}[Cryogenics system components between the surface and the caverns] 
{fig:cryog-eqp-between}
{Cryogenics system components between the surface and the caverns.}
\includegraphics[width=0.8\textwidth]{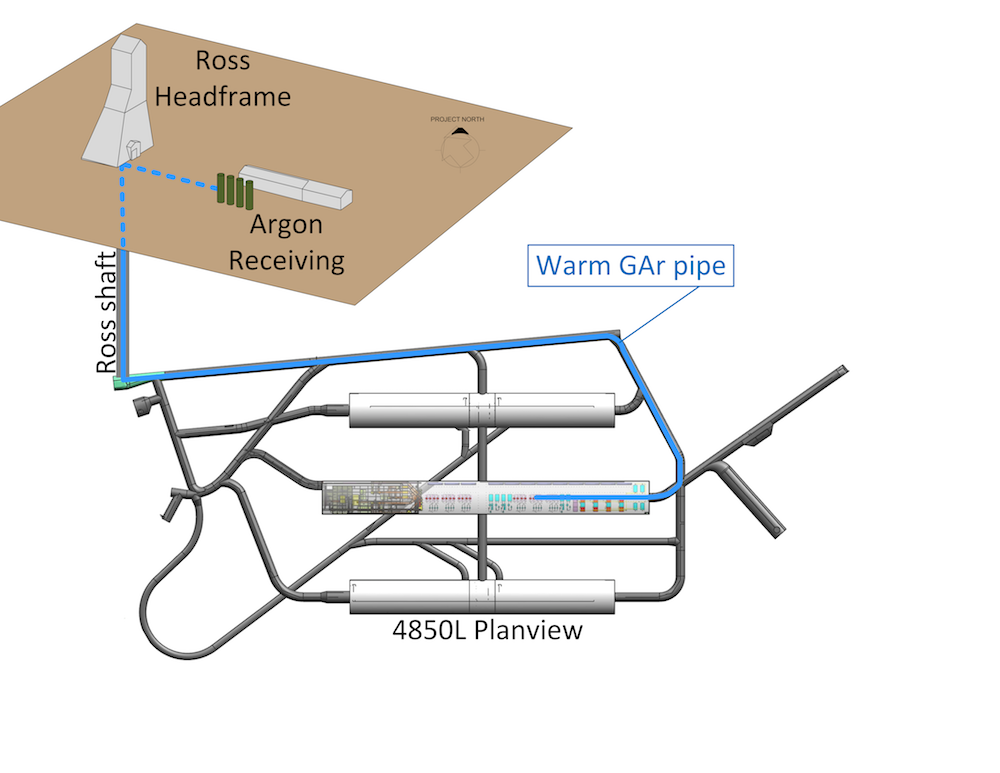}
\end{dunefigure}

The hydrostatic head for the \SI{1.5}{km} vertical gas-only piping is on the order of
\SI{0.05}{MPa}. 
Transferring liquid would increase this number to \SI{20}{MPa}, which would require about 
seven pressure-reducing stations evenly spaced along the vertical drop. This would in turn require that the piping be rerouted down the Oro Hondo ventilation shaft, which would first require  rehabilitation work. This solution would be much more expensive than the selected design that transfers only gas. 

All the \dword{gar} piping connections in the Ross Shaft will be welded or sealed with Grayloc\texttrademark{} metal seal fittings. Welded metal bellows will be used to handle thermal expansion and contraction. The piping material is stainless steel. The frictional pressure drop for the supply pipe is offset by the pressure gained due to the static head from elevation change.

The effort on pipework between surface and cavern is divided between the
cryogenics infrastructure team, who conceptualized the piping,  
and the \dword{fscfbsi} team, who designed the support system and will construct the piping system. Section~\ref{sec:cryo-cryosys-esh-gar-pipe} discusses the \dword{odh} assessment of the design.

\subsection{Nitrogen Refrigeration System}
\label{sec:comp:ln2}

Filling the \dword{dune} far detector modules will require a very large volume of argon gas that must be cooled from $\sim\,\SI{300}{K}$ to \dword{lar} temperature (88.3 $\pm$ 1 K). 
The \dword{lbnf} will procure and install a nitrogen refrigeration system in the \dword{cuc} at the \dword{4850l} that will liquefy nitrogen at a rate sufficient to supply the required liquefaction power to the argon condensers for each cryostat. 
Filling each cryostat with \dword{lar} in a reasonable 
period of time is the driving consideration in determining the sizes of the 
nitrogen refrigerator and argon condensers. 
In addition to providing cooling power to condense the delivered \dword{gar}, during operations 
the refrigerator will cool and recondense the boil-off argon from the cryostats (Section~\ref{subsec:reliquef}).

The nitrogen-refrigeration system design consists of four closed-loop refrigeration plants each comprising  a \coldbox, 
recycle compressor, and expanders. Intermediate \dword{ln} storage tanks (dewars) are connected to the refrigeration system and serve as buffer storage to smooth operational variability between production and usage rates.  The plants are connected with a common stream so that their use can scale as needed. 
Three plants will be used for the initial \cooldown and filling of each of the first two
cryostats (Sections~\ref{sec:cryosys-proc-cool} and~\ref{sec:cryostat-fill}) and the fourth for the final two cryostats. 

The refrigeration system operation is expected to be capable of running  
continuously for at least a year, and then require only 
minor servicing. 
The system will be equipped with automatic controls and a remote monitoring system so that human intervention is minimized during normal operation. To ensure optimum efficiency and minimize downtime, the plan is to award a Maintenance \& Operations (M\&O) subcontract to the vendor supplying the equipment. 

The \dword{lbnf} reference design places the nitrogen compressors underground. A 
water system 
with an evaporative-cooling tower removes heat from the compressor. 
Compression is carried out at close-to-ambient temperature and  
a compressor after-cooler is provided to reject heat. 

The rest of the nitrogen refrigeration system is also underground, comprising \coldbox{}es, expanders and \dword{ln} storage tanks, as well as the supply of nitrogen to charge the system and replenish losses.

\begin{dunefigure}[Nitrogen refrigeration plant diagram]
{fig:nitrogren-refrigerator}
{Nitrogen refrigeration plant diagram}
\includegraphics[width=.7\textwidth]{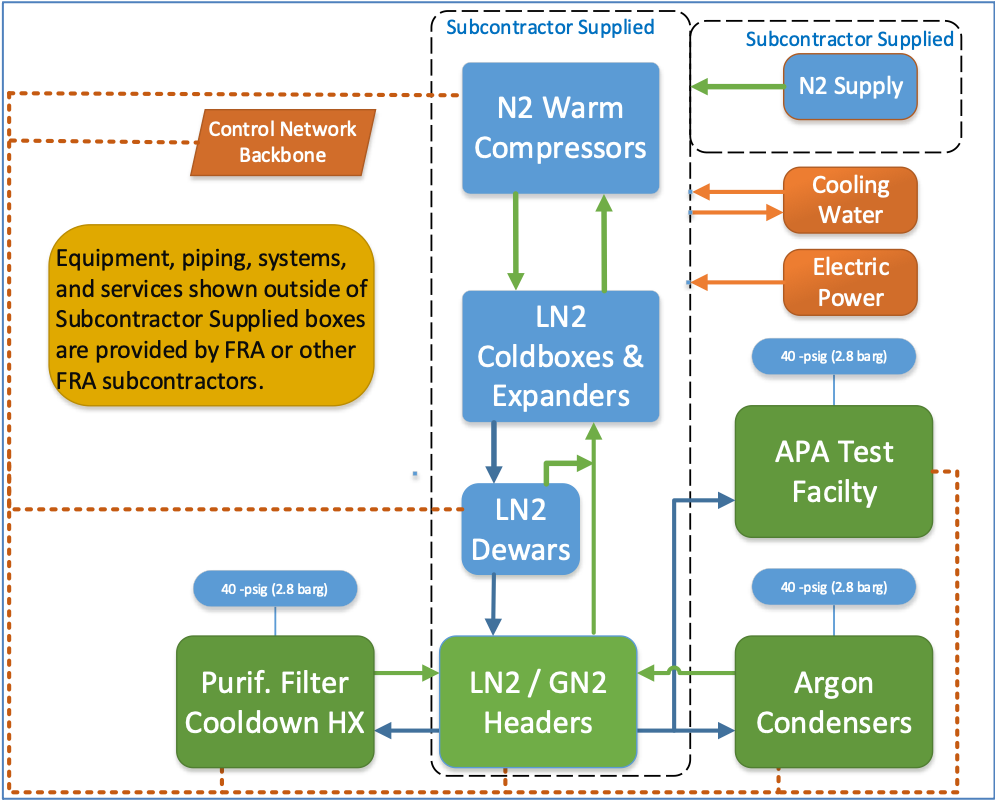}
\end{dunefigure}

Three nitrogen refrigeration plants are needed to provide cooling for the fill and operation of the first two \dwords{detmodule}; 
a fourth unit will be added for the (Phase II) third and fourth modules. All available units are used during the  \cooldown and
fill of each cryostat to minimize the duration of each step. Once all four cryostats are filled
and the required \dword{lar} purity is achieved, three of the four units will remain in use to recondense boil-off argon,
leaving the fourth available as a spare. 
Vacuum-insulated nitrogen pipes will deliver the \dword{ln} to the various user systems and collect the vaporized GN2 back to the refrigeration system to close the cycle. The main user systems are the argon condensers, but the system will also provide cooling power to the \dword{fdc} \coldbox{}es, which are used to test \dword{sphd} 
detector components at cryogenic temperature (in gas phase) prior to their installation in the cryostat,  and to the regeneration system, where 
the regenerated filters are cooled before being put back in service or on stand-by.

One of the key aspects of the design is modularity, which will facilitate the transport underground via the Ross Shaft. To minimize installation time at SURF and the risk of any misalignments, the subcontractor will first assemble the full system at one of their facilities, and label the components prior to delivery.

With four units running, the estimated power requirement is 3,840\,kW and the estimated cooling water usage 3,940\,kW. 
During normal operations, 3,000\,kW of electrical power and cooling water are available for this system. The full four units are needed only during the fill of cryostats FD3 and FD4 (Phase II), during which extra utilities are available because the detectors in these two cryostats are not operational yet. 
Once all four detector modules are operational, three units can support them all while keeping 
the usage within the allowable values.
The Statement of Work for this system is available in~\cite{EDMS2705540}.

\subsection{Argon and Nitrogen Distribution}
\label{ch:cryosys:subsys:an:distrib}

The argon and nitrogen distribution piping consists of a series of vacuum and non-vacuum insulated pipes for transporting argon and nitrogen in both the liquid and gas phases between the \dword{cuc} and the detector caverns. The layout is available in~\cite{EDMS2810467}. 

The nitrogen system acquisition includes both nitrogen generation and distribution~\cite{EDMS2705540}. The argon distribution is acquired independently; the reference design is available in~\cite{EDMS0000196907}, and the Statement of Work in~\cite{EDMS2810641}.

\section{Proximity Cryogenics}
\label{sec:prox-cryogenics}
The Proximity Cryogenics includes the \dword{lar} and \dword{gar} purification and regeneration systems~\cite{EDMS2852844,EDMS2892643}, the argon condensers and
the \dword{lar} circulation pumps~\cite{EDMS2884649}. All items are located within either the \dword{cuc} or the detector caverns.

\subsection{Cryogenics in the CUC}
\label{sec:prox-cryogenics-cuc}

In addition to the nitrogen refrigeration system (\coldbox{}es, recycle compressors and nitrogen generation) and the \dword{ln} storage tanks, the \dword{cuc} at the \dword{4850l} houses the 
argon purification system, which is composed of liquid and gas filtration elements, including particulate filters, and the associated equipment
required to regenerate (i.e., clean) these elements (Section~\ref{subsec:filter-regen}). 
Figure~\ref{fig:cryog-cuc-layout} illustrates the layout.

\begin{dunefigure}[Cryogenics system components in CUC] 
{fig:cryog-cuc-layout}
{Cryogenics system components in the CUC. }
\includegraphics[width=0.95\textwidth]{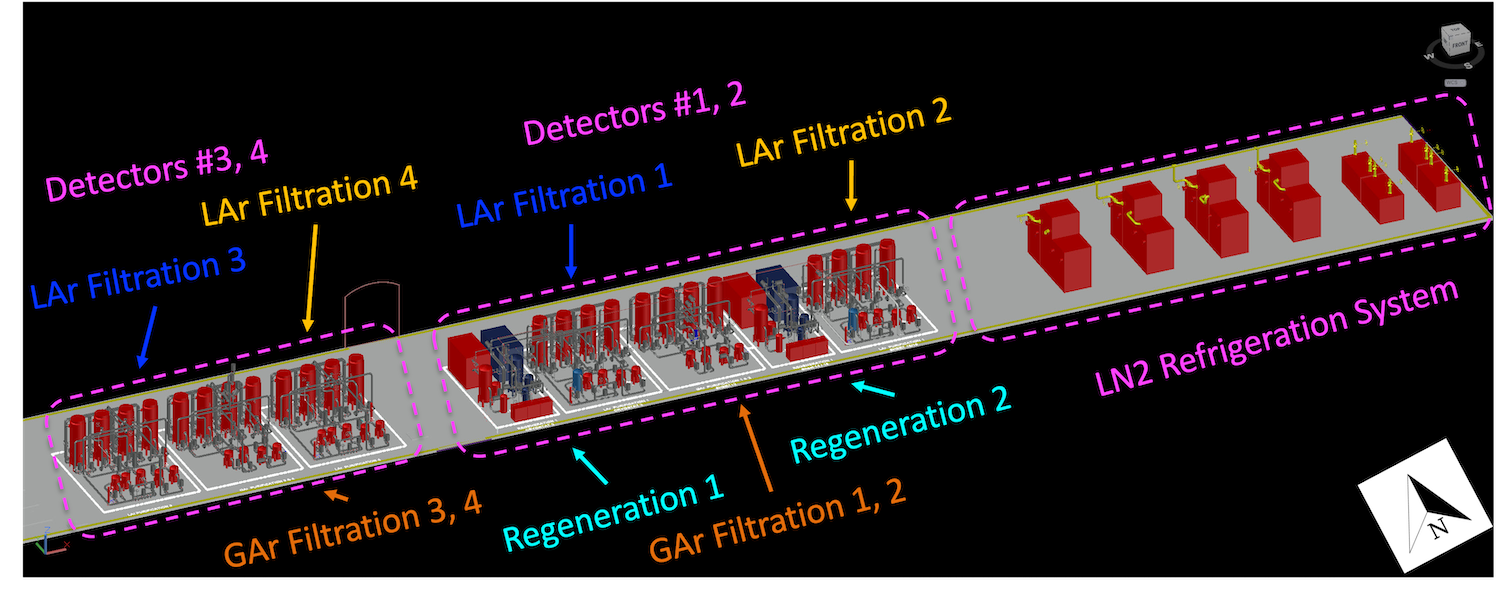}
\end{dunefigure}

The argon purification system also includes interconnecting piping and the necessary
valves and instrumentation. The purifiers themselves 
consist of a molecular sieve and copper pellets 
that remove water and oxygen, respectively, from the argon. The size of each filter, whether for \dword{gar} or \dword{lar}, is selected appropriately to purify 
argon with initial contaminant levels no greater than \SI{5}{ppm} oxygen and \SI{10}{ppm} water. 
The \dword{gar} purifiers
are used during the argon-purge phase; 
one set of these will be used for cryostats \dshort{sphd} and \dshort{spvd}, and another set for cryostats FD3 and FD4. In contrast, the \dword{lar} filters are actively used throughout the experiment's lifetime to achieve
and maintain the contamination level below the required \SI{100}{ppt} for \dshort{sphd} and \SI{50}{ppt} for \dshort{spvd}.  Each cryostat will have its own set of liquid filters.
During operations, the \dword{lar} filters will switch between active filtration and regenerative modes (one filter can be regenerated at a time), to keep the argon filtration process uninterrupted. 

The reference designs for LAr Purification \dshort{sphd} and \dshort{spvd} are available in~\cite{EDMS2810432, EDMS2810433}. 

\subsection{Cryogenics in the Detector Caverns}
\label{sec:prox-cryogenics-detcav}

The bulk of the cryogenics equipment in the detector caverns sits atop the cryostats on a  \SI{12}{m} wide, \SI{60.5}{m} long mezzanine that is installed about \SI{2.3}{m} above each 
cryostat. This cryogenics equipment is planned for installation at the far end of the mezzanine, as shown in Figure~\ref{fig:prox-cryo-detcav} (left). This equipment includes:

\begin{itemize}
    \item \dword{lar} phase separators, through which \dword{lar} is passed and conditioned before returning to the cryostat;
    \item  argon condensers, for initial liquefaction of delivered \dword{gar}, and for reliquefaction of boil-off argon  before it gets pumped to the \dword{lar} purification system;
    \item   nitrogen phase separators; 
    \item a cryostat over-pressure and under-pressure protection system with pressure controls and safety valves;
    \item \dword{plc} racks;   
    \item a nitrogen generator for cryostat insulation purge;
    \item a xenon injection line, connected to the boiloff \dword{gar} manifold, heading to argon condensers; 
    \item a \dword{gar} sampling and measuring system; and 
    \item a set of lines (connected to each cryostat feedthrough and piped to a gas manifold) that allows 
    \begin{itemize}
        \item sampling of the \dword{gar} locally, and
        \item measurement of the concentration of contaminants (oxygen and water) from the cryostat ullage.
    \end{itemize} 
\end{itemize}
 The \dword{gar} sampling occurs during the purge mode to verify progress, and during steady-state operations to ensure that no contaminants enter the cryostat.

The first two cryostats will have three \SI{100}{kW} (or six \SI{50}{kW}) condensers to 
provide the cooling power needed during initial 
\cooldown and filling operations where warm \dword{gar}
is cooled and reliquefied to fill the cryostat. The third and fourth cryostats will need only two condensers each. After filling, only one condenser per cryostat is needed, with the
other providing redundancy.
This will ensure the high availability of the recondensing 
system and minimize the need for venting high-purity argon 
or  down-time for maintenance of
the recondensers and the refrigeration plants. 

Each condenser has a dedicated LAr condenser pump, located in its vicinity, to send the recondensed argon to the liquid filtration in the CUC for purification. This is essential to achieve and maintain the LAr purity inside the cryostat as most of the contaminants are in the ullage space at the top.

Four large pumps recirculate the bulk
of the \dword{lar} at a rate of \SI{10}{kg/s} per pump 
to the \dword{cuc} for filtration. They are installed on the 
floor of the cavern at the near (accessible) end of the cryostat.  
Figure~\ref{fig:prox-cryo-detcav} (right) shows the 
planned configuration. The four pumps each withdraw argon from the cryostat through one of the four side penetrations near
the bottom of the cryostat, each equipped with an in-line safety valve with its seal inside the cryostat
itself. 
The safety valves normally remain open via actuators, but will close down in case of emergency, loss of actuation, or another triggering event. These safety valves have been successfully used for both \dword{pdsp} and \dword{pddp}, and installed at \dword{sbnd}.

\begin{dunefigure}[Proximity cryogenics in the detector cavern] 
{fig:prox-cryo-detcav}
{Proximity cryogenics in the north detector cavern (mezzanine on the left, \dshort{lar} pumps on the right)}
\includegraphics[width=0.59\textwidth]{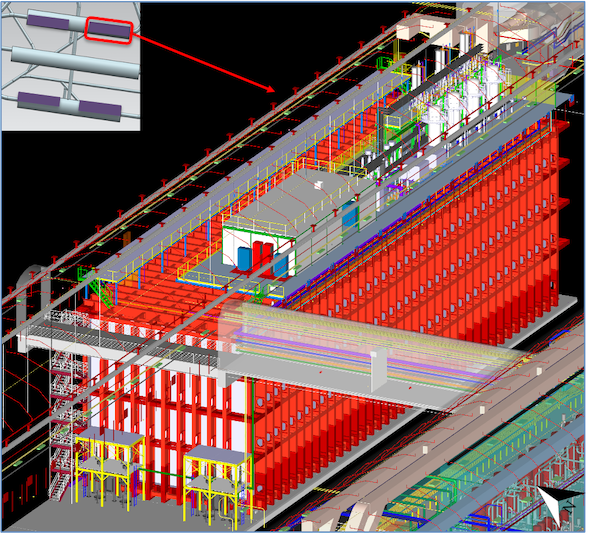}
\includegraphics[width=0.19\textwidth]{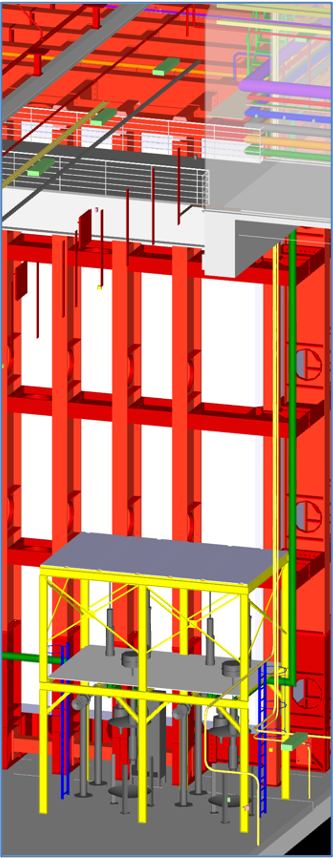}
\end{dunefigure}

\section{Cryogenics inside the Cryostats}
\label{sec:internal-cryo}

The subsystem that distributes \dword{lar} and \dword{gar} inside each cryostat, the internal cryogenics, is active during all modes of operation. This subsystem  consists
of manifolds for the \dword{gar} purge, the \cooldown sprayers (for \dword{sphd}), and  the \dshort{lar} distribution. The reference design for the internal cryogenics is fully described in~\cite{EDMS2211972}. The components are summarized here:

\begin{itemize}
    \item The purge manifold distributes \dword{gar} along both long edges of the cryostat and is supported by the vessel floor. It flows \dword{gar} vertically down during the initial purge with a longitudinal opening that allows the \dword{gar} to flow everywhere. 
    
\item The \cooldown sprayers are located at the top of the \dword{sphd} cryostat, at discrete points staggered along its length. 
The sprayer distribution has been optimized via \dword{cfd} analysis so that 
the available cooling power cycles through the cryostat in a controlled way to cool the volume uniformly~\cite{EDMS2154410}.

\item The liquid distribution manifold runs along both long edges of the cryostat and is supported by the vessel floor. It sprays \dword{lar} vertically up during the fill and steady-state operations of the cryostat. The pipes have calibrated holes along their lengths for even distribution.
\end{itemize}

The distribution mechanism, combined with the fact that the \dword{lar} is returned to the cryostat slightly warmer ($\sim\,\SI{5}{K}$) than the bulk of the liquid (to help the mixing), is very important for obtaining
a uniform \dword{lar} purity.

The internal cryogenics also includes the \dword{gar} boil-off vacuum-insulated pipes that connect the inside of the cryostats to the exterior vacuum-insulated line taking the \dword{gar} 
to the condensers.

\cleardoublepage

\chapter{Cryogenics Procurement}
\label{ch:cryo-procure}

\section{Liquid Argon Procurement and Delivery}
\label{ch:cryo-procure-lar}

Each \dword{dune} \dword{detmodule} requires a cryostat that holds \larmass (metric) of high-purity \dword{lar}.   The standard grade for argon, designated as Grade~4.5 in the gas-supply industry, is specified as having a purity of
 99.995\% (minimum), allowing a maximum concentration of \SI{5.0}{ppm} for O$_2$ 
and \SI{10.5}{ppm} for H$_{2}$O.  Requiring higher-purity product would 
significantly increase cost.  
 Experience has shown that vendors deliver far better purity than the nominal, allowing us to confidently procure the standard grade.

This design report covers only the roughly \SI{37}{kt} of \dword{lar} required to fill the first two cryostats; this requires procurement of that quantity plus some small amount that will inevitably be lost during transport.
Procuring this \dword{lar} will require a steady supply over a period of a few years starting in
the first quarter of 2028 from vendors with the logistics capabilities to deliver the required amount of LAr. 

The \dword{lbnf} has identified three to four qualified \dword{lar} vendors, which together represent 98\% of the current U.S. \dword{lar} production, delivery capacity, and service to the domestic argon market. Other vendors may become available in the coming years.

Planning the supply and logistics 
of \dword{lar} delivery to the \dword{surf} requires consideration of the following issues:

\begin{itemize}
\item  total capacity of commercial air-separation plants within trucking distance of the \dword{surf} site (the peak delivery potential);
\item \dword{dune} detector installation schedule;
\item number and frequency of tanker trucks required and their impact on the local community;
\item number and frequency of railcars and their availability; and
\item availability and cost associated with the delivery of high-purity 
\dword{lar} as opposed to lower-quality commercial-grade argon combined with on-site coarse purification.
\end{itemize}

The current (2023) total argon delivery capacity (taking boil-off and other losses into account) in the U.S. is approximately 
\SI{4.9}{kt/day} and demand is within \SIrange{2}{3}{\%} of this figure. This demand/supply ratio is quite
tight and is likely to remain the same for at least two more years~\cite{EDMS_jrci_rpt}.  Argon markets are growing nationally and new \dword{lar} capacity is being built, most of it on the Gulf Coast and the Chicago area. The average distance from \dword{surf} is about 1,000 miles, therefore  \dword{lbnf} will still look to identify new argon capacity closer to the site. 
At the same time, the \dword{lar} industry has started inter-regional transport of Gulf Coast \dword{lar} to distant national markets, including inter-regional rail-tanker intermodal redistribution facilities. A suitable rail depot has been identified in Tiger Transfer, LLC in Upton, WY, less than 100 miles from \dword{surf}. This could be very advantageous 
as more of the long-distance transportation could be done by rail, reducing the overall over-the-road mileage, which is more affected by weather, thereby increasing reliability of deliveries. It would also add intermediate buffer storage, and may lower the cost of \dword{lar} delivery.

Given the complexities and expense of the long-distance delivery, the uncertainties in growth and costs of \dword{lar} production and transport, the challenges in accessing the \dword{surf} site, and the vendors' limitations in committing assets and people this far in advance, \dword{lbnf} is considering a ``group supply'' scheme with one vendor or a third party coordinating the effort, as well as awarding multiple subcontracts to multiple vendors directly.  Our qualified potential vendors endorse this idea, and  believe that the required \dword{lar} for \dword{lbnf} can be delivered with high reliability and at a reasonable cost.  They are beginning to plan more creatively about the costs for sourcing and delivering \dword{lar} to Lead, SD. 

The most efficient mode of argon delivery appears to be
a combination of rail plus short-haul and 
over-the-road long-haul tank truck with a maximum capacity of \SI{20}{t}. 
The expected number of such deliveries per cryostat is about 1,000 
over a period of eight to 14 months. 

Given the large quantity and logistics associated with this acquisition, the acquisition plan has already been submitted to \dword{doe} for review and approval~\cite{EDMS2818818}. The results from the last RFI and feedback from the consultant are included as well. 
A requirements subcontract has been selected that will enable procurement of the required amounts within the required timeframe for each phase. The current requirements are as follows (current, pre-baseline schedule).

Phase I of the \dword{usproj}, which includes the purge, \cooldown and fill of cryostats \dshort{sphd} and \dshort{spvd} is scheduled for October 2027--July 2030.
The total of \SI{37350}{t} of \dword{lar} will be delivered over two periods, one per cryostat, as detailed in Table~\ref{tab:ph1-lar-deliv}.

\begin{dunetable}[Phase I \dshort{lar} delivery rate and schedule]{p{.11\textwidth}p{.18\textwidth}p{.35\textwidth}p{.22\textwidth}}
{tab:ph1-lar-deliv}
{Phase I \dshort{lar} delivery rate and schedule} 
\textbf{Cryostat} & \textbf{Activity} & \textbf{Rate} & \textbf{Date} \\ \toprowrule

\dshort{sphd} & initial purge and system testing & 40\,t/day, 5 days/week, for 1-2 weeks & Oct 2027 \\ \colhline
\dshort{sphd} & purge/\cooldown & 40\,t/day, 5 days/week, for 
4 mon. & Dec 2027 -- Apr 2028 \\ \colhline
\dshort{sphd} & fill & $\sim$70 t/day, 7 days/week & Apr -- Dec 2028 \\ \colhline

\dshort{spvd} & initial purge and system testing & 40 t/day, 5 days/week, for 1-2 weeks & Aug 2028 \\ \colhline
\dshort{spvd} & purge/\cooldown & 40 t/day, 5 day/week, for 4 mon. & Jan -- Apr 2029 \\ \colhline
\dshort{spvd} & fill & $\sim$45 t/day, 7 day/week & May 2029 -- Jul 2030 \\ 
\end{dunetable}
\FloatBarrier

Phase II of the \dword{usproj}, which similarly includes the purge, \cooldown and fill of cryostats FD3 and FD4, is scheduled for April 2030 to November 2033.
The total of \SI{37550}{t} of \dword{lar} will be delivered over two periods, one per cryostat, as detailed in Table~\ref{tab:ph2-lar-deliv}.

\begin{dunetable}[Phase II \dshort{lar} delivery rate and schedule]{p{.11\textwidth}p{.18\textwidth}p{.35\textwidth}p{.22\textwidth}}
{tab:ph2-lar-deliv}
{Phase II \dshort{lar} delivery rate and schedule} 
 \textbf{Cryostat} & \textbf{Activity} & \textbf{Rate} & \textbf{Date} \\ \toprowrule

FD3 & initial purge and system testing & 40 ton/day, 5 day/week, for 1-2 weeks & Apr 2030 \\ \colhline
FD3 & purge/\cooldown & 40 ton/day, 5 day/week, for 4 mon. & July -- November 2030 \\ \colhline
FD3 & fill & $\sim$50 ton/day, 7 day/week & Nov 2030 -- Nov 2031 \\  \colhline

FD4 & initial purge and system testing & 40 ton/day, 5 day/week, for 1-2 weeks & Sep 2031 \\ \colhline
FD4 & purge/\cooldown & 40 ton/day, 5 day/week, for 4 mon. & Dec 2031 -- Apr 2032 \\ \colhline
FD4 & fill & $\sim$40 ton/day, 7 day/week  & May 2032 -- Nov 2033 \\

\end{dunetable}

Deliveries are to vertical tanks located above ground at the Ross Shaft site in limited space near an existing warehouse.

\section{Nitrogen Refrigeration System Procurement}
\label{ch:cryo-procure-ln}

The nitrogen refrigeration system (Section~\ref{sec:comp:ln2}) will consist of commercially available equipment. To fulfill DUNE performance requirements the equipment will require modifications of a type that is customarily available. The selected vendor will engineer, manufacture, deliver, install and commission the following items, appropriately modified:

\begin{itemize}
    \item refrigeration units (composed of recycle compressors and liquefiers),
    \item cryogenic vacuum-jacketed and uninsulated \lntwo/GN$_2$ distribution pipes and valves,
    \item \lntwo storage tanks,
    \item nitrogen for the initial charging of the equipment and make-up losses, and 
    \item process controls.
\end{itemize}

The nitrogen refrigeration system subcontractor will be required to meet interface specifications and requirements for the part of this system that will connect to other parts of the cryogenics system. The nitrogen refrigeration system is procured in two phases. Phase I, Pre-\dword{feed} Study,
was open competition with down-selection to the top three most technically qualified offerors. Fixed Firm Price (FFP) contracts were awarded to the down-selected offerors. The Phase II subcontract, 
has been awarded to Air Products and Chemicals, Inc. for the remainder of the engineering work, manufacture, installation (assembly and connection), and  commissioning of the equipment. 
This subcontractor is responsible to deliver a system that meets the required performance specifications.  The final engineering is in progress.

\section{Other Procurements}
\label{ch:cryo-procure-other}

Other infrastructure cryogenics procurements include:

\begin{itemize}
    \item surface argon receiving facilities~\cite{EDMS2211568,EDMS2211532,EDMS2211569},
   \item argon distribution system~\cite{EDMS2810641,EDMS2778502,EDMS2211532,EDMS2254565}, 
    \item installation of in-kind contributions~\cite{EDMS2211972,EDMS2211532,EDMS2248818,EDMS2838450}, and
    \item miscellaneous items:
    \begin{itemize}
    \item GAr boil-off and pressure control system~\cite{EDMS2154314,EDMS2519457,EDMS2211532,EDMS2838450},    
    \item GN2 supply to cryostat insulation~\cite{EDMS2771986,EDMS2771977,EDMS2775007pid,EDMS2227399,EDMS2838450},    
    \item xenon injection system~\cite{EDMS2771736,EDMS2477369,EDMS2477451,EDMS2838450}, and  
    \item connections to purity monitors.
\end{itemize}
\end{itemize}
 
The surface argon-receiving facilities and the argon distribution system will both be acquired using the approach ``engineer, manufacture, install and test, and startup.'' Performance specifications have been written and are available in~\cite{EDMS2211568} and~\cite{EDMS2810641}, respectively. In both cases, the selected subcontractor is responsible to deliver a system that meets the required performance specifications.

The in-kind contributions from partners (\dword{unicamp} in Brazil and the \dword{cern} in Switzerland) that require installation include, for both \dshort{sphd} and \dshort{spvd}, the:
\begin{itemize}
    \item argon condenser system,
    \item \dword{lar} purification system,
    \item \dword{gar} purification system,
    \item main \dword{lar} circulation, and
    \item the regeneration system.
\end{itemize}
 
To streamline the installation process and reduce the number of subcontractors working underground, it is planned to select a single subcontractor for these activities using delivery order contracts\footnote{This is a type of contract that provides for an indefinite quantity of supplies or services during a fixed period; also known as IDIQ (indefinite delivery/indefinite quantity) or ``task order.''} for:
\begin{itemize}
    \item engineering, manufacturing, installation, and testing of the argon distribution system; 
    \item installation of equipment provided via in-kind contributions; and 
    \item installation of the miscellaneous items listed above.
\end{itemize}
   The contract for the argon distribution system will come first, and the others will follow according to the installation schedule. This allows the partners' engineering and manufacturing work to proceed and allows the \dword{usproj} team to engage with the already selected subcontractor as relevant information for the installation becomes available.
The miscellaneous items are mostly catalog items; they 
can be procured independently and installed as described above, or their procurement can be included in this acquisition. 

\cleardoublepage

\chapter{Cryogenics System Processes}
\label{sec:cryosys-proc}

The \dword{lbnf} cryogenics systems function in a variety of operational modes, appropriate to different phases of the \dword{dune} far detector. For each cryostat, the operations include:

\begin{itemize}
\item receipt and transfer of cryogens underground (in gas phase) once all systems for a detector module are fully installed in a cryostat and tested,
\item initial purge of the cryostat to replace the air with \dword{gar}, 
\item \dword{gar} circulation inside the cryostat,
\item cryostat \cooldown{} to near \dword{lar} temperature,
\item cryostat fill with \dword{lar},
\item steady-state operations during data-taking, and
\item emptying of the cryostat at the end of the experiment's operations.
\end{itemize}

\begin{figure}[htbp]
\centering
\includegraphics[width=0.8\textwidth]{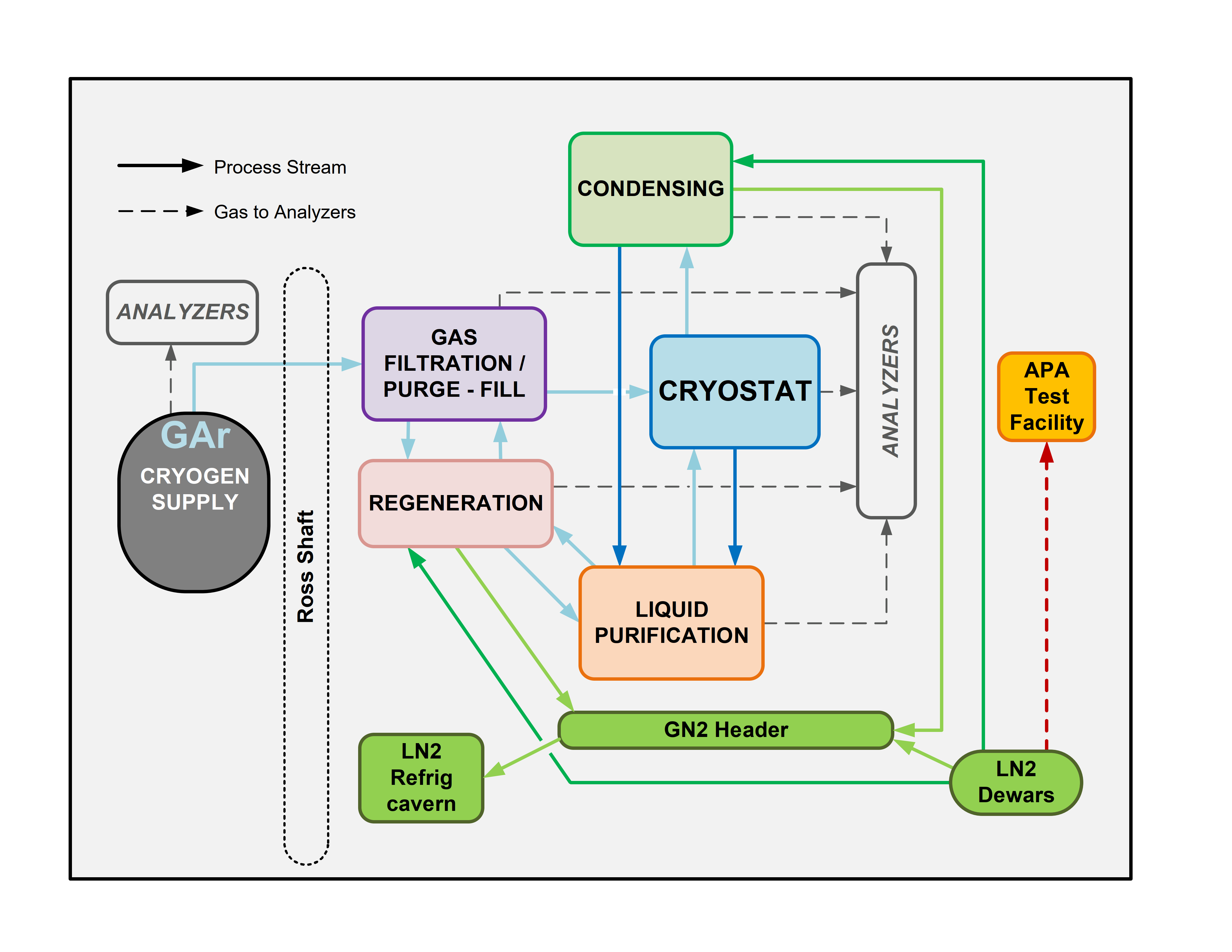}
\caption{Cryogenics system flow block diagram}
\label{fig:v5ch2-LBNF-cryo-process-2014}
\end{figure}

Receipt of the cryogens at the Ross Headframe is discussed in Section~\ref{ch:cryosys:subsys:ic:receiv}. Figures~\ref{fig:cryog-process-flow} and~\ref{fig:v5ch2-LBNF-cryo-process-2014} illustrate the processes.

\begin{dunefigure}[Cryogenics process flow diagram] 
{fig:cryog-process-flow}
{Cryogenics process flow diagram}
\includegraphics[width=\textwidth]{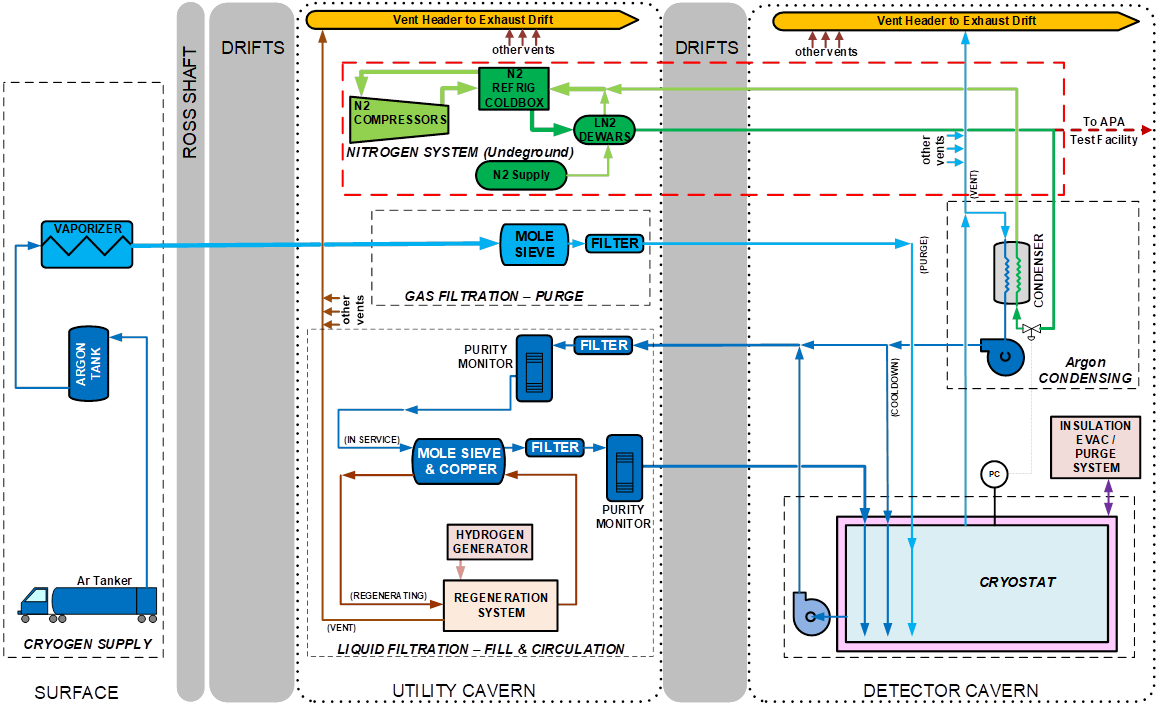}
\end{dunefigure}

\section{Cryostat Initial Purge and \Cooldown}
\label{sec:cryosys-proc-purge-cool}

Cryostat and detector construction procedures will
ensure that the completed cryostat is free
of all debris and loose material that could contaminate the \dword{lar}. Following installation of all 
detector components, the cryostat is verified to be clean before it can be purged and 
cooled with \dword{gar}. 

Once the \dword{lar} arrives at the surface and is vaporized, the \dword{gar} is pushed to the Ross Shaft and down, where it picks up additional pressure in the form of static head. At the \dword{4850l} it is piped to the purification modules 
and then to the purge manifold within the cryostat. The internal cryogenics (Section~\ref{sec:internal-cryo}) provides all the piping and manifolds inside each cryostat for the purge, \cooldown{}, and fill processes. The maximum \cooldown{} rate is given in Table~\ref{table:cryog-params}.


\subsection{Initial Air Piston-Purge} 
\label{sec:cryosys-proc-piston}

First, \dword{gar} will pass through the argon piping to flush out its atmosphere; this process will be repeated ten times (for eleven volume changes total) to reduce contamination
levels in the piping to the ppm level. 
An argon flow/piston-purge technique will be used to remove air from the membrane cryostat. The heavier argon gas is introduced at the
bottom of the cryostat and the exhaust is removed at the top. 
In \dword{sphd} the bottom \dword{gp} (a component of the \dword{tpc} detector) serves an additional role as a flow
diffuser during the initial purge. A matrix of small holes in this 
\dword{gp}, approximately \SI{10}{mm} in diameter at a \SI{50}{mm} pitch,
will provide a uniform flow. 

The design flow velocity of the advancing \dword{gar} volume is set to 1.2 vertical meters per hour, high enough to efficiently overcome the molecular diffusion
of the air downward into the rising argon so that the pure
argon-gas wave front will displace the air rather than just dilute it.
 A \twod \dword{cfd} simulation of the purge process done for a 
\SI{5}{kt} (metric) fiducial-mass cryostat\cite{EDMS2826472} 
shows that after 20 hours
of purge time, and 1.5 volume changes, the air concentration will be
reduced to less than 1\%. It will take 40 hours of elapsed time and three volume
changes to complete the purge process, reducing the residual air to a
few ppm. This simulation includes a representation of the perforated
field cage at the top and bottom of a \dword{sphd}-style detector.  The cathode planes
are modeled as non-porous plates; 
in both designs  they will be
constructed of 
\frfour sheets laminated on both sides by carbon-impregnated Kapton\footnote{DuPont\texttrademark{}, Kapton\textsuperscript{\textregistered} polymide film,  E. I. du Pont de Nemours and Company,  \url{http://www.dupont.com/}.}.

The \dword{cfd} model of the purge process 
has been verified in multiple arrangements:  (1) in an instrumented 
cylinder of 1 m diameter by 2 m height, (2) Liquid Argon Purity 
Demonstrator (LAPD), a vertical cylindrical tank of 3 m diameter by 3 m height,
taking gas-sampling measurements at varying heights and times during 
the purge process, (3) within the 35 ton membrane cryostat, a 
prototype vessel built for LBNE in 2013, of which the results 
are found at~\cite{Montanari:2013/06/13aqa}, 
 (4) within MicroBooNE cryostat, a horizontal cylindrical
tank of 3.8 m diameter by 12.2 m length, and in the two ProtoDUNEs of dimensions 
 \SI{8.6} {\times} \SI{8.6} {\times} \SI{7.9} m$^3$.


\subsection{Contaminant Removal via GAr Recirculation}
\label{sec:cryosys-proc-rmv-h2o}

Both \dwords{detmodule}  contain over ten tons of \frfour 
circuit-board material and \dword{frp} and a
smaller inventory of plastic-jacketed power and signal cables.
These somewhat porous
materials may contain as much as 0.5\% water by weight. Water vapor outgassing from these
materials and adsorbed water from the metallic inner
surfaces of the cryostat and piping system must be removed.

Following the piston purge, we close the \dword{gar} loop. The \dword{gar} circulates for several days to sweep remaining contaminants (water vapor and oxygen) from the bottom of the cryostat to the top, then out. 
The exiting \dword{gar} is routed through the 
recondensers, to the filters, and then 
reintroduced to the bottom of the cryostat, still as vapor. This cycle continues until the contaminant levels are reduced to \dword{ppm} levels. 

Water deep within porous materials will remain;
this is not a problem since
the water diffusion rate in \frfour at room temperature is already 
quite low (\SI{0.3}{\cubic\micro\meter/s}) and decreases as temperature decreases. 

\subsection{\Cooldown}
\label{sec:cryosys-proc-cool}

At this point in \dword{sphd} the cryostat and detector systems must be cooled in a controlled manner. The 
\dword{spvd} does not have a \cooldown requirement (it replaces the delicate wires of \dword{sphd} with etched \dwords{pcb}), so this step does not take place. 
To this end, purified \dword{lar} is introduced at the top of the cryostat by means of atomizing sprayers (Section~\ref{sec:internal-cryo}). The resulting mist distributes itself in the cryostat by means of gravity and convection.  The cooling required for this mode is supplied by the 
\lntwo system to the condensers located on the mezzanine atop the cryostat. 

\dword{cfd} simulation
has shown \cite{EDMS_2154410} that the liquid \cooldown method can
be controlled to stay within the available recondenser capacity. The required cooling rate
is determined by the maximum stress that detector components can
tolerate. For example, in the \dword{sphd} case, the \SI{152}{\micro\meter} thick wires of the \dwords{apa} will cool much more rapidly than the \dword{apa} frames.
A mass flow-control system with a temperature-monitoring system will be used to control the
temperature difference across the cryostat. The temperature difference required, \SI{50}{K} (Table~\ref{table:cryog-params}) is  based on input from the cryostat designer and the requirements of
the \dword{tpc} components and structure.

\section{Cryostat Fill}
\label{sec:cryostat-fill}

Once the \dword{sphd} cryostat and detector are cooled to roughly \SI{90}{K}, it is time to fill the volume with purified \dword{lar}.  For \dword{spvd}, the fill starts right after the purge.  The argon, which arrives at \dword{surf} in liquid form, is transferred from tanks at the receiving facilities above-ground, vaporized, piped down the Ross Shaft to the \dword{4850l}, from there to the \dword{gar} purification filters in the \dword{cuc}, and then to the 
mezzanine above the cryostat. Here it is re-liquefied by the nitrogen-fueled condensers, sent to the \dword{lar} purification system in the \dword{cuc}, transferred back to the mezzanine, and 
injected into the cryostat. The filling of each cryostat will vary in duration, ranging from eight months for the first to about 18 for the fourth, according to the delivery schedule and the power available for argon reliquefaction.

The recirculation pumps can be safely
turned on and liquid argon purification (Section~\ref{subsec:argon-pur}) can begin once the liquid depth in the cryostat reaches about \SI{1.5}{m}. 

Once the fill is complete, the primary operational phase of the experiment can begin. 

\section{Recirculation during Detector Operations}
\label{sec:cryostat-ops}
During operations of the detector module, \dword{lar} will be recirculated continuously through the purification system by means of external pumps.  
Initially, four pumps are used to circulate a large flow, up to \SI{40}{kg/s}, through the purification system; 
this flow is maintained until the required argon purity level is achieved, at which time the flow (and therefore the number of pumps in operation) is reduced to the level sufficient to maintain the achieved purity. 

As each detector module has its own dedicated purification module, each will operate independently of one another.

\section{Ar Gas Recovery, Reliquefaction, and Pressure Control}
\label{subsec:reliquef}

The high-purity \dword{lar} stored in the cryostat will 
be evaporating continuously due to the small but unavoidable heat ingress.  
The argon vapor (boil-off gas) will be recovered, 
recondensed, and returned to the cryostat in a closed system.

During normal operation the expected heat ingress of approximately 
\SI{87.1}{kW} to the argon system will result in 
an evaporation rate of \SI{1900}{kg/hr}. 
In the absence of a pressure-control system an increase in the vapor/liquid ratio within the closed system would raise the internal pressure. To mitigate this problem, the \dword{lbnf} cryogenics system will remove argon vapor from the top of the cryostat 
through two cryogenic feedthroughs. As 
the vapor rises, it cools the cables and feedthroughs, thereby 
minimizing the outgassing. The exiting \dword{gar} will be 
directed to a heat exchanger (the condensers, illustrated in 
Figure~\ref{fig:v5ch2-recondenser-sept-2011}) in which it is 
cooled against a stream of \dword{ln} and condensed 
back to a liquid. As the argon vapor recondenses, its volume 
reduces, which will draw further 
gas into the heat exchanger.  Pressure-control valves on 
the boil-off gas lines will control the flow to the recondensers 
to maintain the pressure within the cryostat at \SI{0.105}{MPa} $\pm$ \SI{0.008}{MPa},
 mitigating the risk of  developing a thermal siphon. Pressure control is discussed further in Section~\ref{sec:press-control}.

The \dword{ln} stream (serving as the coolant for the 
condensers) will come from the closed-loop \dword{ln} refrigeration plant.  
The commercial refrigeration plant uses compression/expansion and heat 
rejection to continuously liquefy and reuse the returning nitrogen 
vapor. The estimated heat loads within the cryostat are listed 
in Table~\ref{tab:cryo-heat-loads}.

 \begin{dunetable}[Estimated heat loads within the cryostat]{lc}{tab:cryo-heat-loads}
{Estimated heat loads within the cryostat.}
{\bf Item} & {\bf Heat Load (kW)} \\ \toprowrule
Insulation heat loss & 48.7  \\
\colhline
Electronics power & 23.7  \\
\colhline
Recirculation-pump power (2 pumps in operation) & 11.0 \\
\colhline
Misc. heat leaks (pipes, filters, etc.) & 3.7 \\
\colhline
{\bf Total} & {\bf 87.1 } \\
\end{dunetable}

\begin{dunefigure}[Liquid argon recondensers] 
{fig:v5ch2-recondenser-sept-2011}
{Liquid argon recondensers}
\includegraphics[width=.80\textwidth]{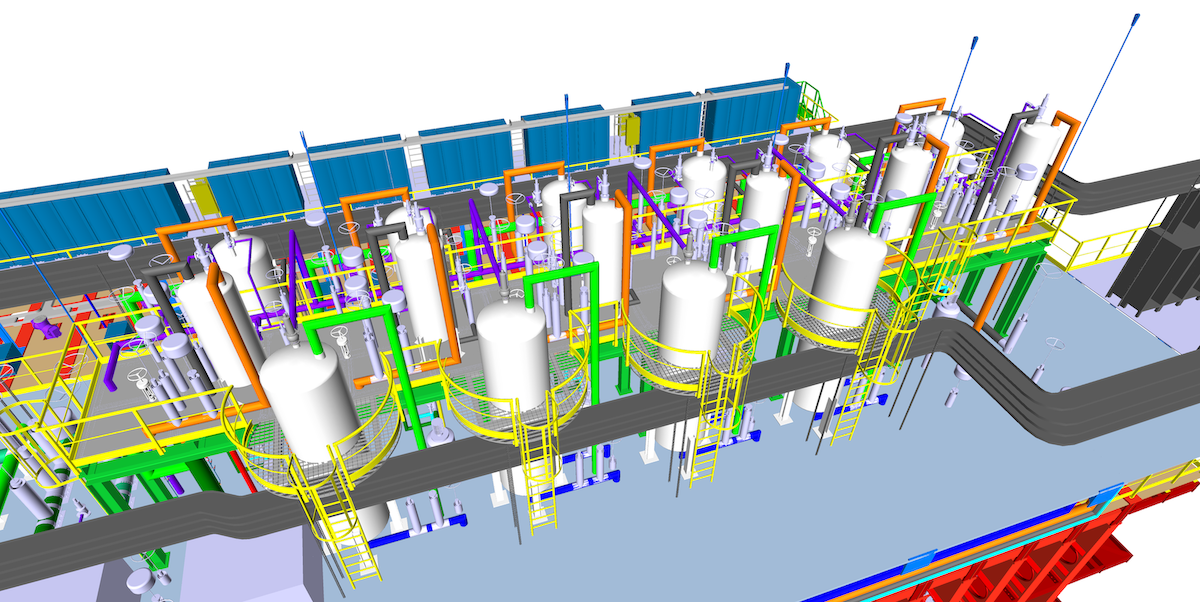} 
\end{dunefigure}

\section{Argon Purification}
\label{subsec:argon-pur}

Each cryostat's \dword{lar} inventory is circulated through a purification 
filter in the \dword{cuc} to achieve and maintain the required purity for the experiment  (Section~\ref{sec:prox-cryogenics-cuc}). 
Four external pumps attached to side penetrations in the cryostat (Section~\ref{sec:prox-cryogenics-detcav}) will continuously circulate the 
\dword{lar} through the filters. Figure~\ref{fig:prox-cryo-detcav} illustrates the external pumps. 

The multiple-pump arrangement will provide a very 
high level of redundancy that will extend the maintenance-free 
operating period of the cryostat.
Maintenance can be done on off-line pumps, which will eliminate pump-related maintenance interruptions to cryostat operations.


The required flow rate of \dword{lar} to be sent for purification 
is expected to decrease over time as the purity of the LAr inside the cryostat increases. The initial maximum flow rate 
will be \SI{93}{\cubic\meter/h} (\SI{411}{gpm}). The \dword{lar} volume in one 
cryostat will turn over every 5.5 days at this rate. 
Longer term, the rate will decrease to  \SI{46}{\cubic\meter/h}
with a turn-over rate of 11 days.  As a point of comparison, \dword{pdsp} has a maximum turn-over rate of about 10.8 days. \dword{pdsp} has achieved an electron lifetime greater than \SI{30}{ms} with this turn-over rate.

A purity monitor will monitor the filter 
effectiveness. When a purification filter becomes saturated it is regenerated to vent the contaminants  (Section~\ref{subsec:filter-regen}). During this time the \dword{lar} flow 
is switched to another purification filter to maintain uninterrupted filtration. 

\dword{dune} will also provide purity monitors in the cryostat to ensure that the purity levels remain 
at or better than their required values. 

\section{Filter Regeneration}
\label{subsec:filter-regen}

The \dword{lar} filters, which consist of a molecular sieve and copper pellets 
that remove contaminants from the argon, are discussed in Section~\ref{sec:prox-cryogenics-cuc}. 
The molecular sieve traps water and the copper pellets trap oxygen. Both will saturate and require  a multi-step regeneration (cleaning) process. 

During regeneration, the filter is first warmed with argon gas heated to an 
elevated temperature (200$\,^\circ$C), driving the water 
trapped in the molecular sieve into the \dword{gar}, and creating water vapor. Hydrogen is then added, 
creating a mixture of up to 1.5\% hydrogen by volume. The hydrogen reacts with the 
oxygen trapped in the copper pellets, releasing more water vapor. All the vapor is vented from the hot circulating gas in a single stream.

The regenerated but still hot filter requires cooling. A heat exchanger (that uses \dword{ln} coolant) in contact with the circulating \dword{gar} cools the gas to cryogenic temperatures, and the gas in turn cools the filter as it circulates through. This completes the regeneration process, at which point the filter is
ready to be switched into service or held cold until needed. 

Two spare purification filters are used with separate heating and 
cooling loops to reduce the usage rate of both electricity and \dword{ln}. Splitting the heating and cooling into separate loops also splits the temperature range seen by the heat exchangers, which decreases mechanical stresses.

\FloatBarrier
\section{Pressure Control}
\label{sec:press-control}
\subsection{Normal Operations}
\label{sec:press-control-ops}

The \dword{gar} pressure-control valves on the feedthroughs are sized and set to control the 
internal cryostat pressure to the experimental operating pressure of \SI{50}{mbarg} (\SI{0.105}{MPa} absolute).
Control systems will take actions to prevent excursions over a millibar.   These actions may 
include stopping the \dword{lar} circulation pumps (to reduce the heat 
ingress to the cryostat), increasing the argon flow rate through 
the condensers, increasing the \dword{ln} flow through the condenser vessels, and/or 
powering down heat sources within the cryostat (e.g., detector electronics).  
Eventually, if the pressure reaches \SI{150}{mbarg}, automatic venting will open partially to start to release the overpressure; at \SI{200}{mbarg} the vents will open fully. The vents close after the pressure event is resolved and the pressure in the ullage is back to operating pressure.  Table~\ref{table:pressure-values} 
gives important pressure values.


\begin{dunetable}
[Important Pressure Values]{ll}
{table:pressure-values}{Important Pressure Values}
Operating pressure &                        \SI{50}{mbarg} \\
\toprowrule
Vessel ullage maximum operating pressure &  \SI{150}{mbarg}\\
\colhline
Vent fully open &                           \SI{200}{mbarg}\\
\colhline
Relief valve set pressure &                 \SI{250}{mbarg}\\
\colhline
Cryostat Design Pressure &                  \SI{350}{mbarg}\\

\end{dunetable}

The ability of the control system to maintain a set pressure is 
dependent on the size of pressure deviations (due to changes in flow, 
heat load, temperature, atmospheric pressure, and so on) and the volume 
of gas in the system.  
The reference design specifies a \dword{gar} depth 
at the top of the cryostat (the ``ullage'') equivalent to 4.5\% of the total 
argon volume for \dword{sphd} (3.5\% for \dword{spvd}); these are typical vapor fractions used for cryogenic 
storage vessels. 
Pressure rise from changes in the heat load is 
gradual, which provides adequate time for the control systems to respond~\cite{EDMS2785497}. 
Two redundant pressure control valves to the recondensers will maintain the required pressure range, each sized to handle \dword{gar} flow during cryostat filling. Two redundant pressure control valves to the vent header will maintain the required pressure range, each sized to handle \dshort{gar} venting scenarios, including loss of recondensing~\cite{EDMS2364392}.

\subsection{Overpressure Control}

In addition to the normal-operation pressure-control system, 
 a cryostat overpressure-protection system is planned.
 This must be a high-integrity, automatic, fail-safe
system capable of preventing catastrophic structural failure
of the cryostat in the case of excessive internal pressure.

The key active components of the planned system are pressure-relief valves (\dwords{prv}) located on the roof of the cryostat that will open rapidly when the differential pressure exceeds a preset value. A pressure-sensing line is used to trigger a pilot valve which in turn opens a PRV. 
A pressurized reservoir of power fluid is provided to each valve to ensure that the valves will operate under all deviation and/or shutdown scenarios. The PRVs 
are self-contained devices provided specially for cryostat protection; 
they are not standard components of the control system. 
Multiple overpressure scenarios are considered in order to size the PRVs~\cite{EDMS2153810}. 

The installation of the \dwords{prv} will ensure that each valve can 
be isolated periodically and tested for correct operation.  
The valves must be removable from service for maintenance 
or replacement without impacting the overall containment envelope 
of the cryostat or the integrity of the over-pressure protection 
system.  This normally requires the inclusion of isolation valves 
upstream and downstream of the pressure-relief valves, which the design includes, and at least
one spare installed relief valve ($2n$ provision) for maintenance.

When the valves open, argon is released, the pressure within the 
cryostat falls, and argon gas discharges into the argon vent riser.  
The valves are designed to close when the pressure drops below 
the preset level.

\subsection{Vacuum-Relief System}

The same \dwords{prv} used for overpressure control are used to prevent catastrophic 
structural failure of the cryostat's primary membrane tank due to low internal pressure.  
Although activation of this system is a non-routine operation and is 
not anticipated to occur during the life of the cryostat, the vent header is appropriately sized for this scenario~\cite{EDMS2154314}.

Potential causes of reduced pressure in the cryostat include 
operation of discharge pumps while the liquid-return inlet 
valves are shut, \dword{gar} condensing in the recondensers 
(a thermo-siphon effect), or a failure of the vent system 
when draining the cryostat.   

The PRVs on the roof of the cryostat 
will monitor the differential pressure between the inside 
and the outside of the cryostat and open when the differential 
pressure exceeds a preset value, pulling in air from the spray chamber to restore a safe pressure. 

\section{Nitrogen Refrigeration System Cycle}
\label{sec:ln-system-cycle}

The nitrogen-refrigeration system, introduced in Section~\ref{sec:comp:ln2}, consists of closed-loop refrigeration plants in the \dword{cuc}. These plants comprise compressors and \coldbox{}es that are used to liquefy the nitrogen, and  storage tanks to store and deliver the liquid nitrogen (\lntwo) to the detector caverns as needed. 
The system serves the argon condensers (Section~\ref{subsec:reliquef}), the argon purification filters (Section~\ref{subsec:filter-regen}) and the integrated \coldbox{}es (for cold APA testing)~\cite{Abi:2020oxb}. Nitrogen is generated underground using membrane technology and supplied to the refrigeration system to initially charge the system, and make up for any losses. \lntwo is delivered from the \coldbox{}es to \lntwo storage tanks with an overall capacity of roughly \SI{30}{m^3}. The tanks will allow for greater than six hours of refrigeration time, a time window that is adequate to cover most power outages, refrigerator performance problems, and  refrigerator switch–overs. 

\lntwo is withdrawn from the storage tanks and supplied via a transfer line to a pressure-reducing valve and phase-separator, both located on the cryostat mezzanine. \lntwo is then withdrawn from the bottom of the phase-separator, at a pressure of \SI{0.62}{barg} (\SI{9}{psig}) and temperature of 
\SI{82}{K} and  temperature of \SI{82}{K}, and directed to the argon recondensers, which is at \SI{89}{K}, i.e., \SI{7}{K} warmer. After flowing through the recondenser the vaporized nitrogen is returned to the refrigeration plant for re-liquefaction. 
The overall liquefaction cycle can be summarized as compressing the nitrogen to high pressure, removing heat from that high pressure nitrogen, and then quickly dropping the pressure of the colder nitrogen. This isentropically reduces both the pressure and temperature of the nitrogen stream, eventually leading to liquefaction. 

In greater detail, the compression, cooling, and expansion are done in stages, with various portions of the nitrogen stream. Initially the nitrogen is compressed through the recycle compressor, this stream is then passed to two additional compression stages, which are each directly coupled to an expander (commonly called ``companders''). After this high-pressure stream enters the heat exchanger it is split into two primary streams, the refrigeration stream, and the liquefaction stream. The refrigeration stream is cooled incrementally in the heat exchanger and then is expanded through the two expanders. This cold gas stream is then passed through the heat exchanger where it is used to cool the liquefaction stream, and then recycled back to the recycle compressor as a warm gas. The liquefaction stream is cooled to near the boiling point in the heat exchanger, and then the pressure is dropped across a valve, which rapidly cools the gas further and creates \lntwo which is transferred to the storage tanks.

The nitrogen refrigeration system (\coldbox) is connected to the \lntwo storage tanks by means of an insulated \lntwo supply line. The insulated \lntwo distribution lines handle the distribution of the \lntwo to the three systems they serve. The return gas (still cryogenic) is collected by insulated return piping and headers. In the event of loss of nitrogen refrigeration (such as a power outage), the \lntwo storage tanks provide  \lntwo to critical argon systems. The gaseous nitrogen is still collected by the return headers, but  without an operating nitrogen refrigeration system, the gas is safely vented into an exhaust drift.

\section{Liquid Argon Removal at the End of Experiment Operations}
\label{sec:liquid-argon-removal}
 
Although removal of the \dword{lar} from the cryostats at the end of life is not in the project 
scope, it is part of the final  disposition of the facility components. \dword{lbnf} has conceptualized a method to 
accomplish this. The \dword{lar} is assumed to be resold to 
suppliers at a fraction of the supply cost.
 
It is expected that storage dewars sized for the task can be carried up and down the skip 
compartments of the shaft (initially used to haul up waste rock from the mine). Because there 
are two skip compartments, an empty vessel can simultaneously be lowered to the \dword{4850l} in one 
skip while a full vessel is raised to the surface in the other. The physical dimensions of 
skip compartment will accommodate a dewar size up to about \SI{3000}{\liter}. 
If  the vessel is pressurized to \SI{50}{psig}, it will contain roughly \SI{3.9}{t} of \dword{lar} (at 95\% full). The pumps 
already present at the cryostats can be used to transfer the \dword{lar} from the cryostat to the 
storage dewar.
 
Assuming that crews work concurrently at the surface and at the \dword{4850l}, one 
optimized conveyance cycle can be fit in approximately one hour.
 This will allow for 18 cycles in an 18-hour day, corresponding to a delivery of \SI{75.6}{t}/day of LAr to the surface. Emptying the \dword{sphd} cryostat will require about 227 days. Emptying the \dword{spvd} cryostat will require about 230 days.
It is expected that about 88\% of the total can be recovered in this process; this takes into account the quantity of liquid below a certain height that cannot be removed using pumps, and a 5\% loss in the transfer of remaining liquid to the dewars. That corresponds to \SI{16.9}{kt} for \dshort{sphd} and \SI{16.1}{kt} for \dshort{spvd}.

\cleardoublepage

\chapter{Process Controls}
\label{ch:cryosys-proc-ctrl}

The \dword{fdc} cryogenics process control system is \dword{plc}-based. It uses a Siemens S7-400 PLC (programmed using the PCS7 software tool) that 
controls, either directly or indirectly, 
all aspects of the \dword{lar} and \lntwo systems. It includes an \dword{odh} subsystem dedicated to safety systems for cryogenic hazards. 
The subcontractor-provided \lntwo system  will 
come with its own \dword{plc}-based control system capable of independently automating nitrogen refrigeration and storage operations; it is part of the overall cryogenics process control system.
The process controls are designed for fully autonomous operations; 
i.e., under normal conditions, no human operators are required. Full details of this system can be found in~\cite{EDMS2154307}. 

\section{System Architecture and Networking}
\label{sec:cryosys-proc-ctrl-arch-net}

The process controls system will make use of the networking infrastructure put in place by the DUNE \dword{daq} Consortium. This includes network switches and cavern-to-cavern cabling. All network connections will be by physical cable (copper or fiber). Networking needs are coordinated through the DAQ Consortium, Facilities Working Group, and Online Computing Coordinator. The process controls architecture is illustrated in Figure~\ref{fig:cryo-ctrl-archit}.

\begin{dunefigure}[Far Site process controls system architecture] 
{fig:cryo-ctrl-archit}
{Far Site process controls system architecture}
\includegraphics[width=\textwidth]{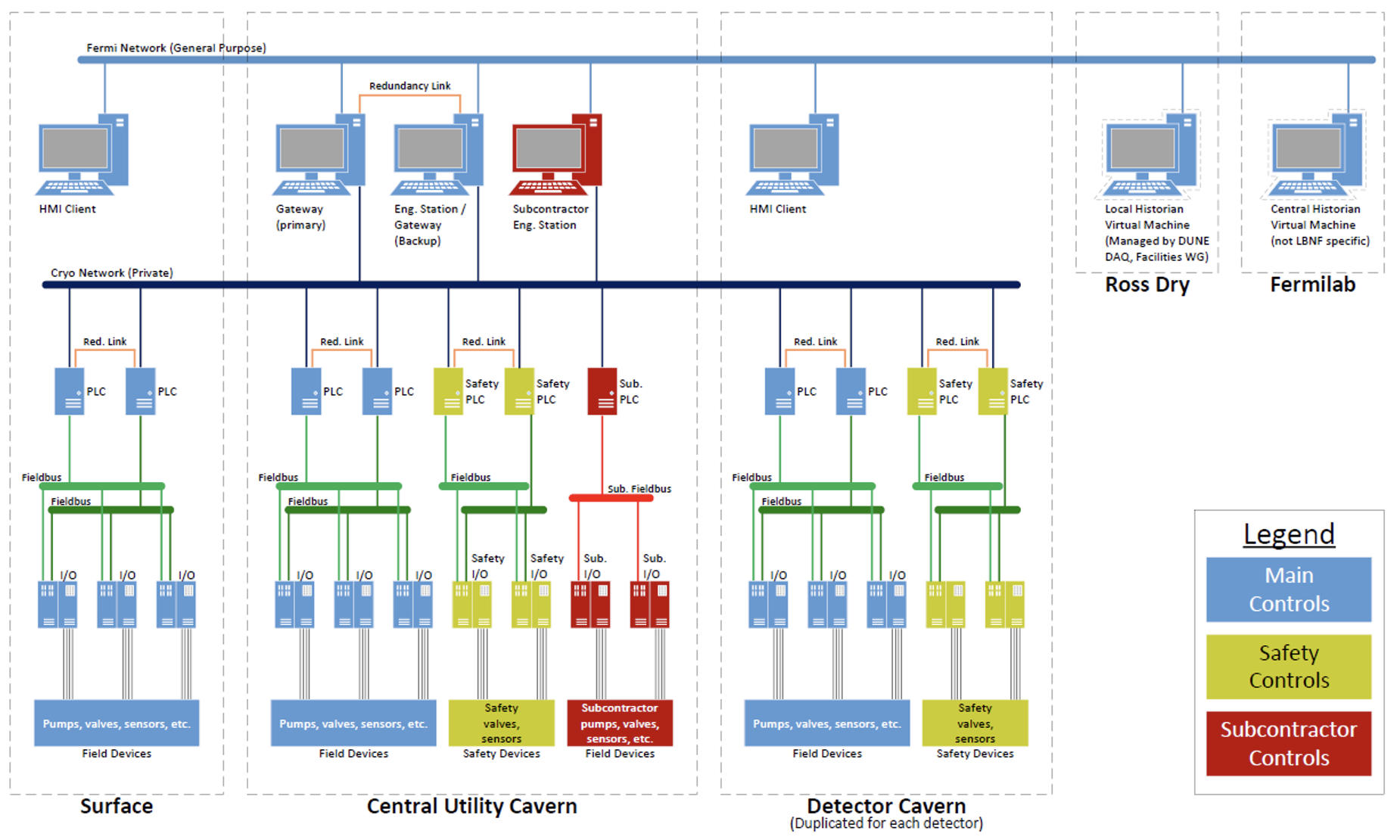}
\end{dunefigure}

The regular Fermilab network 
is extended to the \dword{surf}. The \dword{scada}/engineering station computers will make use of this network for \dword{hmi} clients, connection to historians, and remote control system development.

The process control network is a private VLAN (virtual \dword{lan}). While it makes use of common network switches and cabling, the physical Ethernet ports are dedicated to the VLAN. Computers used as gateways will make use of this network for communication with the PLCs 
via the process control network. It can tolerate interruptions, as it is used for monitoring and operator intervention, not continuous control.

\section{Access to the Process Control System}

The process control system will provide multiple means of local and remote access for engineers, technicians, and shifters to monitor and control the cryogenics systems. 
 Full-time operators are not required.


A control room at the west end of the \dword{cuc}, the most central location with respect to most of the cryogenics equipment, will house the primary process control station, illustrated in Figure~\ref{fig:cryo-ctrl-station-cuc}. 
This will be a dedicated space with the \dword{hmi} displayed across multiple monitors, providing local access to monitor and control the system.
This control station is also home to the gateway computers (gateways) that communicate with the PLCs; one of these computers will serve also as an engineering station for PLC programming. The layout of the process control system is depicted in Figures~\ref{fig:cryo-ctrl-fd-cabinet-layout}, \ref{fig:cryo-ctrl-layout-cuc-layout}, and~\ref{fig:cryo-lar-circ-vdf-layout}.

\begin{dunefigure}[Control station in the CUC] 
{fig:cryo-ctrl-station-cuc}
{Control station in the CUC}
\includegraphics[width=0.6\textwidth]{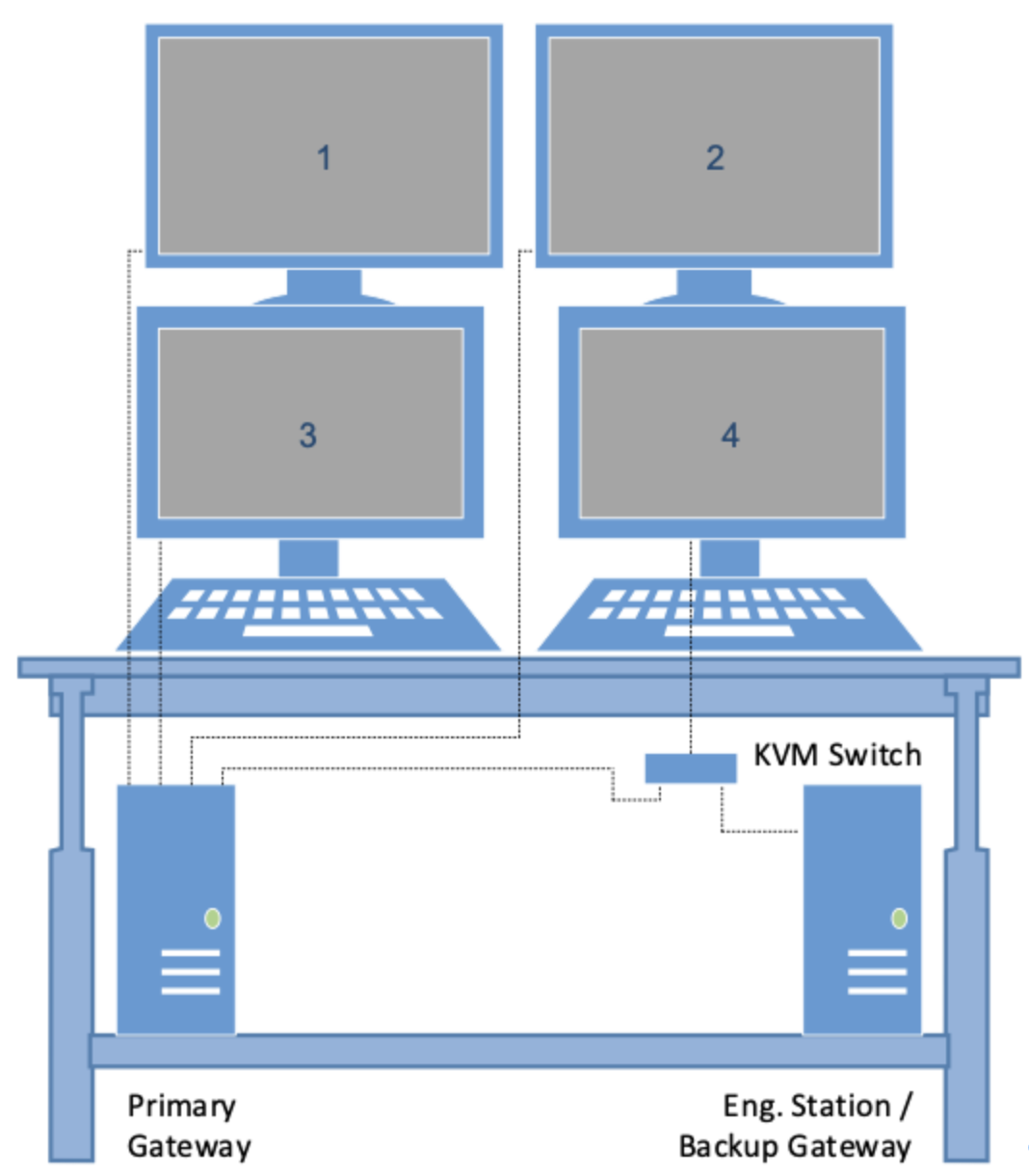}
\end{dunefigure}

\begin{dunefigure}[Control cabinet layout (regular and ODH)] 
{fig:cryo-ctrl-fd-cabinet-layout}
{\dshort{sphd} control cabinet layout (same respective locations on \dshort{spvd}, locations to be used for both regular and \dshort{odh} cabinets)}
\includegraphics[width=\textwidth]{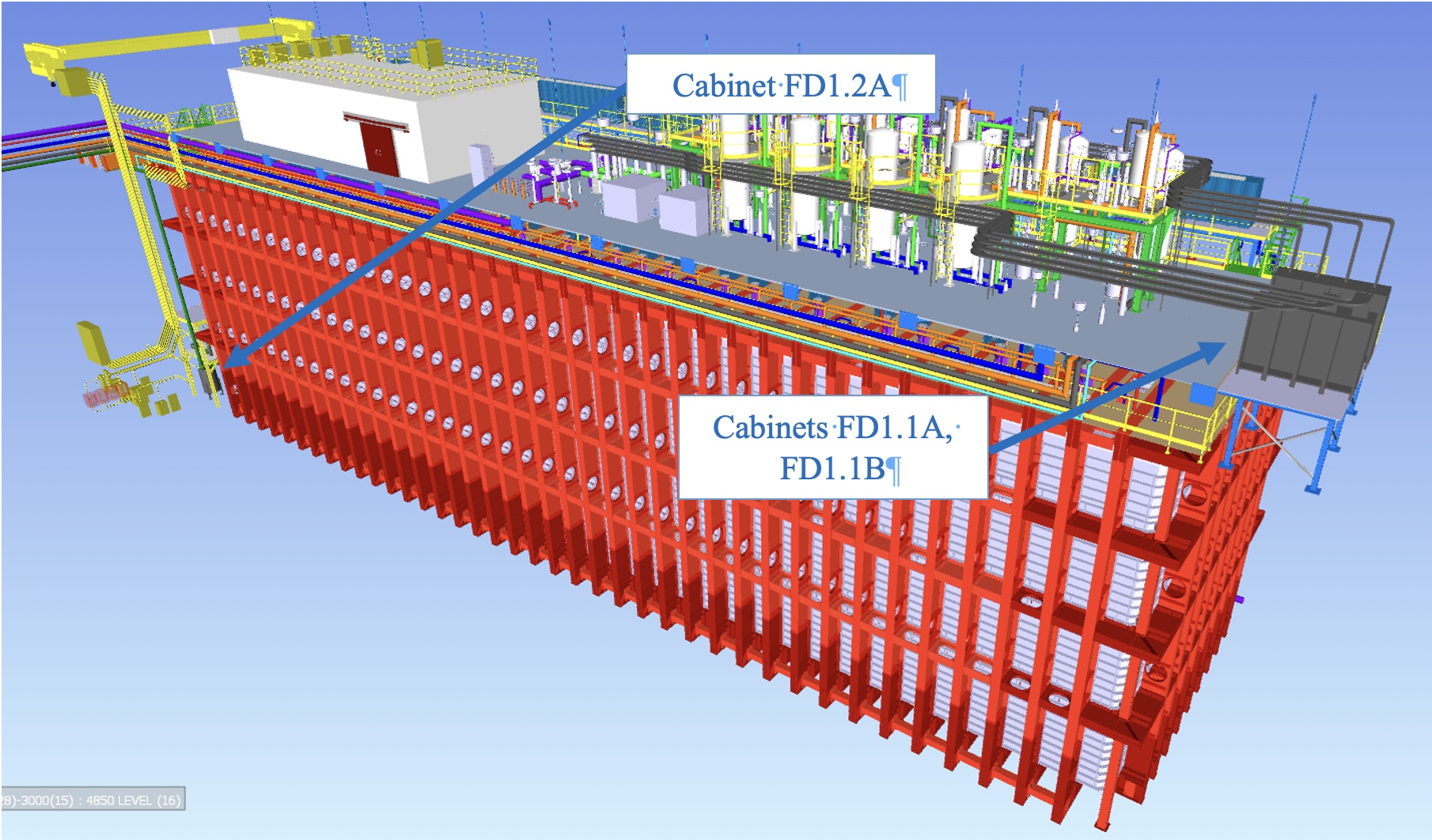}
\end{dunefigure}

\begin{dunefigure}[CUC cabinet layout (regular and ODH)] 
{fig:cryo-ctrl-layout-cuc-layout}
{CUC Control Cabinet Layout (locations to be used for both regular and ODH cabinets)}
\includegraphics[width=\textwidth]{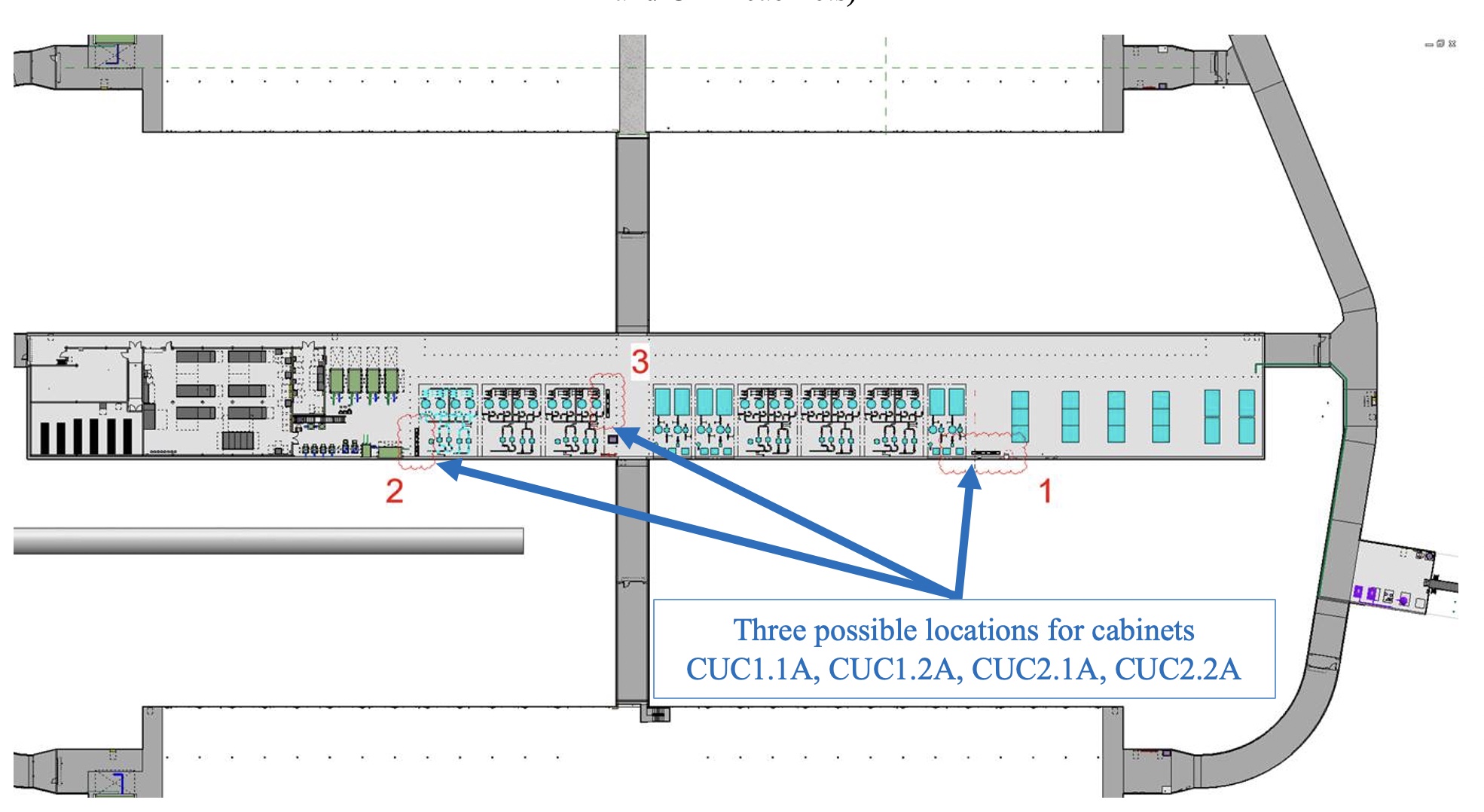}
\end{dunefigure}

\begin{dunefigure}[LAr circulation VFD layout] 
{fig:cryo-lar-circ-vdf-layout}
{LAr circulation VFD layout}
\includegraphics[width=\textwidth]{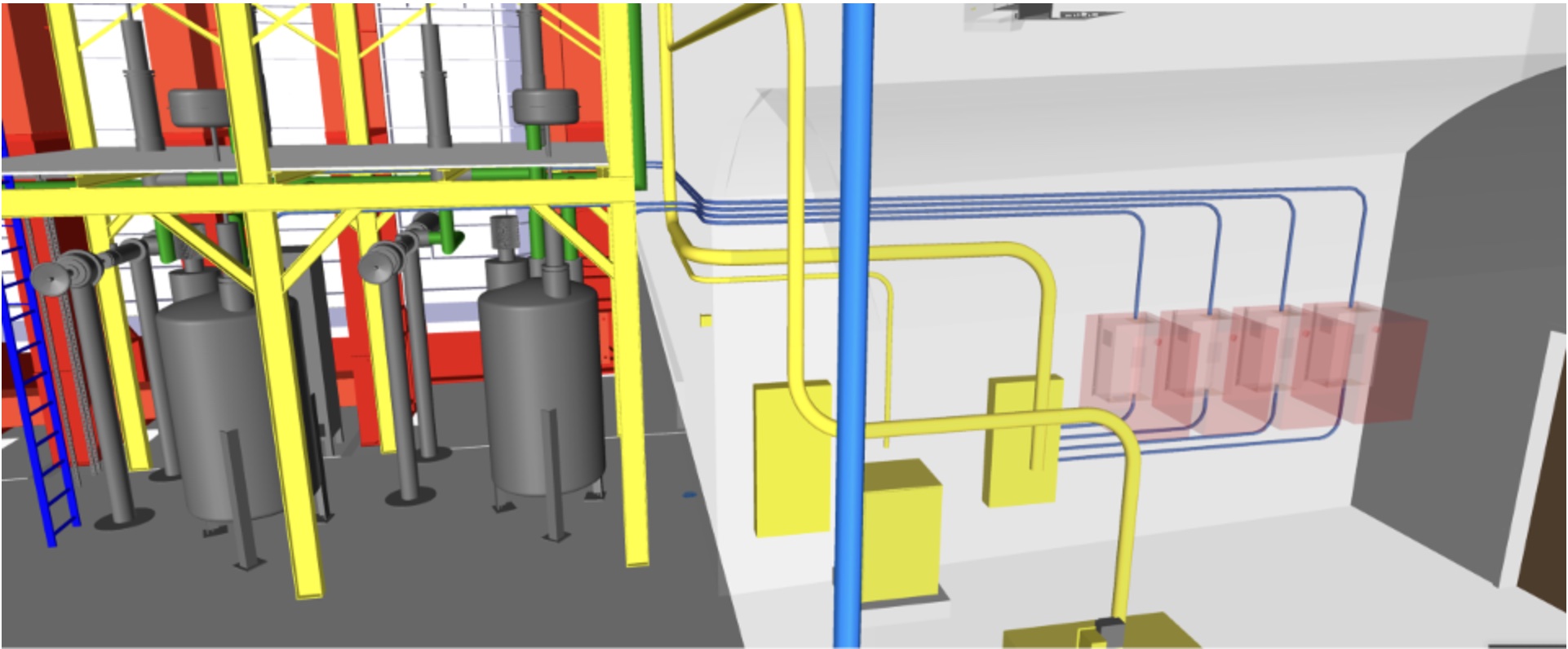}
\end{dunefigure}

The process control system will provide additional local access points in the detector cavern workspaces (small offices on the mezzanines) and on the surface (location TBD). These are shared spaces, not dedicated to cryogenics. They are intended to be used primarily for commissioning, not to be occupied under normal operating conditions. Access to the control system in these spaces may be via an HMI client run on a shared computer.

The process control system will provide remote access via personal computers, phones, and tablets. The \dword{scada} computers will be on the Fermilab network and provide unlimited client connections, 
subject to the Fermilab computer security protocols.

\section{Supervisory Control and Data Acquisition}

The process control system will 
implement Ignition by Inductive Automation, a comprehensive \dword{scada} software platform. It provides 
gateway computers as well as modules that execute HMI, alarm management, and historical data collection functions.


The gateways are the core of the Ignition platform. They
will connect to the PLCs, including those provided by subcontractors, and act as the point of data collection and distribution for the entire control system. The gateways also include \dword{opc} United Architecture (UA) 
 servers, allowing data sharing with other systems (e.g., the DUNE Detector Control System) over the Fermilab network.

The process control system will include \dword{hmi}, using both the Ignition Vision and Perspective modules. Vision is a more traditional HMI, intended for desktops. Perspective is a web-based mobile-responsive HMI.

The HMI will 
enable monitoring and command of all cryogenics equipment, including that provided by subcontractors. No control hardware or software for cryogenics equipment of any kind will exist fully independently of the HMI.
The HMI will not contain any control logic. (\dwords{plc} are not dependent on the HMI to operate.)

The design of the HMI will comply with ANSI/ISA-101.01-2015 (or later), Human Machine Interfaces for Process Automation Systems. This includes implementation of a hierarchy of displays. A sample is given in Table~\ref{table:hmi-hier}.

\begin{dunetable}[Sample HMI process automation hierarchy of displays]
{lp{0.75\textwidth}}
{table:hmi-hier}
{Sample HMI process automation hierarchy of displays}
{Level} &  \textbf{Description} \\ \toprowrule

1 & Entire Cryogenic System Overview
				(only very important parameters, shown on gauges)\\ \colhline
2 & Nitrogen Refrigeration Overview, LAr Filtration Overview
				(important parameters, shown on gauges and trend charts)\\ \colhline
3 & LN2 Storage Tanks, LAr Filter Regeneration Hot/Cold Loop
				(animated P\&ID, most parameters)\\ \colhline
4 & Subsystem Detail\\ 

\end{dunetable}

Table~\ref{table:hmi-user-hier} shows the HMI security scheme that permits access to users as appropriate to their role. Special security access may be implemented for control of subcontractor-provided equipment.

\begin{dunetable}[HMI user access levels]
{ll}
{table:hmi-user-hier}
{HMI user access levels}
{Role} &  \textbf{Description} \\ \toprowrule

Guest	 & Monitoring only\\ \colhline
Operator & Basic Control (i.e., open/close valves, set pump speed)\\ \colhline
Expert & Advanced Control (i.e., set alarm levels and P\&ID tuning parameters) \\
\end{dunetable}

Alarms (Section~\ref{sec:safety-alarms})  will make use of Ignition’s built-in alarm management functionality, which is consistent with ANSI/ISA-18.2-2016, Management of Alarm Systems for Process Industry. Alarms alert personnel to abnormal conditions, but do not impact system operation in any way.
Alarms will implement a consistent level of prioritization. Table~\ref{table:alarmlevels} shows an example.

\begin{dunetable}[Alarm levels]
{ll}
{table:alarmlevels}
{Alarm levels}
{Level} &  \textbf{Description} \\ \toprowrule
Interlocks & Critical Priority\\ \colhline
High-High and Low-Low Alarms & High Priority\\ \colhline
High and Low Alarms & Low Priority\\ \colhline
Input/Output Errors & Diagnostic Priority\\ 
\end{dunetable}


Alarm notification will be handled via the Ignition Alarm Notification Module, which organizes the sequence of events following an alarm into pipelines. Pipelines will be configured by priority to provide alarm notifications via e-mail, text, and or phone call to a limited number of people (i.e., <10 people), namely those engineers and technicians with direct responsibility for far site cryogenic operations. See Table~\ref{table:alarmresponse}.

\begin{dunetable}[Response to alarms]
{lp{0.75\textwidth}}
{table:alarmresponse}
{Response to alarms}
{Level} &  \textbf{Description} \\ \toprowrule
Low Priority &	Notify primary cryogenic engineer if not acknowledged within 10 minutes; notify alternate cryogenic engineer if not acknowledged within 30 minutes\\ \colhline
High Priority &	Notify primary cryogenic engineer immediately;	notify alternate cryogenic engineer if not acknowledged within 10 minutes\\ 
\end{dunetable}

All cryogenics data will be made available to the DUNE Detector Control System so that if DUNE collaborators wish to setup their own alarms and notifications based on cryogenics data, they will have the means to do so.

Historical data collection will be realized by means of the Ignition Tag Historian Module, which filters and stores data in SQL databases.
The SQL databases are not part of Ignition and must be setup separately such that Ignition can point to the desired database. 
Ignition then creates tables in the database and populates them with historical data.
Historical data will be archived to two SQL databases (``historians'') simultaneously, one local and one remote at Fermilab.

\section{Databases for Local and Remote Historians}

The DUNE \dword{daq} Consortium is planning to have a virtual machine cluster at the far site in the computer room at Ross Dry. A virtual machine will be designated for process control with an SQL database to serve as the local historian.
At Fermilab, a virtual machine is to be designated for process control with a SQL database to serve as the remote historian.
Both SQL databases will be managed by a Fermilab Computing DBA (database administrator).

\section{Programmable Logic Controllers (PLC)}

This section applies to Fermilab-provided, standard (not safety-rated) PLCs. For subcontractor provided PLCs, refer to~\cite{EDMS2705540}. The only Fermilab-provided safety PLCs are in the \dword{odh} subsystem (Section~\ref{sec:odh}).

The process control system will include Siemens S7-400 \dwords{plc}. Direct control of all cryogenics equipment will be done exclusively via PLC. PLCs will be installed on the surface, in the \dword{cuc}, and by each \dword{detmodule}, and will control  the cryogenics equipment in their respective areas. Process controls mechanical and electrical drawings for PLC systems can be found in \cite{EDMS2156486}. 
PLCs will contain all control logic necessary to operate all connected equipment. Under normal conditions, PLCs operate the cryogenics systems autonomously (without input from \dword{scada}).

PLCs will be implemented in redundant pairs.
This provides continuous operation in the event of a failure or the need to take one offline (e.g., for updates).
Where possible, redundant PLCs will not be co-located, for example, redundant PLCs will be positioned in cabinets on opposite ends of the CUC. This is to prevent a single hazard from neutralizing both PLCs in a redundant pair.
For each detector module, one PLC is on the mezzanine while the other is in the pit. At the surface, due to the limited space for controls, the redundant PLCs will be near each other.
PLCs will host redundant fieldbus (Profinet) networks for I/O modules.

The engineering station is part of the process control station in the CUC and contains the PLC configuration and programming tools. While the engineering station will contain the PLC programs (control logic) for all PLCs, the programs do not run on the engineering station itself.

\section{Backup Power}
The process control system will make use of the generator provided by \dword{fscf} and its own UPS (uninterruptable power supply) for short-term backup power to cover the gap between loss of power and the generator coming online.

The detector cavern PLCs and I/O are included, as they are necessary for condenser operation to maintain cryostat pressure and minimize loss of LAr. The CUC and surface PLCs and I/O are currently designed with their own UPS systems, but are not necessary for 
maintaining cryostat pressure. Their inclusion is consistent with other experiments. The \dword{vfd}s (for pumps) and heaters are not included.

\section{Oxygen Deficiency Hazard System}
\label{sec:odh}

An \dword{odh} condition is typically the result of a cryogen (e.g., argon or nitrogen) leak. As cryogens transform from liquid to gaseous form, they expand and displace oxygen, creating a dangerous environment as oxygen content decreases. 

Each cryostat is housed within a warm (room-temperature) vessel in one of the detector caverns. Fresh air 
flows in from either end of a given detector cavern, across the warm vessels, and into an exhaust duct near the center of the cavern, as shown in Figure~\ref{fig:cryo-det-cavern-airflow}. 
There are currently no plans for dampers or fans controlled or monitored by the safety PLC system. Air flow and supply is regulated by the FSCF air handling system.

\begin{dunefigure}[Detector cavern airflow] 
{fig:cryo-det-cavern-airflow}
{Detector cavern airflow}
\includegraphics[width=\textwidth]{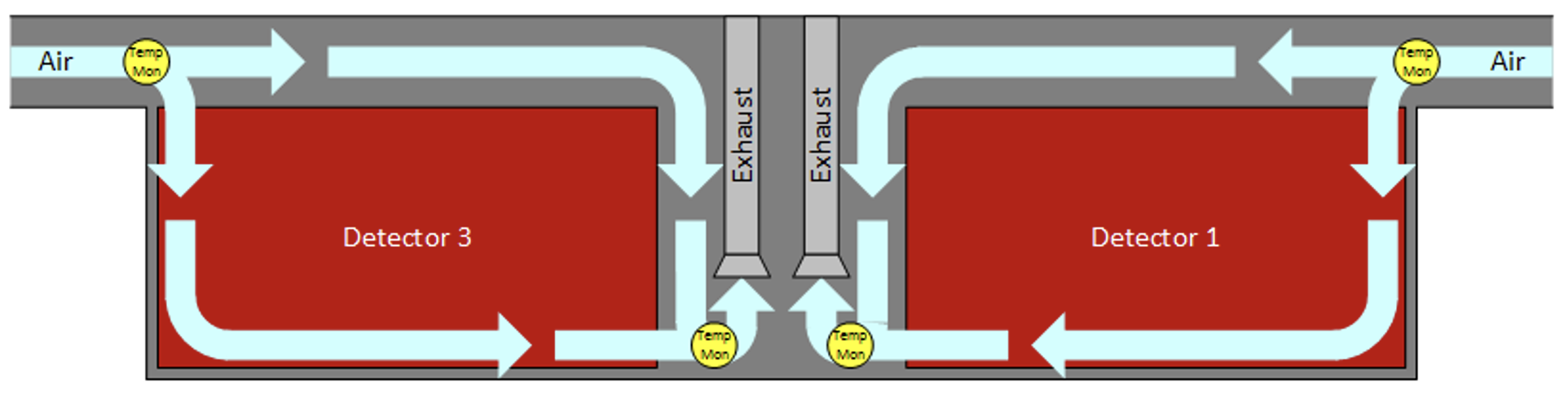}
\end{dunefigure}

\dword{odh} protection is handled by a self-contained subsystem that detects an ODH condition, alerts the occupants of the hazard, and takes action to eliminate the hazard. Details beyond what this section covers are available in~\cite{EDMS2154321}.

\subsection{Software}

The Siemens PCS7 F-Systems Programming tool combined with the Safety Matrix Tool is used to create the safety subsystem. 
The programming language is Continuous Function Chart (CFC), a graphical programming language that extends the standard languages of IEC 61131-3. All logic in the safety program is carried out by certified safety function blocks. The program monitors 
the conditions covered in the following subsections.



\subsection{Gas Monitors}

Twenty gas monitors  (MSA Ultima X5000, Figure~\ref{fig:cryo-ultima} are installed in strategic positions in each of the three caverns (FD1, CUC, FD2).
Each gas monitor provides 
an independent indication if it detects oxygen content falling below preset limits. 
Safety PLCs monitor gas monitor function (e.g., can detect a disconnected sensor) and alarms. 

Each gas monitor provides analog indication of oxygen content over the range of 0\% to 25\%. (Standard, i.e., non-safety, PLCs monitor this for information only.) Additionally, the monitors are capable of \dword{hart} communication over the analog signal, which may be used to alert the control system when the sensor is reaching its end of life (2-3 months prior to sensor failure).

\begin{dunefigure}[MSA Ultima gas monitor] 
{fig:cryo-ultima}
{MSA Ultima gas monitor, part number A-X5000-0-F-0-1-0-16-0-0, replacement sensor part number A-5K-SENS-16-0-F-1-0.}
\includegraphics[width=0.33\textwidth]{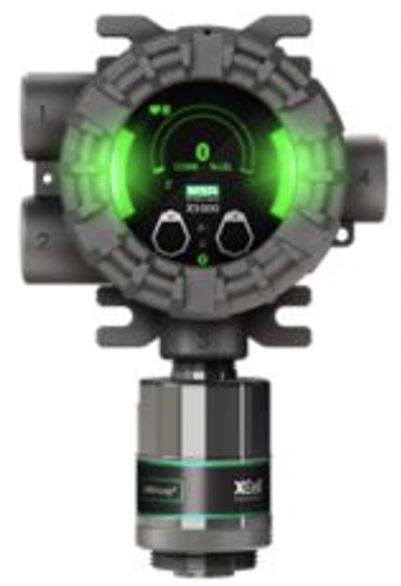}
\end{dunefigure}

\subsection{Temperature Monitors}
Air temperature monitors  (Figure~\ref{fig:cryo-safety-transmitter}) are located near the two lower corners of each detector nearest the center of the cavern. (This location may change to the intake of the exhaust duct.) An additional monitor is placed above each detector. A drop in air temperature at the center of the detector cavern would indicate a leak. 

The temperature monitors 
alarm if air temperature falls below a preset limit. 
Safety PLCs monitor the function of the devices (e.g., to detect an internal fault) and alarms.
Each temperature monitor also provides analog indication in the range of -40 to 450$^\circ$F (-40 to 232$^\circ$C). (The standard PLCs monitor this for information only.)

\begin{dunefigure}[Photo of temperature monitor] 
{fig:cryo-safety-transmitter}
{Temperature monitor; United Electric Controls ``One Series Safety Transmitter,'' part number 2SLP49 TL1-W073}
\includegraphics[width=0.28\textwidth]{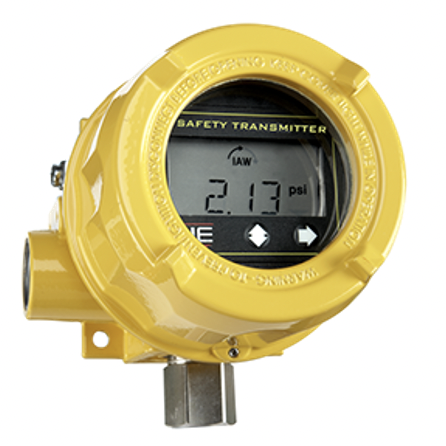}
\end{dunefigure}

\subsection{Vacuum Monitors} 

Cryostat side penetrations 
are each equipped with an isolation valve, and these valves and associated piping are vacuum jacketed.  Loss of vacuum in a jacket would indicate a leak, which could lead to more severe problems, including \dword{odh} and equipment damage.

The ODH system therefore includes multiple vacuum monitors, which are in turn monitored by the safety PLC. Each vacuum space on the cryostat side penetrations has two monitors that 
alarm individually if the pressure exceeds 200\,mTorr. 
Each vacuum monitor also provides analog indication of pressure in the range of 0.1\,mTorr to 1000\,Torr. (The standard PLCs monitor this for information only.)
The vacuum monitors are InstruTech Stinger convection gauge, part number CVM211GGL (Figure~\ref{fig:cryo-convec-vac-gauge}).

\begin{dunefigure}[Photo of convection vacuum gauge] 
{fig:cryo-convec-vac-gauge}
{InstruTech Stinger Convection Vacuum Gauge}
\includegraphics[width=0.33\textwidth]{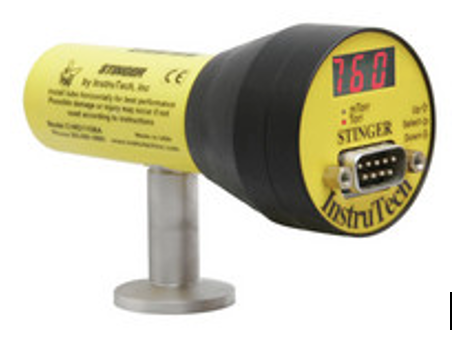}
\end{dunefigure}

\subsection{Seismic Activity Monitors}

Seismic activity can damage equipment and potentially lead to safety hazards, including ODH.
The ODH safety system includes one 
safety seismic switch (Figure~\ref{fig:cryo-safety-seismic-switch}) in the CUC and one per \dshort{detmodule}.
Each seismic switch contains three vibration detection devices in different orientations. These seismic switches will  
alarm if any of their vibration detection devices reads above a preset limit. 

\begin{dunefigure}[Photo of safety seismic switch] 
{fig:cryo-safety-seismic-switch}
{Sensonics model SA-3 safety seismic switch}
\includegraphics[width=0.7\textwidth]{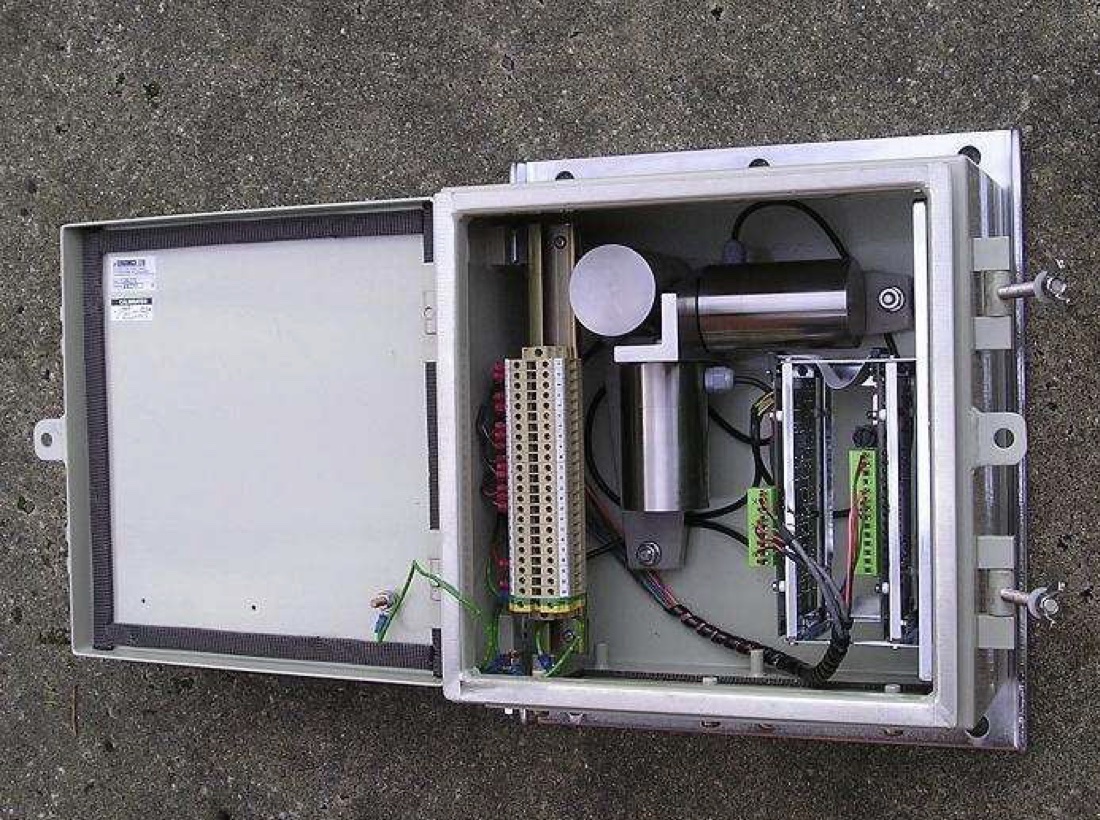}
\end{dunefigure}

\subsection{Safety System Power and Controls}
The \dword{odh} safety subsystem is electrically wired to 
self-contained control circuitry in its own enclosure. 
The overall process controls UPS system provides power to the ODH subsystem via dedicated 120\,VAC circuits that are fed by panels on generator backup. The UPS provides battery power long enough for main power to switch to the generator in the event of an outage.

There is one set of controls in each of the three caverns for the ODH system, 
each housed in a set of two Rittal 8826.500 cabinets. 
In each detector cavern, one cabinet is positioned on the cryogenics mezzanine, and the other is positioned in the pit. In the \dword{cuc}, the cabinets are positioned centrally.
The control cabinets each include one CPU410-5H/F-System PLC. The safety program handles all safety-related functions, including detection of hazards, cycling air from outside, 
alarms, 
and control of the cryostat safety valve. The 
standard PLC handles communication with the gateways and monitoring of non-safety-related  signals.

\subsection{Alarms}
\label{sec:safety-alarms}
Each detector ODH safety system includes four pairs of horns and strobes. 
When enabled, the horns make a warbling noise and the strobes flash, alerting building occupants of a hazard.
Two horn/strobe pairs are located at the mezzanine level and positioned such that at least one strobe is visible from anywhere on the cryostat top and mezzanines. Another horn/strobe pair is located on the west side of the detector cavern, positioned to be visible from the grade and platform levels, and another pair is located on the east side of the detector cavern, also for visibility from the grade and platform levels.

These alarms feed into the \dword{surf} Fire-Life-Safety system that monitors fire and other emergency conditions. The ODH system provides to it 
two distinct signals, one is a warning and another indicates an emergency. 
The Fire-Life-Safety system would then alert the SURF facilities dispatch center. 

The ODH system communicates with the main cryogenics control system 
via a network connection hosted by the main system \dword{opc} server. When necessary, the main system can initiate further mitigation (i.e., stop pumps and close valves).

\cleardoublepage

\chapter{Prototyping Program}
\label{sec:cryo-cryosys-proto-plans}

A prototyping program has been central to the development of the \dword{lbnf} cryostat and cryogenics infrastructure from conceptual to final design. The most 
significant issue to resolve has been whether a membrane cryostat of
the size required by \dword{dune} can achieve the required electron drift lifetime. 

\section{The Liquid Argon Purity Demonstrator}
The \dword{lapd} was an early 
off-project prototype, built to study the concept of achieving
\dword{lar} purity requirements in a non-evacuated vessel. 
The purge process accomplished in the \dword{lapd} was 
repeated ten years ago on the \dword{35t}, 
developed as an \dword{lbne} effort, which confirmed that initial 
evacuation of a membrane cryostat is unnecessary and that 
a \dword{lar} purity level sufficient to enable the electron lifetime 
required by \dword{dune} can be achieved in a membrane cryostat~\cite{Montanari:2013/06/13aqa}.

\section{ProtoDUNE}

A further prototyping program, aimed at testing and demonstrating 
the \dword{dune} detector technologies and engineering procedures at the \SI{1}{kt} (metric) scale, has taken place over the past 
few years as part of the \dword{cern} Neutrino Platform program. Two detectors of this scale, \dword{pdsp} and \dword{pddp}, were installed in a pair of membrane cryostats, each of inner dimensions 
 \SI{8.6} {\times} \SI{8.6} {\times} \SI{7.9} m$^3$.
 \dshort{pdsp} was filled with \dword{lar} in July/August 2018 and operated in a test beam and with cosmics, whereas \dshort{pddp} was filled in August 2019 and operated only with cosmics.

\dshort{pdsp} had the same \spmaxdrift maximum drift length as the full \dword{spmod}. \dshort{pddp} had a \SI{6}{m} maximum drift length, half of the length that had been planned for an eventual \dword{dpmod}\footnote{The second \dword{detmodule} had been planned to be \dword{dp}; it is now planned as \dword{sp} with a vertical drift, with a maximum 6\,m drift length.} See the photos in Figures~\ref{fig:protodunes_northarea} and~\ref{fig:protodune_interior}.

\begin{dunefigure}[ProtoDUNE cryostats at the CERN Neutrino Platform]
{fig:protodunes_northarea}
{ProtoDUNE-SP (foreground) and ProtoDUNE-DP (right rear, red, at an angle) cryostats in the CERN Neutrino Platform in CERN's North Area.}
\includegraphics[width=0.9\linewidth]{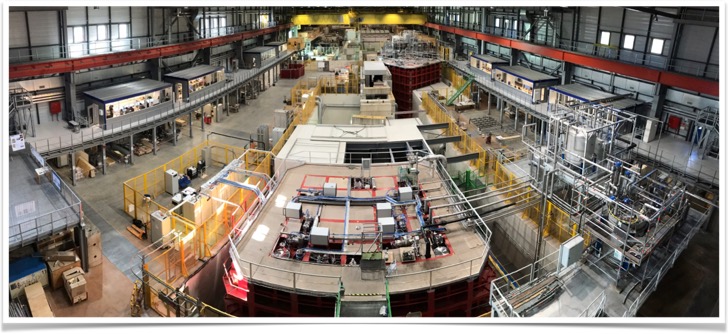}
\end{dunefigure}

\begin{dunefigure}[Interior view of a ProtoDUNE cryostat showing membrane]
{fig:protodune_interior}
{Interior view of one of the ProtoDUNE cryostats, before any detector components were installed, showing the membrane. The gold color is an artifact of the lighting that was set up to protect photo-sensitive detector components during installation.}
\includegraphics[width=0.66\linewidth]{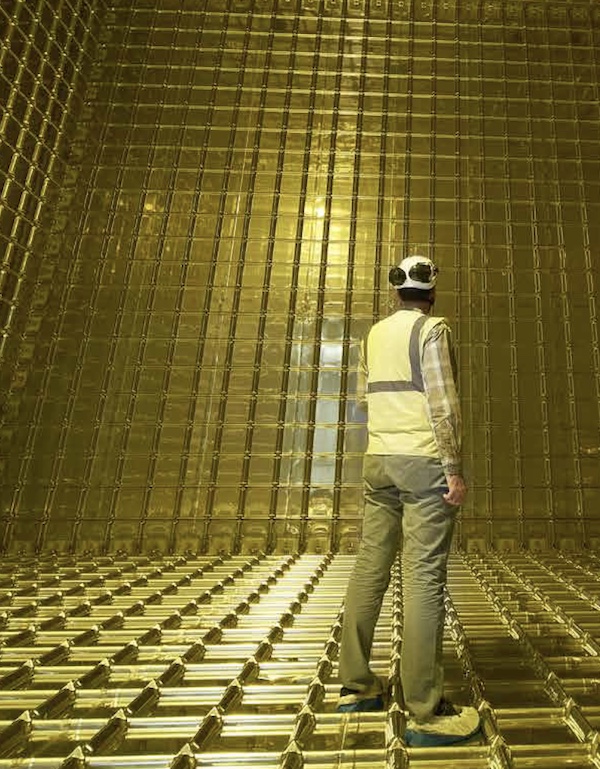}
\end{dunefigure}

\subsection{Cryostats}

The performance of the cryostats is fully described in \cite{2108.01902}; the text in this section comes largely from this paper.

The cryostat inner and outer structures had to satisfy both U.S. and European regulations and standards, as the \dword{pdsp} cryostats were installed and operated in Europe and the \dword{fd} cryostats will operate in the U.S. 
Rigorous \dword{qa} and \dword{qc}  procedures were carried out on all materials and techniques used for the construction.

Validation and certification of the \dword{protodune} cryostat structure had two principal aspects. The first, leak checking, was performed at various stages of the cryostat construction. 
The second  concerned the mechanical behavior of the cryostat in terms of compliance with engineering safety standards and regulations. This validation was also performed multiple times throughout the course of construction, and  included  test campaigns both during and after filling with \dword{lar}. 

Leak testing was performed on the external warm structure, the inner membrane, and all the penetrations, generally 
by spraying helium close to the surface and reading the percentage of helium on the opposite side of the surface using 
a leak detector in \textit{sniffing mode}. A deviation from the environmental background ($\sim2-3 \times 10^{-6}$~mbar~l/s) is considered a possible leak. 
The checks were performed on the warm structure during the assembly of \dword{pdsp} and no leaks were found.  

The cryostat inner membrane was tested twice for leaks. The first test, upon completion of the membrane installation, was the official qualification for leak-tightness by the vendor, and was
performed on all welding lines. The few small leaks found were swiftly repaired. The second leak test 
involved pulling vacuum around localized sections of welding lines and testing them one by one. Overall, about 80\% of the total length of welding lines on the inner membrane was tested and no leaks were found. This gives confidence that the helium leak-check performed by the vendor can be fully trusted when it comes time to verify the cryostat inner membranes for the far detector, and that performing the second test will be unnecessary. Retesting single welded lines on a \dword{fd}-scale cryostat would require months.

The leak-tightness of all feedthrough flanges on the roof was verified. All flanges except one were certified to be leak-tight; the chimney of one temperature profiler was found to be faulty. None of the actions taken fixed the leak entirely, therefore an enclosure was constructed and installed around the leaky flange, defining a buffer volume in which argon gas circulated continuously. 

To fully validate the mechanical performance of the cryostat, a careful stress analysis of each structural component was done, together with pressure tests before and after filling with LAr.  Two \dword{fea} models were developed to evaluate the cryostat design against the required codes and regulations. Predictions from the models were compared to the experimental data collected during the commissioning phase and operation. 

\begin{dunefigure}[Expected deformation of the cryostat]{fig:global_deform}
{Expected deformation (using an exaggerated scale) of the cryostat during standard operations (liquid argon level of 7.4\,m and $\Delta P = +57 $ mbarg. Predictions are obtained using the \emph{shell} FEA model (see text).}\includegraphics[width=0.8\columnwidth]{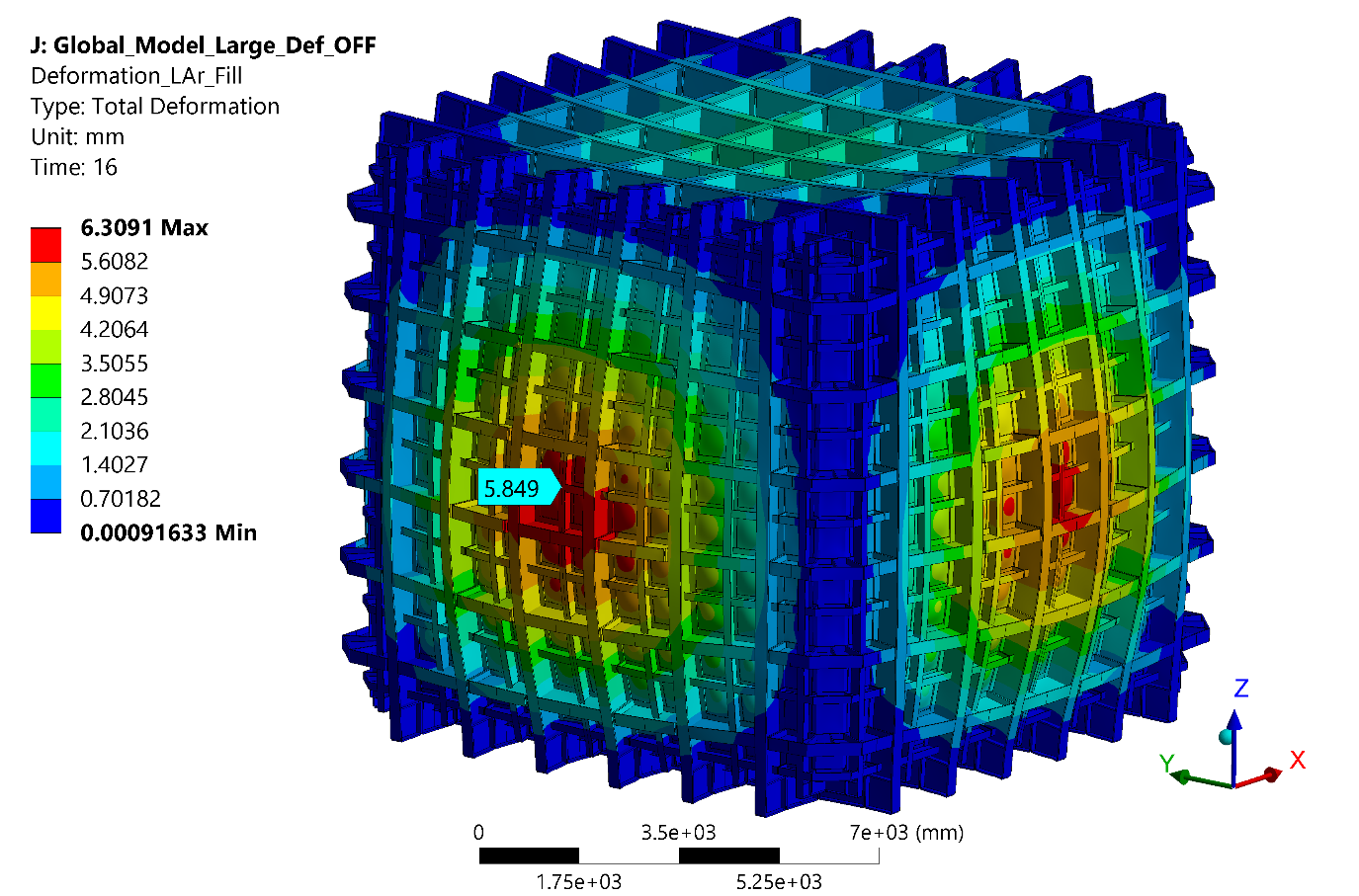}
\end{dunefigure}

All tests performed during the validation campaign have been successful and the \dword{fea} model has reproduced the cryostat behavior in all cases. 

The values of the displacement and strain gauges measured during the fill were compared to the \dword{fea} predictions and found to fulfill the safety requirements. This comparison also validated the \dword{fea} model itself, which was then used to predict the cryostat behavior at the sizing case. The outcome of the simulation was compatible with both EU and U.S. standards, thus providing the final qualification of the inner and outer cryostat mechanical structures. 

\subsection{Cryogenics Systems}

The performance of the cryogenics systems is fully described in \cite{2108.01902}; the text in this section comes largely from this paper.

The \dfirst{qa} and \dfirst{qc} steps  
were performed during the design, construction, installation and commissioning phases, as appropriate. During the construction and installation phases, non-destructive tests (X-rays, He leak tests and pressure tests) were successfully executed. During cold commissioning another series of tests was completed successfully, the main two of which were a check of the I/O signals and a functional test of all valves and equipment.  In addition, tests were run on the Ethernet or hardwired signal lines dedicated to the exchange of information between the cryogenics system and the detector system, the safety system, and the \dword{cern} Central Control room.     Figure~\ref{fig:filling} shows the stability of the level, pressure and outer structure temperature during filling of \dword{pdsp}.

\begin{dunefigure}[Filling trends]{fig:filling}
{LAr level (top), pressure (middle) and cryostat outer structure temperature (bottom) trends during the filling of the \dword{pdsp} cryostat.}
\includegraphics[width=0.9\columnwidth]{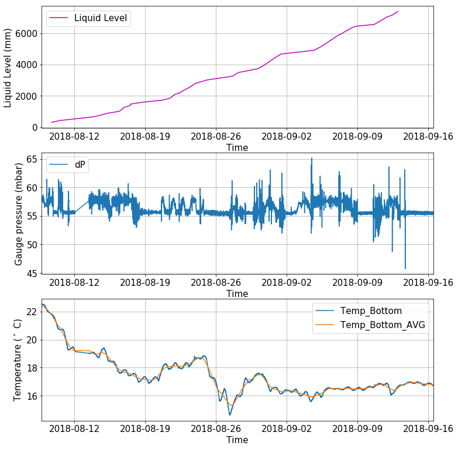}
\end{dunefigure}

A ``mirror'' station (not connected to the actual field equipment) provided a functional test environment for the process control logic. The tests were subsequently carried out on the real system to verify first alarms, interlocks, and the sequential function charts. Operation modes were tested afterwards, during the system commissioning.  

\cleardoublepage

\chapter{Environment, Safety, and Health}
\label{ch:cryo-cryosys-esh}

Both \dword{fnal}  and \dword{surf} \dword{esh}
codes and standards have been guiding and will continue to guide the design, prototyping, 
procurement, and installation phases of the 
far site cryostat and cryogenics infrastructure for \dword{lbnf-dune}. Particular 
attention is paid to critical sections of
Chapter 4240~\cite{feshm} relating to \dword{odh} and 
Chapter 5000~\cite{feshm} standards for 
piping construction and vessel design. The planned 
work process will provide for reviews throughout all 
phases of the project to guarantee stringent adherence 
to the safety requirements. Requirements on the membrane-cryostat
materials and their fabrication are strictly outlined in the 
specification documents. Close communication between the vendors, 
\dword{fnal}'s  and \dword{cern}'s cryogenic and process engineers, and \dword{fnal} and 
\dword{surf} \dword{esh} personnel is and will be maintained at all times.

\begin{figure}[htbp]
\centering
\includegraphics[width=0.95\textwidth]{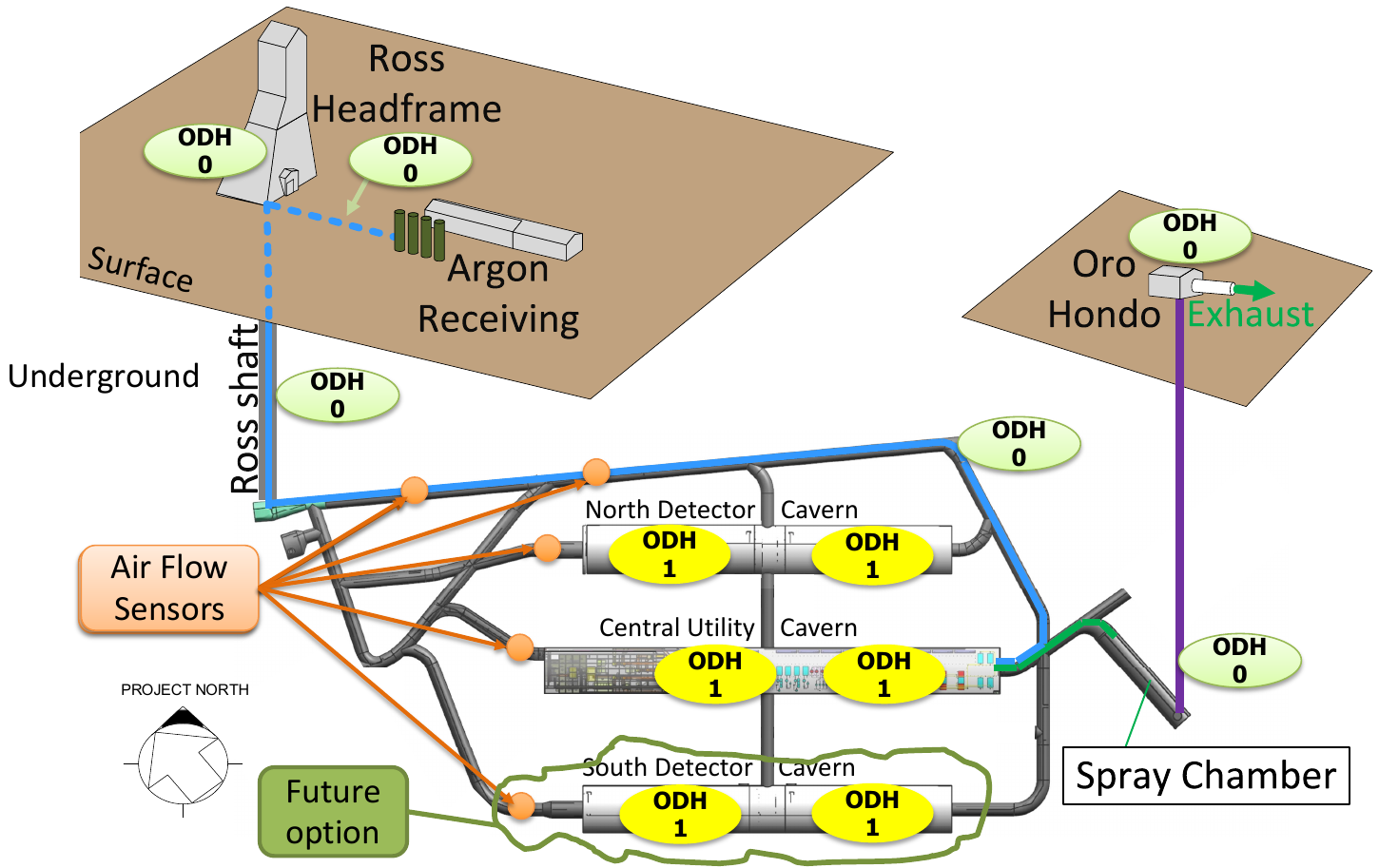}
\caption{\dshort{odh} mapping of the underground caverns at 4850L} 
\label{fig:ODH-mapping-4850L}
\end{figure}

Figure~\ref{fig:ODH-mapping-4850L} shows the \dword{odh} classification for
underground caverns at the \dword{surf}. The detector and \dword{cuc}
are Class 1 \dword{odh} areas~\cite{feshm},
assessed by preliminary \dword{odh} analysis taking into account
potential risks from undetected defects on materials
and equipment, operational causes, etc.
During an \dword{odh} event, workers must leave the area and head towards the Ross or Yates shaft for evacuation.

\begin{figure}[htbp]
\centering
\includegraphics[width=0.95\textwidth]{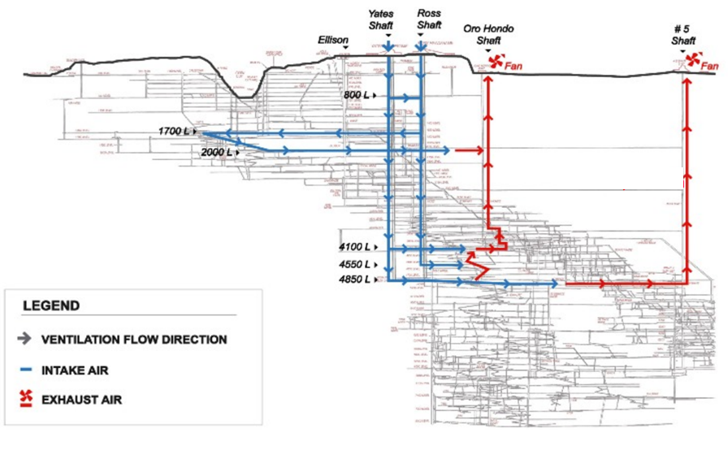} 
\caption{Homestake mine ventilation paths}
\label{fig:ventilation-paths}
\end{figure}

\section{Argon Gas Piping in the Ross Shaft}
\label{sec:cryo-cryosys-esh-gar-pipe}

An \dword{odh} assessment for the piping in the Ross Shaft has 
determined that if any of the pipes
for the cryogenics system were to rupture in the shaft, they would
only  
reduce the oxygen content at the \dword{4850l}  to 20.5\%, thus not reaching the level of \dword{odh} concern. (\dword{osha} defines an area as oxygen deficient if the percentage of oxygen is less than 19.5\% by volume.) The \dword{odh} mapping
of underground cavern area is given in Figure~\ref{fig:ODH-mapping-4850L}.

The facility is designed to draw the fresh air 
in through the Ross and Yates Shafts and exhaust the air 
out of the  \dword{4850l} through the Oro Hondo shaft (see Figure~\ref{fig:ventilation-paths}).
The loss of cavern ventilation for more than a few hours is a safety risk even in the absence of \dword{odh} conditions.

The response for unplanned loss of ventilation is evacuation.

\section{Ventilation in CUC and Detector Caverns}
\label{sec:cryo-cryosys-esh-ventil}

The Ross and Yates shafts supply fresh ventilation air to the \dword{cuc}, detector caverns, and \dword{4850l} drifts. 
 The fresh air enters the \dword{4850l} at which point the flow is split, with some air flowing through the detector caverns, 
 and some flowing through the \dword{cuc} before being 
exhausted out via the Oro Hondo
exhaust shaft.

\chapter{Quality Assurance and Quality Control}

\section{Quality Assurance}

The design, fabrication, installation and commissioning of the cryostat and the cryogenics systems will be performed by subcontractors. The Requests for Proposal will provide the quality requirements for the work to be performed. The bidders will be required to provide their \dword{qa}/\dword{qc} plans with the bid along with how they will meet the inspection and test requirements.  The same quality requirements will apply to the infrastructure, proximity, and internal cryogenics systems.

\subsection{Cryogenics Systems}

During the design phase the supplier will develop mechanical design drawings, mechanical stress, seismic, pressure drop and heat in-leak calculations. Electrical breaks will be identified. A risk analysis will be performed.  Materials will be identified during the design. The design will be verified and validated through a preliminary design review and a detailed final design review.

Prior to fabrication, a manufacturing readiness review will be performed. The welding process, welder qualifications, \dword{nde} personnel qualifications, process, inspection and test procedures will be reviewed. Material certificates for the welded process pipes, seamless process pipes, welded vacuum pipes, stainless steel fittings, bellows, filler material, and machined parts for the process will be provided for review. The control of traceability of material will be demonstrated and verified. The work processes will be reviewed including cleaning, welding, multilayer insulation, evacuation, \dword{pickling} and passivation, and packing and marking.

The following inspections and tests will be performed in process at the manufacturing facility:

\begin{itemize}
    \item visual weld inspection,
    \item radiographic tests (piping and vessels),
    \item longitudinal weld test (vessel),
    \item dye penetrant testing for socket welds that cannot be radiographed,
    \item helium leak tests for flexibles, uninsulated parts, and all spools,
    \item pressure test (piping and vessels),
    \item vacuum retention test,
    \item instruments test (valves, heaters, T-sensors, P-sensors),
    \item pressure safety valves (\dwords{psv}),
    \item bellows/flex test, and
    \item electrical breaks (cold cycle, pressure test, electrical resistance test, He leak test).
\end{itemize}

A factory acceptance test and final inspection will be performed prior to the equipment being released for shipment. The factory acceptance test will consist of:

\begin{itemize}
    \item dimensional inspection,
    \item pressure test, and a
    \item helium leak test.
\end{itemize}

Prior to installation and testing, installation review and test review meetings will be held at \dword{surf}. All equipment received at \dword{surf} will be receipt-inspected for proper documentation, any shipping damage, transport accelerometer verification, packing list, and leakage.

Installation inspection and testing will consist of:

\begin{itemize}
    \item visual weld inspection,
    \item radiographic testing of piping,
    \item pressure test including interfaces, and
    \item final inspection of hardware consisting of final position inspection, positioning and layout control, and interfaces control.
\end{itemize}

The final acceptance test will consist of a warm acceptance test and a cold acceptance test. 

The warm acceptance test will consist of the following:

\begin{itemize}
    \item pressure test, and a
    \item helium leak test.
\end{itemize}
 
The cold acceptance test will consist of the following:
\begin{itemize}
    \item check of documentation, safety devices, all interfaces  installed and ready for test;
    \item cold tests with \lntwo for two hours verifying there is no condensation on the surface, and a
    \item helium leak test.
\end{itemize}

Upon satisfactory completion of the final acceptance test, there will be a final acceptance release meeting to review test reports and documentation (final), operation manuals, and warranty certificate.

\subsection{Protego Valves}

The manufacturer of the Protego\textregistered{} valves shall perform a visual and dimensional check after production to verify that the valves meet the drawing requirements. All seam welds will be inspected by liquid penetrant examination. The valves will be pressure tested  (\SI{3.2}{mbarg}/30 minutes) to verify there is no leakage or deformation. A valve seat tightness test will be performed by the manufacturer to maximum $\SI{28.2}{cm^3/minute}$ at \SI{500}{mbar}. The completed valve will be helium leak-tested to ensure it does not exceed the maximum leakage rate of the specifications (\SI{1.0e-9}{mbar\liter/s} ). A final inspection will be performed prior to release for shipment consisting of a visual check, a check of marking, and release for shipment.  

Upon receipt at \dword{surf}, the valves will be inspected for shipping damage and against the Bill of Lading to ensure the shipment is complete.  Installation will be performed according to manufacturer's instructions. Installation welds will be inspected by liquid penetrant examination.

\subsection{Cryostat}

All equipment for the warm vessel and the cold vessel received at \dword{surf} will be receipt-inspected against the Bill of Lading, and for proper documentation and any shipping damage. All inspections and tests will be performed by qualified personnel using approved procedures.

The warm structure components will be fabricated using qualified personnel for assembly, welding and \dword{nde}. Steel components will be dimensionally inspected prior to welding. Welding will be performed by qualified welders and the welds will be inspected using liquid penetrant and ultrasonic examination based on the fabricator's quality plan. Final inspection verifying all parts will be performed prior to shipment to ensure that subassemblies 
are consistent with the applicable drawings. During installation at \dword{surf}, all welds will be inspected by \dword{nde} and bolting will be verified as meeting the specification.

The installation of the insulation will be verified through visual inspection of the bonding process. The secondary membrane installation will be inspected and tested through a vacuum box test and liquid penetrant testing of the welds. The primary membrane installation will be verified through vacuum box test, helium leak testing and liquid penetrant testing of welds. A global barrier test (vacuum leak test) will be performed on the primary and secondary membranes.

Prior to, during, and upon completion of the filling of the cryostat the following \dword{qc} steps are performed to verify the integrity of the structure during the filling process. Throughout the whole process, strain gauges will be used to measure the deformation of the structural elements, particularly after each pressure increase to validate the results of the \dword{fea} simulations.

\section{Cryostat QC Prior to Purge}

The following lists the steps of the \dword{qc} test to be performed prior to the start of the \dword{gar} purge. While this test does not stress the structure to the design conditions, it still provides an additional \dshort{qc} step in the validation process of the warm vessel, including the roof of the cryostat with all the feedthroughs.

\begin{itemize}
    \item The cryostat \dwords{psv} will be locked out for the duration of the test to avoid opening at \SI{250}{mbarg}.
   \item 
    The cryostat will be pressurized to \SI{250}{mbarg}  to verify the structure for the first steps: \dword{gar} purge, \cooldown and \dword{lar} fill.
    \item 
    The automatic vent valve will be manually reset so that it opens fully when the pressure reaches slightly above \SI{250}{mbarg} (e.g., \SI{270}{mbarg}).   \item A temporary \dword{psv} will be installed in the \dword{gar} supply line for the duration of this test.
    \item Starting from atmospheric, the pressure will be 
    incremented at a rate of about \SI{50}{mbar/hour}.
    \item After every pressure 
    increment, the structure and the pressure stability will be observed for one hour without intervention. 
    Should the pressure decrease, 
    the incrementing will stop and investigation will take place.  If 
    it is stable, pressure increments will continue. 
    \item Strain gauges and displacement sensors will be used to measure the strains and the deformation of the structural elements throughout the 
    process, particularly after each pressure 
    increment, to validate the results of the \dword{fea} simulations. If the readings deviate significantly from the expected values, the incrementing will stop and investigation will take place 
    until the deviation is 
    evaluated. 
    \item During this 
    test, the only source of pressure is the \dword{gar} used to pressurize the cryostat. The supply \dshort{gar} will be regulated as to not exceed the desired pressure at each \SI{50}{mbarg} increment. 
    \item Once  \SI{250}{mbarg} is reached, the cryostat is qualified for \SI{250}{mbarg} of maximum gas pressure and will be depressurised 
    (at about \SI{50}{mbar/hour}) down to \SI{50}{mbarg}.
    \item Once the cryostat pressure falls below \SI{150}{mbarg}, 
    automatic control of the vent valve is re-enabled and reset to \SI{150}{mbarg}.
    \item During 
    the commissioning process, the strain gauge and displacement sensor measurements will be analyzed at least once per day. 
    Any abnormal behavior will be evaluated.
    \item Once the test is completed the cryostat \dwords{psv} are unlocked.
     
\end{itemize}

\section{Cryostat QC During Cryostat Fill}

Strain gauges will measure the deformation of the structural elements throughout the fill process 
to validate the results of the \dword{fea} simulations as the \dword{lar} level  rises. 
Maintaining a stable pressure of about \SI{50}{mbarg}, the argon flow will be decreased at every meter of fill level long enough 
to allow reading of the strain gauges. 
If the readings deviate from the expected values, an investigation will take place and the fill will stop until the deviation is 
evaluated. 
The acceptable deviations will be defined prior to the test.

Every 2.0\,m of liquid rise (at 2.0\,m, 4.0\,m, and so on through 12.0\,m, and again at the nominal \dword{lar} level), i.e., seven times, the gas pressure will be increased in steps of \SI{50}{mbar} up to \SI{200}{mbarg}, then brought back down to \SI{50}{mbarg}. The \SI{200}{mbarg} pressure will be held for one hour in order to check the \dword{fea} calculations and examine the most stressed areas of the warm structure. This process will take about seven hours at each level,  
    adding a total of 49 hours (seven hours $\times$ seven 
    levels) to the fill time, assuming nothing unexpected happens during the test and no unexpected values are observed. Due to the settling of the structure, deformations and loads are likely to differ from those of the \dword{fea} simulations for levels of \dword{lar} up to about \SI{4}{m}.

\section{Cryostat QC After Cryostat Fill}

Once the cryostat is full and stable at the operating pressure of \SI{50}{mbarg}, and risk assessment has been completed, 
the \dword{gar} pressure will be increased to test the system close to its maximum design load: 
full of \dword{lar} and at \SI{200}{mbarg} of \dshort{gar} pressure. 
The procedure has following \dword{qc} steps: 

    \begin{itemize}
    \item The automatic vent valve will be manually reset so that it opens fully when the pressure reaches 
    slightly above \SI{200}{mbarg} (e.g., \SI{230}{mbarg}).
    \item The cryostat \dwords{psv} will remain operational at the nominal set pressure of \SI{250}{mbarg}.
    \item Starting at 
    \SI{50}{mbarg}, the pressure will be 
    increased at a rate of about \SI{50}{mbar/hour}.
    \item Upon reaching \SI{200}{mbarg} the structure will be observed for one hour without intervention, and the stability of the pressure will be monitored. Should the pressure decrease, 
     the incrementing will stop and investigation will take place.  If it is stable, pressure increments will continue. 
   
    \item Strain gauges and displacement sensors will measure the deformation of the structural elements 
    throughout the process, particularly after each pressure increase, to validate the results of the \dword{fea} simulations. If the readings deviate from the expected values, 
     the incrementing will stop and investigation will take place. The acceptable deviations will be defined prior to the test. 
   
    \item During this step the 
    argon contained inside the cryostat is the only source of pressurization.  The condensers will increase and regulate the 
    pressure at the test values  (100, 150, \SI{200}{mbarg}). If the pressure reaches \SI{230}{mbarg} the 
    vent valve will open automatically and release \dword{gar} to the exhaust duct. (This line is sized to release the \dword{gar} overpressure generated by the cryostat boil-off in case 
    refrigeration is lost.) 
    \item Once the \dword{qc} at \SI{200}{mbarg} has been performed, the cryostat will be 
    depressurized at the rate of about \SI{50}{mbar/hour} down to the operating pressure of \SI{50}{mbarg}. 
    \item Once the pressure drops below \SI{200}{mbarg}, the automatic vent  valve will be reset manually to \SI{230}{mbarg}. 
    \item Once the pressure drops below \SI{150}{mbarg}, 
    the vent valve will be 
    reset to open automatically at \SI{150}{mbarg}.
\end{itemize}

The full test (pressurization, hold, depressurization) is expected to take approximately eight hours.
This test provides the final \dword{qc} step in the validation process of the warm and cold vessels, including the roof of the cryostat with all the feedthroughs.

\cleardoublepage

\appendix

\chapter{Refrigeration Load Scenarios}
\label{ch:refrigeration-load-scenarios}

To determine the optimal plant capacity and number of plants 
required, fourteen scenarios were forecast for the \lntwo refrigeration
loads and plant capacity.  Those scenarios are described below and 
a summary is given in Figure~\ref{fig:refrig-load-table}. 
                          
The conclusion points to the requirement of four \SI{100}{kW} plants. Each of 
these plants can achieve a 20\% turn up or turn down. Scenarios 1, 4, 
A and D impose the most severe requirements. In these scenarios, all 
plants available will be required to run at the maximum duty cycle 
to cool down and fill a cryostat, while maintaining purity for
cryostat(s) filled and purified earlier. These scenarios will
also require frequent filter regeneration.

\begin{description}
\item[Scenario 1]
The initial operation will be the purging, cooling and filling of the 
first cryostat, condensing gaseous argon in the cavern by heat exchange 
via the recondensers. The surface and cavern LAr and  \lntwo dewars 
will be operational and the cooling load for the dewars will come 
directly from the refrigeration plant. The cavern pipework and vessels 
will be cold, the LAr in the cryostat will be circulating at high flow 
rate through the purification plant, and the cryostat will be cold.
The cryostat cool-down rate is constrained by three variables: 1) The 
size of the piping from the surface to bottom of Ross shaft, 2) The 
size of the \lntwo refrigeration units, and 3) the cooling power 
available via the recondensers.  All three variables have been 
matched for the physical constraints of a 40 kt module at 4850L 
using the Ross shaft. The refrigerators and condensers have been 
sized to accommodate the long-term refrigeration load associated 
with the cryostats.  As the LAr is circulated to achieve the 
operational purity the filtration plant will need to be regularly 
regenerated. This will mean that the associated refrigeration 
load will normally be present.

\item[Scenario 2]
Once the first cryostat is filled with LAr, the cool-down load will 
reduce to zero and the cryogenic plant will run for several months 
purifying the LAr inventory.

\item[Scenario 3]
When the LAr in the cryostat reaches the required purity level, the 
circulation flow rate will be reduced and the detector electronics 
will be turned on.  At this stage the recondenser refrigeration 
load falls such that only one recondenser is required and the 
rest of units can operate as spare units.

\item[Scenario 4] The first cryostat continues to operate in normal
experimental mode 
while  the second cryostat is being purged, cooled down and 
filled with LAr. Again a very large burden is placed on the 
recondensers due to the gas condensation and rate of liquefaction.

\item[Scenario 5] The second cryostat is full and LAr is circulated 
at high flow rate through the purification plant. The first cryostat
continues to operate as normally.

\item[Scenario 6]
Both cryostats are operating in normal experimental mode. A spare 
recondenser is available on each cryostat to facilitate maintenance.

\item[Scenario 7]
It is assumed that a total failure of the refrigeration plant has 
occurred. All noncritical heat sources are isolated and liquid 
nitrogen from the \lntwo vessels in the central utility cavern 
is utilized to recondense the inventory of high purity LAr. Nitrogen 
refrigeration must be reestablished before the liquid nitrogen reservoir 
is exhausted or the high purity argon will need to be vented. In the 
locked-down state, the recirculation pumps and the purification 
plants are shut down.

\item[Scenario A]
The first and second cryostats continue to operate in normal
experimental mode while  the third cryostat is being purged, 
cooled down and filled with LAr. Again a very large burden 
is placed on the recondensers due to the gas condensation 
and rate of liquefaction.

\item[Scenario B]
The third cryostat is full and LAr is circulated at high 
flow rate through the purification plant. The first and 
second cryostats continue to operate as normally.

\item[Scenario C]
Three cryostats are operating in normal experimental mode. A spare 
recondenser is available on each cryostat to facilitate maintenance.

\item[Scenario D]
The three cryostats continue to operate in normal experimental
mode while the fourth cryostat is being purged, cooled down 
and filled with LAr. Again a very large burden is placed 
on the recondensers due to the gas condensation 
and rate of liquefaction.

\item[Scenario E]
The fourth cryostat is full and LAr is circulated at high 
flow rate through the purification plant. The three cryostats 
previously filled continue to operate as normally.

\item[Scenario F]
All four cryostats are operating in normal experimental mode. A spare 
recondenser is available on each cryostat to facilitate maintenance.

\item[Scenario G]
This is the same condition as Scenario 7, but now all four cryostats
are in the LAr inventory protection mode.

\end{description}

\begin{dunefigure}[Refrigeration Loads] 
{fig:refrig-load-table}
{Refrigeration Loads. Note that the heat load contribution of the \lntwo storage and \lntwo piping is assumed by the Nitrogen Refrigeration System.}
\includegraphics[width=1.0\textwidth]{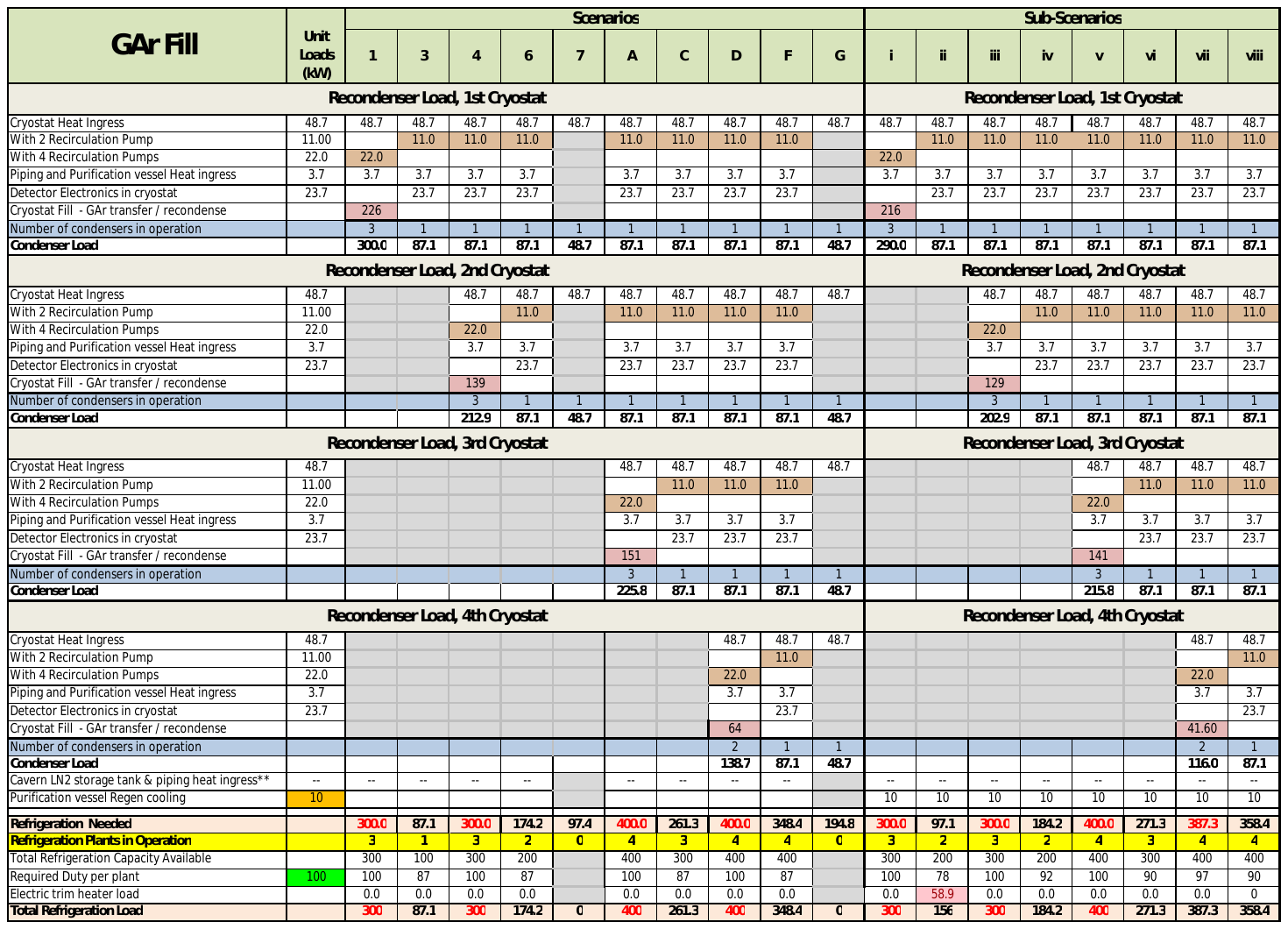}
\end{dunefigure}

\cleardoublepage

\cleardoublepage
\printglossaries

\cleardoublepage
\cleardoublepage
\renewcommand{\bibname}{References}
\bibliographystyle{utphys} 
\bibliography{common/tdr-citedb}

\providecommand{\href}[2]{#2}\begingroup\raggedright\begin{thebibliography}{10}

\bibitem{EDMS2838447}
DOE, ``{Letter regarding P5 plan for upgrading beyond Phase I}.''
\newblock \url{https://edms.cern.ch/document/2838447}.

\bibitem{lar-fd-req}
{LBNF/DUNE-US Project}, ``{LAr-FD Requirements spreadsheets}.''.
  \url{https://edms.cern.ch/document/2391830}.

\bibitem{bib:docdb117}
{LBNF/DUNE-US Project}, ``{LBNF/DUNE-US Project Management Plan},'' Tech. Rep.
  117, 2022.
\newblock Available upon request.

\bibitem{Adams_2020}
D.~Adams {\em et~al.}, ``Design and performance of a 35-ton liquid argon time
  projection chamber as a prototype for future very large detectors,''
  \href{http://dx.doi.org/10.1088/1748-0221/15/03/p03035}{{\em Journal of
  Instrumentation} {\bfseries 15} no.~03, (Mar, 2020) P03035--P03035}.
  \url{https://doi.org/10.1088%2F1748-0221%2F15%2F03%2Fp03035}.

\bibitem{adams2020protodunesp}
D.~Adams {\em et~al.}, ``The {ProtoDUNE}-{SP} {LArTPC} electronics production,
  commissioning, and performance,''
  \href{http://dx.doi.org/10.1088/1748-0221/15/06/p06017}{{\em Journal of
  Instrumentation} {\bfseries 15} no.~06, (Jun, 2020) P06017--P06017}.
  \url{https://doi.org/10.1088%2F1748-0221%2F15%2F06%2Fp06017}.

\bibitem{EDMS2826471}
LBNF/DUNE, ``{LBNF Cryostat FNAL Cryo Safety Panel Review April 2020}.''
\newblock \url{https://edms.cern.ch/document/2826471}.

\bibitem{EDMS2211532}
LBNF/DUNE, ``{Piping and Instrumentation Diagrams (P\&ID) }.''
\newblock \url{https://edms.cern.ch/document/2211532}.

\bibitem{EDMS2775007}
LBNF/DUNE, ``{PIDs for Cryostat 1 Systems}.''
\newblock \url{https://edms.cern.ch/document/2775007}.

\bibitem{EDMS2775008}
LBNF/DUNE, ``{PIDs for Cryostat 2 Systems}.''
\newblock \url{https://edms.cern.ch/document/2775008}.

\bibitem{EDMS2775002}
LBNF/DUNE, ``{PIDs for Cryostat Support Equipment}.''
\newblock \url{https://edms.cern.ch/document/2775002}.

\bibitem{EDMS2211569}
LBNF/DUNE, ``{Receiving Facilities Model}.''
\newblock \url{https://edms.cern.ch/document/2211569}.

\bibitem{EDMS2447338}
LBNF/DUNE, ``{Receiving Facilities Layout}.''
\newblock \url{https://edms.cern.ch/document/2447338}.

\bibitem{EDMS2211567}
LBNF/DUNE, ``{Receiving Facilities Simulations}.''
\newblock \url{https://edms.cern.ch/document/2211567}.

\bibitem{EDMS2705540}
LBNF/DUNE, ``{Nitrogen System Statement of Work}.''
\newblock \url{https://edms.cern.ch/document/2705540}.

\bibitem{EDMS2810467}
LBNF/DUNE, ``{Layout of Piping for Argon and Nitrogen Distribution
  Underground}.''
\newblock \url{https://edms.cern.ch/document/2810467}.

\bibitem{EDMS0000196907}
LBNF/DUNE, ``{Argon distribution and vents}.''
\newblock \url{https://edms.cern.ch/nav/CERN-0000196907}.

\bibitem{EDMS2810641}
LBNF/DUNE, ``{Argon Distribution Statement of Work}.''
\newblock \url{https://edms.cern.ch/document/2810641}.

\bibitem{EDMS2852844}
LBNF/DUNE, ``{Statement of Work for Regeneration System 1}.''
\newblock \url{https://edms.cern.ch/document/2852844}.

\bibitem{EDMS2892643}
LBNF/DUNE, ``{Statement of Work Regeneration System 2}.''
\newblock \url{https://edms.cern.ch/document/2892643}.

\bibitem{EDMS2884649}
LBNF/DUNE, ``{Statement of Work Liquid Argon Circulation Pumps Package for the
  Cryostat 1}.''
\newblock \url{https://edms.cern.ch/document/2884649}.

\bibitem{EDMS2810432}
LBNF/DUNE, ``{Liquid Argon Purification 1 Statement of Work}.''
\newblock \url{https://edms.cern.ch/document/2810432}.

\bibitem{EDMS2810433}
LBNF/DUNE, ``{Liquid Argon Purification 2 Statement of Work}.''
\newblock \url{https://edms.cern.ch/document/2810433}.

\bibitem{EDMS2211972}
LBNF/DUNE, ``{Internal Cryogenics Reference Design}.''
\newblock \url{https://edms.cern.ch/document/2211972}.

\bibitem{EDMS2154410}
E.~Voirin, ``{DUNE Cool Down - CFD Simulation - LAr Sprayers},'' tech. rep.,
  Fermilab, 2020.
\newblock \url{https://edms.cern.ch/document/2154410}.

\bibitem{EDMS_jrci_rpt}
J.~Campbell and Associates, ``{Report to Fermilab on RFI of May 2021 for LBNF
  Liquid Argon Supply},'' 2023.

\bibitem{EDMS2818818}
LBNF/DUNE, ``{LAr Acquisition Plan}.''
\newblock \url{https://edms.cern.ch/document/2818818}.

\bibitem{EDMS2211568}
LBNF/DUNE, ``{Surface Argon Receiving Facilities Performance Specifications}.''
\newblock \url{https://edms.cern.ch/document/2211568}.

\bibitem{EDMS2778502}
LBNF/DUNE, ``{Distribution Lines with Tie Points and Interfaces}.''
\newblock \url{https://edms.cern.ch/document/2778502}.

\bibitem{EDMS2254565}
LBNF/DUNE, ``{3D model Cryogenics in CUC, Interconnecting Piping, Circulation
  pumps in DC}.''
\newblock \url{https://edms.cern.ch/document/2254565}.

\bibitem{EDMS2248818}
LBNF/DUNE, ``{LBNF Valve and Instrumentation List }.''
\newblock \url{https://edms.cern.ch/document/2248818}.

\bibitem{EDMS2838450}
LBNF/DUNE, ``{FD1, FD2, and CUC Cryogenics - Navisworks format only}.''
\newblock \url{https://edms.cern.ch/document/2838450}.

\bibitem{EDMS2154314}
LBNF/DUNE, ``{LBNF Detector Cavern Relief Header Evaluation}.''
\newblock \url{https://edms.cern.ch/document/2154314}.

\bibitem{EDMS2519457}
LBNF/DUNE, ``{PSV Calculations}.''
\newblock \url{https://edms.cern.ch/document/2519457/1}.

\bibitem{EDMS2771986}
LBNF/DUNE, ``{GN2 for Cryostat Insulation Scope and Notes}.''
\newblock \url{https://edms.cern.ch/document/2771986}.

\bibitem{EDMS2771977}
LBNF/DUNE, ``{GN2 Cryostat Insulation Purge sizing}.''
\newblock \url{https://edms.cern.ch/document/2771977}.

\bibitem{EDMS2775007pid}
LBNF/DUNE, ``{Cryostat Piping and Instrumentation Diagram}.''
\newblock \url{https://edms.cern.ch/file/2775007/1/17._F10028334.17_Cryo.pdf}.

\bibitem{EDMS2227399}
LBNF/DUNE, ``{GN2 supply for cryostat insulation purge}.''
\newblock \url{https://edms.cern.ch/document/2227399}.

\bibitem{EDMS2771736}
LBNF/DUNE, ``{Xenon injection station scope and notes}.''
\newblock \url{https://edms.cern.ch/document/2771736}.

\bibitem{EDMS2477369}
LBNF/DUNE, ``{Xenon doping station PFD}.''
\newblock \url{https://edms.cern.ch/document/2477369/}.

\bibitem{EDMS2477451}
LBNF/DUNE, ``{Xenon amount and cost (link requires authorization)}.''
\newblock \url{https://edms.cern.ch/file/2477451/3/Xenon_amount_and_cost.xlsx}.

\bibitem{EDMS2826472}
E.~Voirin, ``{Argon Piston Purge of ProtoDUNE - CFD Simulation},'' project doc,
  2015.
\newblock \url{https://edms.cern.ch/document/2826472/1}.

\bibitem{Montanari:2013/06/13aqa}
{\bfseries LBNE} Collaboration, D.~Montanari {\em et~al.}, ``{First Scientific
  Application of the Membrane Cryostat technology.},''.
\url{http://lss.fnal.gov/archive/2013/conf/fermilab-conf-13-076-ppd.pdf}.

\bibitem{EDMS_2154410}
D.~Montanari and J.~Freitag, ``{DUNE Cool Down - CFD Simulation}.''
\newblock \url{https://edms.cern.ch/document/2154410}.

\bibitem{EDMS2785497}
LBNF/DUNE, ``{Avail time for Cryostat pressure recovery from loss of power}.''
\newblock \url{https://edms.cern.ch/document/2785497}.

\bibitem{EDMS2364392}
LBNF/DUNE, ``{Cost Estimates for pressure control systems in the detector
  cavern}.''
\newblock \url{https://edms.cern.ch/document/2364392}.

\bibitem{EDMS2153810}
LBNF/DUNE, ``{Cryogenics - Cryostat relief valve information}.''
\newblock \url{https://edms.cern.ch/document/2153810}.

\bibitem{Abi:2020oxb}
{\bfseries DUNE} Collaboration, B.~Abi {\em et~al.}, ``{Deep Underground
  Neutrino Experiment (DUNE), Far Detector Technical Design Report, Volume III.
  DUNE far detector technical coordination},''
  \href{http://dx.doi.org/10.1088/1748-0221/15/08/T08009}{{\em JINST}
  {\bfseries 15} no.~08, (2020) T08009},
  \href{http://arxiv.org/abs/2002.03008}{{\ttfamily arXiv:2002.03008
  [physics.ins-det]}}.

\bibitem{EDMS2154307}
LBNF/DUNE, ``{Cryogenic Control System}.''
\newblock \url{https://edms.cern.ch/document/2154307}.

\bibitem{EDMS2156486}
LBNF/DUNE, ``{Process Controls Electrical and Mechanical Drawings}.''
\newblock \url{https://edms.cern.ch/document/2156486}.

\bibitem{EDMS2154321}
LBNF/DUNE, ``{ODH System}.''
\newblock \url{https://edms.cern.ch/document/2154321}.

\bibitem{2108.01902}
{DUNE Collaboration} {\em et~al.}, ``{Design, construction and operation of the
  ProtoDUNE-SP Liquid Argon TPC},'' 2021.
\newblock \url{https://arxiv.org/abs/2108.01902}.

\bibitem{feshm}
Fermilab, ``{Fermilab ES\&H Manual}.''
  \url{https://publicdocs.fnal.gov/cgi-bin/ListBy?topicid=49}.

\end{thebibliography}\endgroup

\end{document}